\documentclass[11pt]{article}
\usepackage{nicefrac}
\usepackage[utf8]{inputenc}
\usepackage{amssymb}
\usepackage{amsmath}

\usepackage{cite}

\usepackage[left=1.25in,right=1.25in,top=1.0in,bottom=2.10in]{geometry}

\def\SBigg#1{{\hbox{$\left#1\vbox to19.5\p@{}\right.\n@space$}}}

\newcommand*\laplace{\mathop{}\!\mathbin\bigtriangleup}

\allowdisplaybreaks 

\begin{document}

\title{Many-particle quantum hydrodynamics: \\ exact equations and pressure tensors} 
\author{Klaus Renziehausen$^*$
, Ingo Barth$^*$
        \\ \\
\textit{\small{Max Planck Institute of Microstructure Physics, Weinberg 2, 06120 Halle (Saale), Germany}} \\
\small{$^*$E-mail: ksrenzie@mpi-halle.mpg.de; barth@mpi-halle.mpg.de}}

\date{}
\maketitle 

\begin{abstract}%
\noindent In the first part of this paper, the many-particle quantum hydrodynamics (MPQHD) equations for a system containing many particles of different sorts are derived exactly from the many-particle Schrödinger equation. It includes the derivation of the many-particle continuity equations (MPCE), many-particle Ehrenfest equations of motion (MPEEM), and many-particle quantum Cauchy equations (MPQCE) for any of the different particle sorts and for the total particle ensemble. The new point in our ana\-lysis is that we consider a set of arbitrary particles of different sorts in the system. In MPQCEs, there appears a quantity called pressure tensor. In the second part of this paper, we analyze two versions of this tensor in depth -- the Wyatt pressure tensor and the Kuzmenkov pressure tensor. 
There are different versions because there is a gauge freedom for the pressure tensor similar to that for potentials. We find that the interpretation of all quantities contributing to the Wyatt pressure tensor is understandable but for the Kuzmenkov tensor, it is difficult. Furthermore, the transformation from Cartesian coordinates to cylindrical coordinates for the Wyatt tensor can be done in a clear way, but for the Kuzmenkov tensor, it is rather cumbersome.   
\end{abstract}

\section{Introduction}
Quantum hydrodynamics (QHD) is a concept that was developed already 1926 by Madelung \cite{Madelung_1926_1,Madelung_1926_2}. He transformed the Schrödinger equation for a single particle into the cor\-res\-pon\-ding QHD equations. It was further developed by Bohm in 1952 \cite{Bohm_1952_1,Bohm_1952_2}. The motivation to name this field QHD is that by applying it one finds differential equations with a similar form like well-known differential equations in classical hydrodynamics, like the continuity equation or the Navier-Stokes equation  $\textnormal{(\hspace{-0.15 cm} \cite{Landau_2000}, p. 2 and 45)}$. Such equations related to QHD were analyzed for systems where the wave function was a single or quasi-single particle wave function in several papers \cite{Madelung_1926_1,Madelung_1926_2, Bohm_1952_1,Bohm_1952_2, Takabaysi_1952,Takabaysi_1955,Bohm_1955,Harvey_1966,Bialynicki-Birula_1971,Rosen_1974,Deb_1978,Sonego_1991,Wallstrom_1994,Recami_1998,Wyatt_2000_a,Wyatt_2000_b,Wyatt_2002,Lopreore_1999,Lopreore_2000,Lopreore_2002,Koide_2013}. First ideas for many-particle quantum hydrodynamics (MPQHD) were already discussed by Bohm \cite{Bohm_1952_1}. In addition, MPQHD was analyzed using the energy-density functional method \cite{Deb_1979_1,Holzwarth_1978,Holzwarth_1979}, a time-dependent Hartee-Fock ansatz \cite{Himi_1984,Wong_1975,Wong_1976}, and a non-stationary non-linear Schrödinger equation ansatz for quantum plasma physics \cite{Shukla_2010,Shukla_2011}. Futhermore, in 1999, Kuzmenkov and Maksimov developed a method where equations for mass, momentum, and energy balance for MPQHD were derived for exact non-relativistic many-particle wave functions without regarding the particle spin  \cite{Kuzmenkov_1999}. Later, the method was further developed by Kuzmenkov and his colleagues to investigate spin effects \cite{Kuzmenkov_2001_1,Andreev_2007} and Bose-Einstein-Condensates \cite{Andreev_2008}. Moreover, applications of this method were discussed \mbox{e.\ g.} related to electrons in graphene \cite{Andreev_2012_a} and plasma effects \cite{Andreev_2012_b, Andreev_2013, Andreev_2014a, Andreev_2014b, Trukhanova_2013,Trukhanova_2015}. In particular, in \cite{Andreev_2014a} it is briefly mentioned how to apply MPQHD when several sorts of particles are present, and the MPQHD equations stated in \cite{Andreev_2014a, Andreev_2014b, Trukhanova_2013,Trukhanova_2015} describe the special case of two particle sorts in a plasma. In  \cite{Andreev_2014a, Andreev_2014b,Trukhanova_2015}, these two sorts are electrons and a single ion sort, and in \cite{Trukhanova_2013}, these two sorts are  electrons and positrons. \newline 
In Sec.\ \ref{Basic Physics} of this paper, we are aiming at developing further the methods described in \cite{Kuzmenkov_1999,Andreev_2014a} by deriving rigorously  the MPQHD equations for the case that the particle ensemble includes several sorts of particles -- in particular, in our general ansatz we do not restrict the number of the particle sorts and we do not specify the types of the particle sorts. As we want to focus there on the main points, we neglect spin effects in our calculations, and at the end of Sec.\ \ref{Basic Physics} we just briefly mention the effects of external electromagnectic fields. In addition, in our calculations in Sec.\ \ref{Basic Physics}, we mention a quantity called the pressure tensor. One can find different versions of this pressure tensor in literature \cite{Wong_1976,Deb_1979_1,Sonego_1991,Kuzmenkov_1999}; an explanation for this variety can be found in \cite{Sonego_1991}. In Sec.\ \ref{Tensor transformations}, we pick up the pressure tensor version given in \cite{Kuzmenkov_1999} and name it the ``Kusmenkov pressure tensor''. In addition, the discussion about QHD in \cite{Wyatt_2005}, p.\ 30f., is our motivation to introduce another pressure tensor version called the ``Wyatt pressure tensor''. We analyze how these tensors can be interpreted physically. Moreover, we discuss for which of these two tensors a transformation from Cartesian coordinates into cylindrical coordinates can be done more easily. 
\section{Basic physics of exact MPQHD} \label{Basic Physics}
Here, the basic physics for many-particle quantum hydrodynamics (MPQHD) is analyzed: A particle ensemble consisting of different particle sorts is examined, and a many-particle continuity equation (MPCE), a many-particle Ehrenfest equation of motion (MPEEM) and a many-particle quantum Cauchy equation (MPQCE) is derived each for the total ensemble of particles and for a particular sort of particles. The MPCEs are equations related to the mass conservation, the MPEEMs are equations related to the time evolution of mass flux densities and the MPQCEs are equations related to the momentum balance. For these derivations, several quantities have to be defined first. 
\subsection{Definitions}
We assume that there are ${N_S}$ sorts of particles, and that $\textnormal{A,B,C}$ stands for any number $\in \{1,2,\ldots, {N_S}\}$ which is related to one sort of particles. 
For brevity, we denote any $A$-th sort of particles also as the sort of particles $\textnormal{A}$ or just as the sort $\textnormal{A}$.  The $N(\textnormal{A})$ particles of any sort $\textnormal{A}$ shall be indistinguishable. In particular, each particle of the sort $\textnormal{A}$ has the same mass $m_\textnormal{A}$ and the same charge $e_\textnormal{A}$. In this publication, spin effects were not considered, for a more general analysis with spin effects one would have to consider that each particle of the sort $\textnormal{A}$ has a spin $s_\textnormal{A}$.\newline 
The ansatz to treat particles of the same sort as indistinguishable does not diminish the generality of the following analysis for this reason: If there are sorts of particles where the individual particles can be distinguished, this can be implemented in the calculations below by treating each of these particles as a whole sort of particles of its own. In this sense, the following analysis is valid both for distinguishable and for indistinguishable particles. \newline 
Moreover, all the subsequent analysis in this paper is correct for these three cases: \mbox{1.\ All} particles are fermions. 2.\ All particles are bosons. 3.\ The particles of some sorts are fermions, and the particles of the remaining sorts are bosons. We mention that in  \cite{Kuzmenkov_1999}, one can find a discussion where the question is analyzed how the property of the particles being either bosons or fermions influences quantum hydrodynamics. Hereby, the many-particle wave function is decomposed within the Hartee-Fock-method as a sum over many-particle eigenfunctions in the occupation number space. As a result, for such a decomposition of the many-particle wave function one needs to make a distinction of the cases that the particles are bosons or fermions -- but since we will not make a decomposition of the many-particle wave function into its eigenfunctions within the analysis in our paper, all equations in our paper are valid both for bosons and for fermions. \newline 
The position vector of the $i$-th particle of the sort $\textnormal{A}$ (so $i \in {1,2,\ldots, N(\textnormal{A})}$) is $\vec q_i^{\hspace{0.05 cm}\textnormal{A}}$; this particle is called $(\textnormal{A},i)$-particle. Moreover,
\begin{eqnarray}
\vec Q &=&  \left ( \vec q_1^{\hspace{0.05 cm}1}, \vec q_2^{\hspace{0.05 cm}1}, \ldots, \vec q_{N(1)}^{\hspace{0.05 cm}1}, \vec q_1^{\hspace{0.05 cm}2}, \ldots,  \vec q_{N(2)}^{\hspace{0.05 cm}2}, \ldots, \vec q_{1}^{\hspace{0.05 cm}N_S}, \ldots, \vec q_{N({N_S})}^{\hspace{0.05 cm}{N_S}} \right) \label{Particle Coordinate Set}
\end{eqnarray}
is the complete set of particle coordinates, and $\Psi(\vec Q,t)$ is the normalized total wave function of the system. \newline  
The particles shall be exposed only to forces arising from a real-valued two-particle potential (e. \hspace{-0.3 cm} g.\ a Coulomb potential)
\begin{eqnarray}
V_{ij}^{\textnormal{AB}} = \left \{ \begin{array}{cccc} V^{\textnormal{AB}}(|\vec q_i^{\hspace{0.05cm} \textnormal{A}} - \vec q_j^{\hspace{0.05cm}  \textnormal{B}}|) & \textnormal{for} \; (i \neq j)   &  \textnormal{or} & (\textnormal{A} \neq \textnormal{B}) \\ 
0 &  \textnormal{for} \; (i=j) &  \textnormal{and} &  (\textnormal{A}   =  \textnormal{B})                                \end{array} \right. , \label{Potential} 
\end{eqnarray}
where we regard that the two-particle potential does not couple a particle with itself by the distinction of cases in the equation above. \newline 
This two-particle potential has the symmetry properties 
\begin{eqnarray}
V_{ij}^{\textnormal{AB}} &=&  V_{ji}^{\textnormal{BA}}, \label{V_symmetry} \\
\nabla_i^\textnormal{A} V_{ij}^{\textnormal{AB}} &=& - \nabla_{\hspace{-0.05 cm}j}^\textnormal{B} V_{ji}^{\textnormal{BA}}, \label{V_antisymmetry}
\end{eqnarray}
where $\nabla_i^\textnormal{A}$ is the Nabla operator relative to the coordinate  $\vec q_i^{\hspace{0.05cm}\textnormal{A}}$. Later, we will explain what happens if external fields are present. \newline
The canonical momentum operator $\hat {\vec p}_i^{\hspace{0.05 cm}\textnormal{A}}$ relative to the coordinate $\vec q_i^{\hspace{0.05cm} \textnormal{A}}$ is 
\begin{eqnarray}
\hat {\vec p}_i^{\hspace{0.05 cm} \textnormal{A}} &=& \frac{\hbar}{\mathrm{i}} \nabla_i^\textnormal{A}. \label{momentum operator}
\end{eqnarray}
Then, the Schr\"odinger equation of the system is given by
\begin{eqnarray}
\mathrm{i} \hbar \; \frac{\partial \Psi (\vec Q,t)}{\partial t} &=& \hat H (\vec Q) \; \Psi (\vec Q,t) \label{Schroedinger equation}
\end{eqnarray}
with a Hamiltonian
\begin{eqnarray}
\hat H (\vec Q) &=&  \sum_{\textnormal{A} = 1}^{N_S} \sum_{i = 1}^{N(\textnormal{A})} \frac {\left(\hat {\vec p}_i^{\hspace{0.05 cm} \textnormal{A}}\right)^2}{2 m_\textnormal{A}} + \frac{1}{2} \sum_{\textnormal{A} = 1}^{{N_S}} \sum_{i = 1}^{N(\textnormal{A})} \sum_{\textnormal{B} = 1}^{{N_S}} \sum_{j=1}^{N(\textnormal{B})} V_{ij}^{\textnormal{A}\textnormal{B}}. \label{Hamiltion operator with momentum operator}  
\end{eqnarray}
The next quantity we define is the volume element $\textnormal {d} \vec Q$ for all particles; it is given by:
\begin{eqnarray}
\textnormal {d} \vec Q &=& \prod_{\textnormal{A}=1}^{N_S} \left( \prod_{i=1}^{N(\textnormal{A})} \textnormal {d} \vec q_i^{\hspace{0.05 cm}\textnormal{A}} \right). \label{Total Volume Element}
\end{eqnarray} 
The volume element $\textnormal {d} \vec Q_{i}^{\textnormal{A}}$ for all coordinates except for coordinate $\vec q_i^{\hspace{0.05 cm}\textnormal{A}}$ is then defined by: 
\begin{eqnarray}
\textnormal {d} \vec Q_{i}^{\textnormal{A}} &=& \frac{\textnormal {d} \vec Q}{\textnormal {d} \vec q_i^{\hspace{0.05 cm}\textnormal{A}}}. \label{Restricted Total Volume Element}
\end{eqnarray}
Note that ${d} \vec q_i^{\hspace{0.05 cm}\textnormal{A}}$ is a volume element and not a vector, so, its appearance in the denominator is correct. We now define the total particle density $D(\vec Q,t)$ by 
\begin{eqnarray}
D(\vec Q ,t) =  \left |\Psi(\vec Q,t) \right |^2. \label{Def_total_particle_density}
\end{eqnarray} 
For the case of a single particle, Eqn.\ (\ref{Def_total_particle_density}) is equivalent to the equation for the particle density in a single particle system in quantum mechanics textbooks (\hspace{-0.15 cm} \cite{Greiner_2001}, p.\ 38f.\ and \cite{Gustafson_2011}, p.\ 4). \newline
Using the definitions above and the indistinguishability of the particles of each sort, we introduce the total one-particle mass density $\rho_m^{\textnormal{tot}}(\vec q,t)$:
\begin{eqnarray}
\rho_m^{\textnormal{tot}}(\vec q,t) &=& \sum_{\textnormal{A}=1}^{N_S} m_\textnormal{A} \sum_{i=1}^{N(\textnormal{A})} \int \textnormal{d} \vec Q \; \delta (\vec q - \vec q_i^{\hspace{0.05 cm}\textnormal{A}} ) \; D(\vec Q,t) \\
                                    &=& \sum_{\textnormal{A}=1}^{N_S} N(\textnormal{A}) \hspace{0.075cm} m_\textnormal{A}  \int \textnormal{d} \vec Q_1^\textnormal{A} \; D(\vec Q_1^\textnormal{A}(\vec q),t). \label{Total one-particle density}
\end{eqnarray}
Moreover, $\vec Q_i^\textnormal{A}(\vec q)$ means that in the particle coordinate set $\vec Q$ given by Eqn.\ (\ref{Particle Coordinate Set}), the coordinate vector  $\vec q_i^{\hspace{0.05 cm}\textnormal{A}}$ is set to $\vec q$. \newline 
Because of Eqn.\ (\ref{Total one-particle density}), it is clear that the one-particle mass density of the $\textnormal{A}$-th sort $\rho_m^{\textnormal{A}}(\vec q,t)$ is given by: 
\begin{eqnarray}
\rho_m^{\textnormal{A}}(\vec q,t) &=& N(\textnormal{A}) \hspace{0.075cm} m_\textnormal{A} \int \textnormal{d} \vec Q_1^\textnormal{A} \; D(\vec Q_1^\textnormal{A}(\vec q),t),  \label{One-particle density of sort k}
\end{eqnarray}
and it holds
\begin{eqnarray}
\rho_m^{\textnormal{tot}}(\vec q,t) &=& \sum_{\textnormal{A}=1}^{N_S} \rho_m^{\textnormal{A}}(\vec q,t).  \label{summing up particle densities}
\end{eqnarray}
Here, we introduce mass densities instead of just particle densities because the use of mass densities makes the MPEEMs and MPQCEs more compact. For the same 
reason, we introduce in the following mass current densities instead of just particle current densities. \newline 
Thus, as a next quantity, we define the total particle mass current density $\vec j^{\textnormal{tot}}_m(\vec q,t)$ as: 
\begin{eqnarray}
\vec j^{\textnormal{tot}}_m(\vec q,t) &=& \sum_{\textnormal{A}=1}^{N_S} \sum_{i=1}^{N(\textnormal{A})} \int \textnormal{d} \vec Q \;  \delta (\vec q - \vec q_i^{\hspace{0.05 cm}\textnormal{A}} ) \; \Re
\left [ \Psi^*(\vec Q,t) \; \hat {\vec p}_i^{\hspace{0.05 cm} \textnormal{A}} \; \Psi(\vec Q,t) \right]. \label{Total particle current density with momentum operator}
\end{eqnarray}
Regarding the definition of the canonical momentum operator $\hat {\vec p}_i^{\hspace{0.05 cm} \textnormal{A}}$ of the $(\textnormal{A},i)$-particle in Eqn.\ (\ref{momentum operator}) and the indistinguishability of the particles of each sort, we can transform Eqn.\ (\ref{Total particle current density with momentum operator}) into 
\begin{eqnarray}
\vec j^{\textnormal{tot}}_m(\vec q,t) &=& \hbar \sum_{\textnormal{A}=1}^{N_S} \sum_{i=1}^{N(\textnormal{A})} \int \textnormal{d} \vec Q \;  \delta (\vec q - \vec q_i^{\hspace{0.05 cm}\textnormal{A}} ) \; \Im
\left [ \Psi^*(\vec Q,t) \; \nabla_i^\textnormal{A} \Psi(\vec Q,t) \right] \label{Total particle current density} \\
&=& \hbar \sum_{\textnormal{A}=1}^{N_S} N(\textnormal{A}) \int \textnormal{d} \vec Q \;  \delta (\vec q - \vec q_1^{\hspace{0.05 cm}\textnormal{A}} ) \; \Im
\left [ \Psi^*(\vec Q,t) \; \nabla_1^\textnormal{A} \Psi(\vec Q,t) \right]. \label{Total particle current density 2} 
\end{eqnarray}
For the case of a single particle system, Eqn.\ (\ref{Total particle current density 2}) turns into the definiton of the particle current density (\hspace{-0.15 cm} \cite{Greiner_2001}, p.\ 144f.\ and \cite{Gustafson_2011}, p.\ 24). \newline  
Furthermore, Eqns.\ (\ref{Total particle current density with momentum operator})--(\ref{Total particle current density 2}) make clear that the mass current density $\vec j^{\textnormal{A}}_m(\vec q,t)$ of all the particles of the sort $\textnormal{A}$ is given by: \noindent
\begin{eqnarray}
\vec j^{\textnormal{A}}_m(\vec q,t) &=& \sum_{i=1}^{N(\textnormal{A})} \int \textnormal{d} \vec Q \; \delta (\vec q - \vec q_i^{\hspace{0.05 cm}\textnormal{A}} ) \; \Re
\left [ \Psi^*(\vec Q,t) \; \hat {\vec p}_i^{\hspace{0.05cm} \textnormal{A}} \Psi(\vec Q,t) \right] \label{k_flux_density with momentum operator}  \\
&=& \hbar \hspace{0.1cm} N(\textnormal{A}) \int \textnormal{d} \vec Q \;  \delta (\vec q - \vec q_1^{\hspace{0.05 cm}\textnormal{A}} ) \; \Im
\left [ \Psi^*(\vec Q,t) \; \nabla_1^\textnormal{A} \Psi(\vec Q,t) \right], \label{k_flux_density}
\end{eqnarray}
so, it holds 
\begin{eqnarray}
\vec j^{\textnormal{tot}}_m(\vec q,t) &=& \sum_{\textnormal{A}=1}^{N_S} \vec j^{\textnormal{A}}_m(\vec q,t). \label{summing up flux densities}
\end{eqnarray}
Moreover, we note that because of Eqn.\ (\ref{k_flux_density with momentum operator}), the quantity $\vec j^{\textnormal{A}}_m(\vec q,t)$ can be interpreted as the expectation value of this operator $\hat {\vec j}^{\textnormal{A}}_m(\vec Q,\vec q)$ \cite{Kuzmenkov_1999}: \newpage \noindent
\begin{eqnarray}
\hat {\vec j}^{\textnormal{A}}_m(\vec Q,\vec q) &=& \frac{1}{2} \sum_{i=1}^{N(\textnormal{A})}\left[ \delta (\vec q - \vec q_i^{\hspace{0.05 cm}\textnormal{A}} ) \hat{\vec p}_i^{\hspace{0.05 cm} \textnormal{A}} +   \hat{\vec p}_i^{\hspace{0.05 cm} \textnormal{A}} \delta (\vec q - \vec q_i^{\hspace{0.05 cm}\textnormal{A}} ) \right], \label{Def_operator_j} \\
\vec j^{\textnormal{A}}_m(\vec q,t) &=& \int \textnormal{d} \vec Q \; \Psi^*(\vec Q,t) \; \hat {\vec j}^{\textnormal{A}}_m(\vec Q,\vec q) \; \Psi(\vec Q,t). 
\end{eqnarray}
As the next step, for the particles of the sort $\textnormal{A}$, we define the mean particle velocity $\vec v^\textnormal{A}(\vec q,t)$ for all positions $\vec q$, where 
$\rho_m^{\textnormal{A}}(\vec q,t) \neq 0$:
\begin{eqnarray}
\vec v^\textnormal{A}(\vec q,t) &=& \frac{\vec j^{\textnormal{A}}_m(\vec q,t)}{\rho_m^{\textnormal{A}}(\vec q,t)}. \label{mean particle velocity def}
\end{eqnarray}
For all positions $\vec q_0$, where $\rho_m^{\textnormal{A}}(\vec q_0,t) = 0$, we define:
\begin{eqnarray}
\vec v^\textnormal{A}(\vec q_0,t) &=& \lim_{\vec q \rightarrow \vec q_0} \frac{\vec j^{\textnormal{A}}_m(\vec q,t)}{\rho_m^{\textnormal{A}}(\vec q,t)}. \label{mean particle velocity def rho 0}
\end{eqnarray}
Now, we use the representation \cite{Madelung_1926_2,Bohm_1952_1}
\begin{eqnarray}
\Psi(\vec Q,t) &=& a(\vec Q,t) \exp \left[\frac{\mathrm{i} S(\vec Q,t)}{\hbar}\right] \label{Bohm representation}
\end{eqnarray}
for the wave function $\Psi(\vec Q,t)$, where $a(\vec Q,t)$, $S(\vec Q,t)$ are real-valued, continuous, and differentiable functions,  
and they define the velocity $\vec w_i^\textnormal{A}(\vec Q,t)$ of the $(\textnormal{A},i)$-th particle by 
\begin{eqnarray}
\vec w_i^\textnormal{A}(\vec Q,t) &=& \frac{1}{m_\textnormal{A}}  \nabla_i^\textnormal{A} S(\vec Q,t). \label{local velocity}
\end{eqnarray}
Note that for the velocity of the $(\textnormal{A},i)$-th particle, we assigned the letter $w$, and for the mean particle velocity for particles of the sort $\textnormal{A}$, we assigned the letter $v$ because then the MPQHD equations will be similar to the classical hydrodynamic equations in textbooks. These equations can be found e.\hspace{-0.1 cm} g.\ in \cite{Landau_2000}, p.\ 2f., 11f., and 44f. As Madelung realized already in 1927 \cite{Madelung_1926_2} for the case of a single-particle system, the direct consequence of the definition (\ref{local velocity}) for the velocity  $\vec w_i^{\textnormal A}(\vec Q,t)$ of the $(\textnormal{A},i)$-th particle is that the rotation of this velocity relative to the coordinate $\vec q_i^{\hspace{0.05 cm} \textnormal A}$ va\-nishes: 
\begin{eqnarray}
\nabla_i^\textnormal{A} \times \vec w_i^{\textnormal A}(\vec Q,t) = \vec 0.
\end{eqnarray} 
The definition above for $\vec w_i^\textnormal{A}(\vec Q,t)$ can now be used to do the following transformation of the term  $\Im
\left [ \Psi^*(\vec Q,t) \nabla_1^\textnormal{A} \Psi(\vec Q,t) \right]$ appearing in Eqn.\ (\ref{Total particle current density 2}) for $\vec j^{\textnormal{tot}}_m(\vec q,t)$: \newpage \noindent
\begin{eqnarray}
\Im  \left [ \Psi^*(\vec Q,t) \nabla_1^\textnormal{A} \Psi(\vec Q,t) \right] &=&  \Im  \left [ \underbrace{ a(\vec Q,t)  \nabla_1^\textnormal{A} a(\vec Q,t)}_{\in \; \mathbb{R}}  + \frac{\mathrm{i}}{\hbar} \underbrace{a(\vec Q,t)^2 \nabla_1^\textnormal{A} S(\vec Q,t)}_{= \; D(\vec Q,t) \; m_\textnormal{A} \; \vec w_1^\textnormal{A}(\vec Q,t)} \right] \nonumber \\
&=& \frac{m_\textnormal{A}}{\hbar} D(\vec Q,t) \; \vec w_1^\textnormal{A}(\vec Q,t). \label{Imaginary Transformation}
\end{eqnarray}
With this transformation, we find the following form for $\vec j^{\textnormal{tot}}_m(\vec q,t)$:
\begin{eqnarray}
\vec j^{\textnormal{tot}}_m(\vec q,t) &=& \sum_{\textnormal{A}=1}^{N_S}  \; N(\textnormal{A}) \hspace{0.075cm} m_\textnormal{A} \int \textnormal{d} \vec Q_1^\textnormal{A} \; D(\vec Q_1^\textnormal{A} (\vec q),t) \; \vec w_1^\textnormal{A} (\vec Q_1^\textnormal{A} (\vec q),t). \label{tot_flux_density_bohm_representation}
\end{eqnarray}
Thus, the mass current density $\vec j^{\textnormal{A}}_m(\vec q,t)$ for particles of the sort $\textnormal{A}$ is given by:
\begin{eqnarray}
\vec j^{\textnormal{A}}_m(\vec q,t) &=& N(\textnormal{A})  \hspace{0.075cm} m_\textnormal{A} \int \textnormal{d} \vec Q_1^\textnormal{A} \; D(\vec Q_1^\textnormal{A} (\vec q),t) \; \vec w_1^\textnormal{A} (\vec Q_1^\textnormal{A} (\vec q),t). \label{k_flux_density_bohm_representation}
\end{eqnarray}
Eqn.\ (\ref{k_flux_density_bohm_representation}) is a logical result for $\vec j^{\textnormal{A}}_m(\vec q,t)$ because it can be explained in the following way: \newline
For the situation that the $(\textnormal{A},1)$-particle is located at $\vec q$ and we average over the positions of all the other particles, it is intuitive that the corresponding mass flux density of the $(\textnormal{A},1)$-particle is given by the integral of the term   $D(\vec Q_1^\textnormal{A} (\vec q),t) \;  \vec w_1^\textnormal{A} (\vec Q_1^\textnormal{A} (\vec q),t)$ over the infinitesimal $d \vec Q_1^\textnormal{A}$ multiplied by $m_\textnormal{A}$. Since the $N(\textnormal{A})$ particles of the sort $\textnormal{A}$ cannot be distinguished from each other, the flux density $\vec j^{\textnormal{A}}_m(\vec q,t)$  for the $N(\textnormal{A})$ particles of the sort $\textnormal{A}$ is then just $N(\textnormal{A})$ times this integral.  \newline
As a further quantity, we define the relative velocity $\vec u_i^\textnormal{A} (\vec Q,t)$ of the $(\textnormal{A},i)$-particle as: 
\begin{eqnarray}
\vec u_i^\textnormal{A} (\vec Q,t) &=& \vec w_i^\textnormal{A}(\vec Q,t) - \vec v^\textnormal{A}(\vec q_i^{\hspace{0.05 cm} \textnormal{A}},t). \label{definition relative velocity}
\end{eqnarray}
The motivation to name it relative velocity is that $\vec u_i^\textnormal{A} (\vec Q,t)$ is the velocity of the $(\textnormal{A},i)$-particle relative to $\vec v^\textnormal{A}(\vec q_i^{\hspace{0.05 cm} \textnormal{A}},t)$. Moreover, $\vec u_i^\textnormal{A} (\vec Q,t)$ has the following property: \newline 
The $(\textnormal{A},i)$-particle shall be in the position $\vec q_i^{\hspace{0.05cm}\textnormal{A}} = \vec q$, so $\vec Q = \vec Q_i^\textnormal{A}(\vec q)$, and we average $\vec u_i^\textnormal{A} (\vec Q,t)$ over all positions which the other particles can occupy. Hereby, we weigh $\vec u_i^\textnormal{A} (\vec Q,t)$ with the probability $D (\vec Q,t)$ that the positions of all particles are given by $\vec Q$. This average for the relative velocity $\vec u_i^\textnormal{A} (\vec Q,t)$ vanishes. In the following calculation, the vanishing of this average is shown, and we use in this calculation Eqns.\ (\ref{One-particle density of sort k}) and (\ref{k_flux_density_bohm_representation}): 
\begin{eqnarray}
&& \int  \textnormal{d} \vec Q \; \delta (\vec q - \vec q_i^{\hspace{0.05 cm} \textnormal{A}}) \; D(\vec Q,t) \; \vec u_i^\textnormal{A} (\vec Q,t) \; \hspace{0.08cm} = \; \nonumber \int  \textnormal{d} \vec Q_i^\textnormal{A} \;  D(\vec Q_i^\textnormal{A} (\vec q),t) \; \vec u_i^\textnormal{A} (\vec Q_i^\textnormal{A} (\vec q),t) \nonumber \\
&=& \nonumber \int  \textnormal{d} \vec Q_1^\textnormal{A} \;  D(\vec Q_1^\textnormal{A} (\vec q),t) \; \vec u_1^\textnormal{A} (\vec Q_1^\textnormal{A} (\vec q),t) \; \; = \; \nonumber \int  \textnormal{d} \vec Q_1^\textnormal{A} \;  D(\vec Q_1^\textnormal{A} (\vec q),t) \; \left( \vec w_1^\textnormal{A}(\vec Q_1^\textnormal{A}(\vec q),t) - \vec v^{\hspace{0.05 cm} \textnormal{A}}(\vec q,t) \right) \nonumber \\
&=& \underbrace{\int  \textnormal{d} \vec Q_1^\textnormal{A} \;  D(\vec Q_1^\textnormal{A} (\vec q),t) \; \vec w_1^\textnormal{A} (\vec Q_1^\textnormal{A}(\vec q),t)}_{= \; \frac{1}{N(\textnormal{A}) \hspace{0.025cm} m_\textnormal{A}} \vec j_m^{\textnormal{A}} (\vec q,t)} \; - \;
\underbrace{\left [ \int  \textnormal{d} \vec Q_1^\textnormal{A} \;  D(\vec Q_1^\textnormal{A} (\vec q),t) \right]}_{= \; \frac{1}{N(\textnormal{A}) \hspace{0.025cm}  m_\textnormal{A}} \rho_m^\textnormal{A}(\vec q,t)} \vec v^{\hspace{0.02 cm} \textnormal{A}} (\vec q,t) \nonumber  \\
&=& \frac{1}{N(\textnormal{A}) \hspace{0.05cm}  m_\textnormal{A}} \left( \vec j_m^\textnormal{A} (\vec q,t)  -  \rho_m^\textnormal{A}(\vec q,t) \; \vec v^{\hspace{0.02 cm} \textnormal{A}} (\vec q,t) \right) = \vec 0. 
\end{eqnarray}
In this context, we call the relative velocities $\vec u_i^\textnormal{A} (\vec Q,t)$ also the fluctuating velocities -- but note that this is a fluctuation relative to coordinate dependencies and not to time dependencies. \newline
Moreover, we can define new quantities related to the total ensemble of particles analogously to the velocities $\vec v^\textnormal{A}(\vec q,t)$ and  $\vec u_i^\textnormal{A} (\vec Q,t)$.\newline 
The first one is the mean particle velocity $\vec v^{\hspace{0.05 cm} \textnormal{tot}}(\vec q,t)$ for the total particle ensemble. For all positions $\vec q$ where $\rho_m^{\textnormal{tot}}(\vec q,t) \neq 0$, it is: 
\begin{eqnarray}
\vec v^{\hspace{0.05 cm} \textnormal{tot}}(\vec q,t) &=& \frac{\vec j^{\textnormal{tot}}_m(\vec q,t)}{\rho_m^{\textnormal{tot}}(\vec q,t)},  \label{total mean particle velocity}
\end{eqnarray}
and for all positions $\vec q_0$, where $\rho_m^{\textnormal{tot}}(\vec q_0,t) = 0$, it is:
\begin{eqnarray}
\vec v^{\hspace{0.05 cm} \textnormal{tot}}(\vec q_0,t) &=&  \lim_{\vec q \rightarrow \vec q_0} \frac{\vec j^{\textnormal{tot}}_m(\vec q,t)}{\rho_m^{\textnormal{tot}}(\vec q,t)}.
\end{eqnarray}
The second one is another relative velocity  of the $(\textnormal{A},i)$-particle named $\vec {\mathfrak{u}}^\textnormal{A}_i(\vec Q,t)$: 
\begin{eqnarray}
\vec {\mathfrak{u}}^\textnormal{A}_i(\vec Q,t) := \vec w_i^\textnormal{A}(\vec Q,t)  - \vec v^{\hspace{0.05 cm} \textnormal{tot}}(\vec q_i^{\hspace{0.05 cm}\textnormal{A}},t).  \label{alternative definition relative velocity}
\end{eqnarray}
Note that $\vec {\mathfrak{u}}^\textnormal{A}_i(\vec Q,t)$ is the relative velocity of the $(\textnormal{A},i)$-particle to $\vec v^{\hspace{0.05cm} \textnormal{tot}}(\vec q_i^{\hspace{0.05cm} \textnormal{A}},t)$, while $\vec u_i^\textnormal{A} (\vec Q,t)$ is the relative velocity of this particle to $\vec v^\textnormal{A}(\vec q_i^{\hspace{0.05cm} \textnormal{A}},t)$. We emphasize that this is an expansion relative to \cite{Kuzmenkov_1999,Andreev_2014a}, where just one kind of relative particle velocity was defined. 
\subsection{Derivation of the MPCE}
Now, we derive the many-particle continuity equation (MPCE) both for all particles and for particles of a certain sort $\textnormal{A}$. This can be done in an 
analogous way to the derivation of the continuity equation for a single particle wave function in quantum mechanics textbooks $\textnormal{(\hspace{-0.01cm}\cite{Greiner_2001}, p.\ 144f.\ and \cite{Gustafson_2011}, p.\ 24)}$. \newline
Therefore, we calculate the time derivative of $\rho_m^{\textnormal{A}}(\vec q,t)$ by inserting the Schrödinger equation (\ref{Schroedinger equation}) into Eqn.\ (\ref{One-particle density of sort k}):
\begin{eqnarray}
\frac{\partial \rho_m^{\textnormal{A}}(\vec q,t)}{\partial t} &=& N(\textnormal{A}) \hspace{0.075cm}  m_\textnormal{A} \int \textnormal{d} \vec Q_1^\textnormal{A} \; \left( \frac{\partial \Psi^*(\vec Q_1(\vec q),t)}{\partial t} \; \Psi(\vec Q_1(\vec q),t) + \Psi^*(\vec Q_1(\vec q),t) \; \frac{\partial \Psi(\vec Q_1(\vec q),t)}{\partial t} \right) \nonumber \\
&=& \frac{2 N(\textnormal{A}) \hspace{0.075cm}  m_\textnormal{A}}{\hbar} \int \textnormal{d} \vec Q \; \delta ( \vec q - \vec q_1^{\hspace{0.05cm}\textnormal{A}}) \;  \Im \left [ \Psi^*(\vec Q,t) \; \hat H(\vec Q)  \; \Psi(\vec Q,t) \right]. \label{derivation CE intermediate result}
\end{eqnarray}
We evaluate the imaginary part appearing in Eqn.\ (\ref{derivation CE intermediate result})  with Eqns.\ (\ref{momentum operator}) and (\ref{Hamiltion operator with momentum operator}): 
\begin{eqnarray}
\Im \left ( \Psi^*\; \hat H  \; \Psi \right) &=& \Im \left( - \sum_{\textnormal{B}=1}^{N_S} \sum_{j=1}^{N(\textnormal{B})} \Psi^* \frac{\hbar^2}{2 m_\textnormal{B}} \bigtriangleup_j^\textnormal{B} \Psi  +  \frac{1}{2} \sum_{\textnormal{B}=1}^{N_S} \sum_{j=1}^{N(\textnormal{B})} \sum_{\textnormal{C}=1}^{N_S} \sum_{k=1}^{N(\textnormal{C})} \underbrace{\Psi^* V_{jk}^{\textnormal{BC}} \Psi}_{\in \; \mathbb{R}} \right) \nonumber \\
&=& - \sum_{\textnormal{B}=1}^{N_S} \sum_{j=1}^{N(\textnormal{B})} \frac{\hbar^2}{2 m_\textnormal{B}} \Im \Biggl [ \nabla_j^\textnormal{B} \left ( \Psi^* \nabla_j^\textnormal{B} \Psi \right ) - \underbrace{\left(\nabla_j^\textnormal{B} \Psi^*\right) \left(\nabla_j^\textnormal{B} \Psi\right)}_{\in \; \mathbb{R}} \Biggl ] \nonumber \\
&=& - \sum_{\textnormal{B}=1}^{N_S} \sum_{j=1}^{N(\textnormal{B})} \frac{\hbar^2}{2 m_\textnormal{B}} \Im \left[ \nabla_j^\textnormal{B} \left ( \Psi^* \nabla_j^\textnormal{B} \Psi \right ) \right], \label{derivation CE intermediate result 2}
\end{eqnarray}
where $\bigtriangleup_j^\textnormal{B}$ is the Laplace operator relative to the coordinate $\vec q_j^{\hspace{0.05 cm} \textnormal{B}}$. \newline 
As a next step, we insert Eqn.\ (\ref{derivation CE intermediate result 2}) into Eqn.\ (\ref{derivation CE intermediate result}), and after that, the summand for the case $\{\textnormal{B}=\textnormal{A}, j=1\}$ is extracted out of the double sum over $\textnormal{B},j$. We can then transform the integration over the coordinate $\vec q_j^{\hspace{0.05cm}\textnormal{B}}$ for all the remaining summands with the divergence theorem into an integral over the system boundary surface where the wave function vanishes. So, these remaining summands vanish, and only the extracted summand of the double sum for the case $\{\textnormal{B}=\textnormal{A}, j=1\}$ remains:
\begin{eqnarray}
\frac{\partial \rho_m^{\textnormal{A}}(\vec q,t)}{\partial t} &=& - \hbar \hspace{0.03cm} N(\textnormal{A}) \hspace{0.075cm}  m_\textnormal{A} \sum_{\textnormal{B}=1}^{N_S} \sum_{j=1}^{N(\textnormal{B})} \frac{1}{m_\textnormal{B}} \int \textnormal{d} \vec Q \; \delta ( \vec q - \vec q_1^{\hspace{0.05cm}\textnormal{A}}) \; \Im \left[ \nabla_j^\textnormal{B} \left ( \Psi^* \nabla_j^\textnormal{B} \Psi \right ) \right] \nonumber \\ \nonumber
\\ 
&=& - \hbar  \hspace{0.03cm} N(\textnormal{A}) \hspace{0.075cm}  m_\textnormal{A} \Biggl \{  \frac{1}{m_\textnormal{A}} \int \textnormal{d} \vec Q_1^\textnormal{A} \int  \textnormal{d} \vec q_1^{\hspace{0.05 cm}\textnormal{A}} \; \delta ( \vec q - \vec q_1^{\hspace{0.05cm}\textnormal{A}})  \; \Im \left [  \nabla_1^\textnormal{A} \left (\Psi^* \nabla_1^\textnormal{A} \Psi \right) \right] \nonumber  \\
&&  + \; \underset{\{ \textnormal{B}  ,j \} \neq \{  \textnormal{A}, 1 \}}{\sum_{\textnormal{B}=1}^{{N_S}} \sum_{j=1}^{N(\textnormal{B})}}  \frac{1}{m_\textnormal{B}} \int \textnormal{d} \vec Q_{j}^\textnormal{B} \; \delta ( \vec q - \vec q_1^{\hspace{0.05cm}\textnormal{A}}) \; \Im \Biggl [ \underbrace{\int  \textnormal{d} \vec q_j^{\hspace{0.05 cm}\textnormal{B}} \; \nabla_j^\textnormal{B}  \left ( \Psi^* \nabla_j^\textnormal{B} \Psi \right )}_{= \; \vec 0} \Biggl ] \Biggl \}  \nonumber \\
&=& - \; \hbar  \hspace{0.03cm} N(\textnormal{A}) \int  \textnormal{d} \vec Q \; \delta ( \vec q - \vec q_1^{\hspace{0.05cm}\textnormal{A}}) \; \Im \left [ \nabla_1^\textnormal{A} \left ( \Psi^* \nabla_1^\textnormal{A} \Psi \right) \right].
\label{derivation CE intermediate result 3}
\end{eqnarray}
Finally, regarding the $\delta$-function in Eqn.\ (\ref{derivation CE intermediate result 3}), we can substitute the outer Nabla operator  $\nabla_1^\textnormal{A}$ in the imaginary part by a $\nabla$ operator related to the coordinate $\vec q$ in the following manner:  
\begin{eqnarray}
\frac{\partial \rho_m^{\textnormal{A}}(\vec q,t)}{\partial t} &=& - \nabla \left \{ \hbar  \hspace{0.03cm} N(\textnormal{A}) \int \textnormal{d} \vec Q \; \delta (\vec q - \vec q_1^{\hspace{0.05cm}\textnormal{A}}) \; \Im \left [\Psi^*(\vec Q,t) \nabla_1^\textnormal{A} \Psi(\vec Q,t) \right] \right \}.  
\end{eqnarray}
Because of Eqn.\ (\ref{k_flux_density}), which describes the mass flux density $\vec j^{\textnormal{A}}_m(\vec q,t)$ for all particles of the sort $\textnormal{A}$, the equation above is then the MPCE for these particles: 
\begin{eqnarray}
\frac{\partial \rho_m^{\textnormal{A}}(\vec q,t)}{\partial t} &=& - \nabla \vec j^{\textnormal{A}}_m(\vec q,t). \label{CE particle sort k} 
\end{eqnarray}
By summing up the MPCE for particles of a certain sort $\textnormal{A}$ over all sorts of particles, we get, using Eqns.\ (\ref{summing up particle densities}) and (\ref{summing up flux densities}), the MPCE for all particles: 
\begin{eqnarray}
\frac{\partial \rho_m^{\textnormal{tot}}(\vec q,t)}{\partial t} &=&  - \nabla \vec j^{\textnormal{tot}}_m(\vec q,t). \label{CE all particles} 
\end{eqnarray}
We note here that Eqns.\ (\ref{CE particle sort k}), (\ref{CE all particles}) are MPCEs, where mass densities and mass flux densities appear. Corresponding MPCEs for particle densities and particle flux densities can be derived.
\subsection{Derivation of the MPEEM}  
As our next task, we start with the derivation of the many-particle Ehrenfest equation of motion (MPEEM) both for all particles and for particles of a certain sort $\textnormal{A}$. Therefore, we calculate the time derivative of the flux density $\vec j^{\textnormal{A}}_m(\vec q,t)$ for this sort $\textnormal{A}$. We do not regard the indistinguishability of the particles of each sort form the beginning of the following calculation but we will take this point into account later because by applying this approach, some details in this derivation can be treated more systematically. Thus, we start with the time derivation of Eqn.\ (\ref{k_flux_density with momentum operator}) instead of Eqn.\ (\ref{k_flux_density}) and transform it, using  Eqn.\ (\ref{momentum operator}) for the momentum operator: 
\begin{eqnarray}
\frac{ \partial \vec j^{\textnormal{A}}_m(\vec q,t)}{\partial t} &=& \sum_{i=1}^{N(\textnormal{A})} \int \textnormal{d} \vec Q \; \delta (\vec q - \vec q_i^{\hspace{0.05 cm}\textnormal{A}} ) \; \frac{\partial}{\partial t}  \Re
\left [ \Psi^*(\vec Q,t) \; \hat {\vec p}_i^{\hspace{0.05cm} \textnormal{A}} \Psi(\vec Q,t) \right] \nonumber \\
&=& \sum_{i=1}^{N(\textnormal{A})} \int \textnormal{d} \vec Q \; \delta (\vec q - \vec q_i^{\hspace{0.05cm}\textnormal{A}} )  \frac{\partial}{\partial t} \Im \left [ \hbar \; \Psi^*(\vec Q,t) \nabla_i^\textnormal{A} \Psi(\vec Q,t) \right]. \label{Time derivative of total particle current density}
\end{eqnarray}
Now, we transform the time derivative term in Eqn.\ (\ref{Time derivative of total particle current density}). Here,  $q_{j\beta}^\textnormal{B}$ are the Cartesian components of the vector $\vec q_{j}^{\hspace{0.05 cm} \textnormal{B}}$. So, the index $\beta$ is an element of the set $K_{\textnormal{Ca}} = \{x,y,z\}$:
\begin{eqnarray} 
\hbar \; \frac{\partial}{\partial t} \Im \left [ \Psi^* \nabla_i^\textnormal{A} \Psi \right] &=&  \hbar \; \Im \left [ \left(\frac{\partial \Psi^*}{\partial t} \right) \nabla_i^{\textnormal{A}} \Psi + \Psi^* \nabla_i^\textnormal{A} \left( \frac{\partial \Psi}{\partial t} \right) \right]  \nonumber \\
                        &=& \hbar \; \Im \left [ \left( \frac{1}{\mathrm{i} \hbar} \hat H \Psi \right)^* \nabla_i^\textnormal{A} \Psi + \Psi^* \nabla_i^\textnormal{A} \left ( \frac{1}{\mathrm{i} \hbar} \hat H \Psi    \right) \right] \label{Ehrenfest_intermediate_result} \\
                        &=& \Re \left \{ \left[ -\sum_{\textnormal{B}=1}^{N_S} \sum_{j=1}^{N(\textnormal{B})} \frac{\hbar^2}{2 m_\textnormal{B}} \left( \bigtriangleup_j^\textnormal{B} \Psi^* \right) + \frac{1}{2} \sum_{\textnormal{B}=1}^{N_S} \sum_{j=1}^{N(\textnormal{B})} \sum_{\textnormal{C}=1}^{N_S} \sum_{k=1}^{N(\textnormal{C})} \Psi^* V_{jk}^{\textnormal{BC}} \right] (\nabla_i^\textnormal{A} \Psi) \right. \nonumber \\
                        &&  \left. + \; \Psi^*\nabla_i^\textnormal{A} \left[ \sum_{\textnormal{B}=1}^{N_S} \sum_{j=1}^{N(\textnormal{B})} \frac{\hbar^2}{2 m_\textnormal{B}}  \bigtriangleup_j^\textnormal{B} \Psi - \frac{1}{2} \sum_{\textnormal{B}=1}^{N_S} \sum_{j=1}^{N(\textnormal{B})} \sum_{\textnormal{C}=1}^{N_S} \sum_{k=1}^{N(\textnormal{C})} V_{jk}^{\textnormal{BC}} \Psi \right] \right \} \nonumber \\
                        &=& \Re \left \{ \sum_{\textnormal{B},j} \frac{\hbar^2}{2 m_\textnormal{B}} \left[ \Psi^* \nabla_i^\textnormal{A} \left( \bigtriangleup_j^\textnormal{B} \Psi \right) - \left( \bigtriangleup_j^\textnormal{B} \Psi^* \right)   \left( \nabla_i^\textnormal{A} \Psi \right) \right] \right . \nonumber \\
                        && \left.  + \; \frac{1}{2} \sum_{\textnormal{B},\textnormal{C},j,k} \Psi^* \left [ V_{jk}^{\textnormal{BC}} \nabla_i^\textnormal{A} \Psi - \nabla_i^\textnormal{A} \left (V_{jk}^{\textnormal{BC}} \Psi \right) \right] \right \} \nonumber \\
                        &=& \Re \left \{ \sum_{\textnormal{B},j} \frac{\hbar^2}{2 m_\textnormal{B}} \sum_{\beta \in K_{\textnormal{Ca}}} \left[ \Psi^* \nabla_i^\textnormal{A} \left( \frac{\partial^2  \Psi}{\partial q_{j \beta}^{\textnormal{B}} \partial q_{j \beta}^{\textnormal{B}} } \right)  - \left( \frac{\partial^2  \Psi^*}{\partial q_{j \beta}^{\textnormal{B}} \partial q_{j \beta}^{\textnormal{B}}} \right)   \left( \nabla_i^\textnormal{A} \Psi \right) \right] \right \} \nonumber \\
                        && - \; \frac{1}{2} \left | \Psi \right |^2 \sum_{\textnormal{B},\textnormal{C},j,k} \nabla_i^\textnormal{A} V_{jk}^{\textnormal{BC}} \nonumber \\
                        &=& \Re \left \{ \sum_{\textnormal{B},j} \frac{\hbar^2}{2 m_\textnormal{B}}  \sum_{\beta \in K_{\textnormal{Ca}}} \left \{ \frac{\partial}{\partial q_{j \beta}^\textnormal{B}} \left[ \Psi^* \nabla_i^\textnormal{A} \left ( \frac {\partial \Psi}{\partial q_{j \beta}^\textnormal{B}} \right) \right] -  \left ( \frac {\partial \Psi^*}{\partial q_{j \beta}^\textnormal{B}} \right)  \nabla_i^\textnormal{A} \left ( \frac {\partial \Psi}{\partial q_{j \beta}^\textnormal{B}} \right) \right. \right. \nonumber \\
                        &&  \left. \left. -  \frac {\partial }{\partial q_{j \beta}^\textnormal{B}} \left[ \left( \frac{\partial \Psi^*}{\partial q_{j\beta}^\textnormal{B}} \right) \left( \nabla_i^\textnormal{A} \Psi \right) \right] + \left(\frac{\partial \Psi^*}{\partial q_{j \beta}^\textnormal{B}} \right) \frac {\partial }{\partial q_{j \beta}^\textnormal{B}} \left( \nabla_i^\textnormal{A} \Psi \right) \right \} \right \} \nonumber \\
                        && - \; \frac{1}{2} \; D \sum_{\textnormal{B},\textnormal{C},j,k} \left ( \delta_{ij} \delta_{\textnormal{A}\textnormal{B}} \nabla_i^\textnormal{A} V_{ik}^{\textnormal{A}\textnormal{C}} +  \delta_{ik} \delta_{\textnormal{AC}} \nabla_i^\textnormal{A} V_{ji}^{\textnormal{BA}} \right)                                                          \nonumber \\
                        &=& \Re \left \{ \sum_{\textnormal{B},j}  \frac{\hbar^2}{2 m_\textnormal{B}}  \sum_{\beta \in K_{\textnormal{Ca}}} \frac{\partial}{\partial q_{j \beta}^\textnormal{B}} \left[ \Psi^* \nabla_i^\textnormal{A} \left ( \frac {\partial \Psi}{\partial q_{j \beta}^\textnormal{B}} \right) - \left( \frac{\partial \Psi^*}{\partial q_{j\beta}^\textnormal{B}} \right) \left( \nabla_i^\textnormal{A} \Psi \right)  \right]  \right \}  \nonumber \\
                        && - \; D \sum_{\textnormal{B},j}  \nabla_i^\textnormal{A} V_{ij}^{\textnormal{A}\textnormal{B}}. 
\end{eqnarray}
With this result, we get the following intermediate result for $\frac{ \partial \vec j^{\textnormal{A}}_m(\vec q,t)}{\partial t}$: 
\begin{eqnarray}
\frac{ \partial \vec j^{\textnormal{A}}_m(\vec q,t)}{\partial t} &=& \sum_{i=1}^{N(\textnormal{A})}   \sum_{\textnormal{B}=1}^{{N_S}} \sum_{j=1}^{N(\textnormal{B})} \int \textnormal{d} \vec Q \; \delta (\vec q - \vec q_i^{\hspace{0.05cm}\textnormal{A}} )  \frac{\hbar^2}{2 m_\textnormal{B}} \nonumber \\ 
&& \times \; \Re \left \{ \sum_{\beta \in K_{\textnormal{Ca}}} \frac{\partial}{\partial q_{j \beta}^\textnormal{B}} \left[ \Psi^* \nabla_i^\textnormal{A} \left ( \frac {\partial \Psi}{\partial q_{j \beta}^\textnormal{B}} \right) - \left( \frac{\partial \Psi^*}{\partial q_{j\beta}^\textnormal{B}} \right) \left( \nabla_i^\textnormal{A} \Psi \right)  \right]  \right \}  \nonumber \\
&& - \; \sum_{i=1}^{N(\textnormal{A})} \sum_{\textnormal{B}=1}^{{N_S}} \sum_{j=1}^{N(\textnormal{B})} \int \textnormal{d} \vec Q \; \delta (\vec q - \vec q_i^{\hspace{0.05cm}\textnormal{A}} )  \; D \; \nabla_i^\textnormal{A} V_{ij}^{\textnormal{AB}}. \label{intermediate result for time derivation} 
\end{eqnarray}
The term in the last line of Eqn.\ (\ref{intermediate result for time derivation}) is the force density $\vec f^{\hspace{0.05cm}\textnormal{A}}(\vec q,t)$ for all particles of the sort $\textnormal{A}$; it is caused by the two-particle potential $V_{ij}^{\textnormal{AB}}$: 
\begin{eqnarray}
\vec f^{\hspace{0.05 cm} \textnormal{A}}(\vec q,t) &=& \sum_{i=1}^{N(\textnormal{A})} \left [- \; \sum_{\textnormal{B}=1}^{N_S} \sum_{j=1}^{N(\textnormal{B})} \int \textnormal{d} \vec Q \; \delta (\vec q - \vec q_i^{\hspace{0.05 cm \textnormal{A}}}) \; D(\vec Q,t) \nabla_i^\textnormal{A} V_{ij}^{\textnormal{AB}} \right].  \label{Total force density}                              
\end{eqnarray}  
Please regard that in spite of Eqn.\ (\ref{V_antisymmetry}) the summands in the triple sum are not antisymmetric relative to a permutation of $\{\textnormal{A},i\} \leftrightarrow \{\textnormal{B},j\}$ because of the argument $\vec q - \vec q_i^{\hspace{0.05 cm \textnormal{A}}}$ in the $\delta$-function. This is related to the fact  
that the term in squared brackets in Eqn.\ (\ref{Total force density}) is the force density for the $(\textnormal{A},i)$-th particle. \newline
As the next step, we do a case-by-case analysis by splitting in Eqn.\ (\ref{Total force density}) the sum over the sort of particles $\textnormal{B}$ into a sum over  
the summands for $\textnormal{B} \neq \textnormal{A}$ and the remaining summand for $\textnormal{B} = \textnormal{A}$: 
\begin{eqnarray}
\vec f^{\hspace{0.05 cm} \textnormal{A}}(\vec q,t) &=& - \sum_{i=1}^{N(\textnormal{A})} \sum_{\textnormal{B}=1, \textnormal{B} \neq \textnormal{A}}^{N_S} \sum_{j=1}^{N(\textnormal{B})} \int \textnormal{d} \vec Q \; \delta (\vec q - \vec q_i^{\hspace{0.05 cm \textnormal{A}}}) \; D(\vec Q,t) \nabla_i^\textnormal{A} V_{ij}^{\textnormal{AB}} \nonumber \\ 
&& \left. - \; \sum_{i=1}^{N(\textnormal{A})} \sum_{j=1}^{N(\textnormal{A})}  \int \textnormal{d} \vec Q \; \delta (\vec q - \vec q_i^{\hspace{0.05 cm \textnormal{A}}}) \; D(\vec Q,t) \nabla_i^\textnormal{A} V_{ij}^{\textnormal{AA}} \right. . \label{Total force density 2a}   
\end{eqnarray}  
The first line in Eqn.\ (\ref{Total force density 2a}) is related to the case  $\textnormal{B} \neq \textnormal{A}$, and the second line in this equation is related to the case $\textnormal{B} = \textnormal{A}$. \newline 
The interaction of a particle with itself is excluded because in Eqn.\ (\ref{Potential}), we defined $V_{ii}^{\textnormal{AA}} = 0$. Thus, a particle of the sort $\textnormal{A}$ can interact with $(N(\textnormal{A})-1)$ particles of its own sort $\textnormal{A}$ and with $N(\textnormal{B})$ particles of the sort $\textnormal{B}$ if $\textnormal{B} \neq \textnormal{A}$. Therefore, we evaluate the indistinguishability between particles of one sort in the following manner: \newline
We substitute the sum over $i$ each in the first line and in the second line of Eqn.\ (\ref{Total force density 2a}) by its summand for $i=1$ multiplied by $N(\textnormal{A})$. Moreover, in the first line, we substitute the sum over $j$ by its summand for $j=N(\textnormal{B})$ multiplied by $N(\textnormal{B})$, and in the second line, we substitute the sum over $j$ by its summand for $j=N(\textnormal{A})$ multiplied by $(N(\textnormal{A})-1)$. Combining these substitutions with 
the definition (\ref{Potential}) for the potential terms $V_{ij}^{\textnormal{AB}}$ leads to this result:
\begin{eqnarray}
\vec f^{\hspace{0.05 cm} \textnormal{A}}(\vec q,t) &=& - \; N(\textnormal{A}) \left[ \sum_{\textnormal{B}=1, \textnormal{B} \neq \textnormal{A}}^{N_S} N(\textnormal{B}) \int \textnormal{d} \vec Q_{1}^\textnormal{A} \; D(\vec Q_1^\textnormal{A}(\vec q),t) \nabla V^{\textnormal{AB}}(|\vec q - \vec q_{N(\textnormal{B})}^{\hspace{0.05cm}B}|) \right] \nonumber \\
&& - \; N(\textnormal{A}) \left (N(\textnormal{A}) - 1 \right) \int \textnormal{d} \vec Q_{1}^\textnormal{A} \; D(\vec Q_1^\textnormal{A}(\vec q),t) \nabla V^{\textnormal{AB}}(|\vec q - \vec q_{N(\textnormal{A})}^{\hspace{0.05cm}\textnormal{A}}|).  \label{Total force density 2b}   
\end{eqnarray}
The reader might wonder why we choose the last summand for $j=N(\textnormal{A})$ or  $j=N(\textnormal{B})$, respectively, for the substitution of the sums over $j$ in Eqn.\ (\ref{Total force density 2a}) and not, in a more obvious approach, the first summand for $j=1$. The reason for this is that we have to make sure that we choose two different particles of the sort $\textnormal{A}$ as representative particles for the consideration of the two-particle interaction between the particles of the sort $\textnormal{A}$ since the interaction of a particle with itself is excluded. \newline
Here, the transformation of the second line of Eqn.\ (\ref{Total force density 2a}) into the term appearing in the second line of Eqn.\ (\ref{Total force density 2b}) 
can be interpreted in this manner: The $(\textnormal{A},1)$-particle and the $(\textnormal{A},N(\textnormal{A}))$-particle are chosen as representative particles for the two-particle interaction between the particles of the sort $\textnormal{A}$. This choice is appropriate because for $N(\textnormal{A}) > 1$, the $(\textnormal{A},1)$-particle and the $(\textnormal{A},N(\textnormal{A}))$-particle are different particles. Though, the $(\textnormal{A},1)$-particle and the $(\textnormal{A},N(\textnormal{A}))$-particle are the same particle for the special case $N(\textnormal{A})=1$, but this case is still evaluated in Eqn.\ (\ref{Total force density 2b}) correctly, since for $N(\textnormal{A})=1$, the factor $(N(\textnormal{A}) - 1 )$ in the second line of Eqn.\ (\ref{Total force density 2b}) is zero -- so we regard rightly that for $N(\textnormal{A})=1$ the single particle of the sort $\textnormal{A}$ cannot interact with other particles of this sort $\textnormal{A}$. \newline
This explanation above gives a clear reason why we choose $j=N(\textnormal{A})$ instead of $j=1$ for the transformation of the term appearing in the second line of Eqn.\ (\ref{Total force density 2a}). However, we could have used $j=1$ for the transformation of the term appearing in the first line of Eqn.\ (\ref{Total force density 2a}) because it is related to interactions between particles of different sorts. But there we still used the last summand for $j=N(\textnormal{B})$ because, as a consequence,  we can now combine the two lines in  Eqn.\ (\ref{Total force density 2b}) in this compact final result for $\vec f^{\hspace{0.05 cm} \textnormal{A}}(\vec q,t)$: 
\begin{eqnarray}
\vec f^{\hspace{0.05 cm} \textnormal{A}}(\vec q,t) &=& - \; N(\textnormal{A}) \left[ \sum_{\textnormal{B}=1}^{N_S} \left (N(\textnormal{B}) - \delta_{\textnormal{AB}} \right) \hspace{-0.1 cm} \int  \hspace{-0.1 cm} \textnormal{d} \vec Q_{1}^\textnormal{A} \; D(\vec Q_1^\textnormal{A}(\vec q),t) \nabla V^{\textnormal{AB}}(|\vec q - \vec q_{N(\textnormal{B})}^{\hspace{0.05cm}B}|) \right] \hspace{-0.1cm} .  \label{Total force density 2}                              
\end{eqnarray}
As an additional result, we can now calculate the total force density $\vec f^{\hspace{0.05 cm} \textnormal{tot}}(\vec q,t)$ for all particles; it is given by: 
\begin{eqnarray} 
\vec f^{\hspace{0.05 cm} \textnormal{tot}}(\vec q,t) &=& \sum_{\textnormal{A}=1}^{N_S} \vec f^{\hspace{0.05 cm} \textnormal{A}}(\vec q,t) \label{summing up force density}   \\ 
&=& - \sum_{\textnormal{A}=1}^{N_S} N(\textnormal{A}) \; \times \nonumber \\
&&  \times \; \left[ \sum_{\textnormal{B}=1}^{N_S} \left (N(\textnormal{B}) - \delta_{\textnormal{AB}} \right) \int \textnormal{d} \vec Q_{1}^\textnormal{A} \; D(\vec Q_1^\textnormal{A}(\vec q),t) \nabla V^{\textnormal{AB}}(|\vec q - \vec q_{N(\textnormal{B})}^{\hspace{0.05cm}\textnormal{B}}|) \right].  \label{Force density for all k}                              
\end{eqnarray}
In \cite{Andreev_2014a}, a particle ensemble for two sorts of particles, namely electrons and one ion sort, is examined, and hereby, the Coulomb force density for the electrons is shown. Eqn.\ (\ref{Total force density 2}) is a generalization of this result for an arbitrary number $N_S$ of sorts of particles and any 
two-particle potential which can be described by Eqn.\ (\ref{Potential}). \newline   
Having discussed the force density term $\vec f^{\hspace{0.05 cm} \textnormal{A}}(\vec q,t)$ in the intermediate result (\ref{intermediate result for time derivation}) for $\frac{ \partial \vec j^{\textnormal{A}}_m(\vec q,t)}{\partial t}$, we will now analyze the remaining term. \newline 
Therefore, we define a vector ${\vec x}_{j}^{\hspace{0.03cm} \textnormal{B}}(\vec Q,\textnormal{A},i,\alpha) $ 
\begin{eqnarray}
{\vec x}_{j}^{\hspace{0.03cm} \textnormal{B}}(\vec Q,\textnormal{A},i,\alpha) &=& \frac{\hbar^2}{2 m_\textnormal{B}} \Re \left [ \Psi^* \frac{\partial}{\partial q_{i \alpha}^\textnormal{A}} \left (\nabla_j^\textnormal{B} \Psi \right) - \nabla_j^\textnormal{B} \Psi^* \left (\frac{\partial \Psi}{\partial q_{i \alpha}^\textnormal{A}} \right) \right] \label{definition vector x}
\end{eqnarray}
with a $\beta$-component 
\begin{eqnarray}
x_{j\beta}^\textnormal{B}(\vec Q,\textnormal{A},i,\alpha) &=& \frac{\hbar^2}{2 m_\textnormal{B}} \Re \left [ \Psi^* \frac{\partial}{\partial q_{i \alpha}^\textnormal{A}} \left ( \frac{\partial \Psi}{\partial q_{j \beta   }^\textnormal{B}} \right)  - \left( \frac{\partial \Psi^* }{\partial q_{j \beta}^\textnormal{B}} \right) \Biggl (\frac{\partial \Psi}{\partial q_{i \alpha}^\textnormal{A}} \Biggl ) \right].
\end{eqnarray}
For the following calculation, it is advantageous to choose the notation above for the vector ${\vec x}_{j}^{\hspace{0.03cm} \textnormal{B}}(\vec Q,\textnormal{A},i,\alpha)$ because in this calculation, gradient terms $\nabla_{j}^\textnormal{B} {\vec x}_{j}^{\hspace{0.03cm} \textnormal{B}}(\vec Q,\textnormal{A},i,\alpha)$ of this vector appear. Therefore, we emphasize the dependence of the vector ${\vec x}_{j}^{\hspace{0.03cm} \textnormal{B}}(\vec Q,\textnormal{A},i,\alpha)$ on $j$ and $\textnormal{B}$ by writing $j$ as a subscript and $\textnormal{B}$ as a superscript. Moreover, the remaining terms $\vec Q,\textnormal{A},i$ and $\alpha$ are listed as additional parameters in brackets for the vector ${\vec x}_{j}^{\hspace{0.03cm} \textnormal{B}}(\vec Q,\textnormal{A},i,\alpha)$. \newline
With the definition (\ref{definition vector x}), the $\alpha$-component of $\frac{ \partial \vec j^{\textnormal{A}}_m(\vec q,t)}{\partial t}$ is given by: 
\begin{eqnarray}
\frac{ \partial j^{\textnormal{A}}_{m,\alpha}(\vec q,t)}{\partial t} &=&  f_\alpha^{\textnormal{A}}(\vec q,t) + \sum_{i=1}^{N(\textnormal{A})} \sum_{\textnormal{B}=1}^{{N_S}} \sum_{j=1}^{N(\textnormal{B})} \int \textnormal{d} \vec Q \; \delta (\vec q - \vec q_i^{\hspace{0.05cm}\textnormal{A}} ) \nonumber \\ 
&& \times \; \sum_{\beta \in K_{\textnormal{Ca}}} \frac{\partial}{\partial q_{i \beta}^\textnormal{B}} \left \{  \frac{\hbar^2}{2 m_\textnormal{B}} \Re \left [ \Psi^* \frac{\partial}{\partial q_{i \alpha}^\textnormal{A}} \left ( \frac{\partial \Psi}{\partial q_{j \beta}^\textnormal{B}} \right)  - \left( \frac{\partial \Psi^* }{\partial q_{j \beta}^\textnormal{B}} \right) \Biggl (\frac{\partial \Psi}{\partial q_{i \alpha}^\textnormal{A}} \Biggl ) \right] \right \} \nonumber \\
&=&  f_\alpha^{\textnormal{A}}(\vec q,t) + \sum_{i=1}^{N(\textnormal{A})} \sum_{\textnormal{B}=1}^{{N_S}} \sum_{j=1}^{N(\textnormal{B})} \int \textnormal{d} \vec Q \; \delta (\vec q - \vec q_i^{\hspace{0.05cm}\textnormal{A}} ) \; \nabla_{j}^\textnormal{B} {\vec x}_{j}^{\hspace{0.03cm} \textnormal{B}}(\vec Q,\textnormal{A},i,\alpha).
\end{eqnarray}
Now a case-by-case analysis is done for two different summand types in the triple sum over $i,j$, and $\textnormal{B}$: For the first type, the tuple $\{ \textnormal{A},i \}$ is not equal to the tuple $\{ \textnormal{B}, j \}$,  and for the second type, these two tuples are equal. By separating these two summand types, we get:
\begin{eqnarray}
\frac{ \partial j^{\textnormal{A}}_{m,\alpha}(\vec q,t)}{\partial t} &=&   f_\alpha^{\textnormal{A}}(\vec q,t) \nonumber \\
&& + \; \sum_{i=1}^{N(\textnormal{A})} \underset{\{\textnormal{B},j \}\neq\{ \textnormal{A},i\}}{ \sum_{\textnormal{B}=1}^{{N_S}} \sum_{j=1}^{N(\textnormal{B})}} \int \textnormal{d} \vec Q_{\hspace{-0.02 cm}j}^{\textnormal{B}} \; \delta (\vec q - \vec q_i^{\hspace{0.05cm}\textnormal{A}}) \int  \textnormal{d} \vec q_j^{\hspace{0.05 cm}\textnormal{B}} \; \nabla_j^\textnormal{B} {\vec x}_{j}^{\hspace{0.03cm} \textnormal{B}}(\vec Q,\textnormal{A},i,\alpha)  \nonumber \\
&& + \;   \sum_{i=1}^{N(\textnormal{A})}  \int \textnormal{d} \vec Q_{\hspace{-0.02 cm}i}^{\textnormal{A}} \int  \textnormal{d} \vec q_i^{\hspace{0.05 cm}\textnormal{A}}  \; \delta (\vec q - \vec q_i^{\hspace{0.05cm}\textnormal{A}}) \; \left [ \nabla_j^\textnormal{B} {\vec x}_{j}^{\hspace{0.03cm} \textnormal{B}}(\vec Q,\textnormal{A},i, \alpha) \right]_{ \{ \textnormal{B}, j \} =  \{ \textnormal{A}, i \} }.
\end{eqnarray}
In the middle line of the equation above, a volume integral appears over the coordinate $\vec q_j^{\hspace{0.05 cm}\textnormal{B}}$. Using the divergence theorem, 
this integral can be converted into an integral of the system boundary surface where the wave function vanishes  -- so, the full term in the middle line vanishes. However, the term in the last line does not vanish because the integrand for the integral over the coordinate $\vec q_i^{\hspace{0.05 cm}\textnormal{A}}$  contains the $\delta$-function $\delta (\vec q - \vec q_i^{\hspace{0.05cm}\textnormal{A}})$. This context leads to: 
\begin{eqnarray}
\frac{ \partial j^{\textnormal{A}}_{m,\alpha}(\vec q,t)}{\partial t} &=&  f_\alpha^{\textnormal{A}}(\vec q,t)  \nonumber  \\
&& \hspace{-2.75cm} - \hspace{-0.15 cm} \sum_{\beta \in K_{\textnormal{Ca}}} \hspace{-0.15 cm} \frac{\partial}{\partial q_\beta} \left \{  \hspace{-0.1cm}  - \hspace{-0.1 cm} \sum_{i=1}^{N(\textnormal{A})} \hspace{-0.1 cm} \int \hspace{-0.1 cm} \textnormal{d} \vec Q \; \delta (\vec q - \vec q_i^{\hspace{0.05cm}\textnormal{A}}) \;  \hspace{-0.1cm} \frac{\hbar^2}{2 m_\textnormal{A}}   \Re  \hspace{-0.05cm} \left [ \Psi^* \frac{\partial}{\partial q_{i \alpha}^\textnormal{A}} \left ( \hspace{-0.05cm} \frac{\partial \Psi}{\partial q_{i \beta   }^\textnormal{A}} \hspace{-0.05cm} \right)  - \left( \hspace{-0.05cm} \frac{\partial \Psi^* }{\partial q_{i \beta}^\textnormal{A}} \hspace{-0.05cm} \right) \Biggl ( \hspace{-0.05cm} \frac{\partial \Psi}{\partial q_{i \alpha}^\textnormal{A}} \hspace{-0.05cm}  \Biggl ) \hspace{-0.05cm} \right]  \hspace{-0.1cm} \right  \} \hspace{-0.1cm} . \label{MMBE first version without indistinguishability}                                                        
\end{eqnarray}
Then, we take into account the indistinguishability of the particles of one sort and find the following result for $\frac{ \partial j^{\textnormal{A}}_{m,\alpha}(\vec q,t)}{\partial t}$:  \newpage \noindent
\begin{eqnarray}
\frac{ \partial j^{\textnormal{A}}_{m,\alpha}(\vec q,t)}{\partial t} &=& f_\alpha^{\textnormal{A}}(\vec q,t) \nonumber  \\
&& \hspace{-2.75cm} - \hspace{-0.15 cm} \sum_{\beta \in K_{\textnormal{Ca}}} \hspace{-0.15 cm} \frac{\partial}{\partial q_\beta} \left \{  \hspace{-0.1cm} - N(\textnormal{A}) \hspace{-0.1 cm}  \int \hspace{-0.1 cm} \textnormal{d} \vec Q \; \delta (\vec q - \vec q_1^{\hspace{0.05cm}\textnormal{A}}) \; \hspace{-0.1cm}  \frac{\hbar^2}{2 m_\textnormal{A}}  \Re \hspace{-0.05cm}   \left [ \Psi^* \frac{\partial}{\partial q_{1 \alpha}^\textnormal{A}} \left ( \hspace{-0.05cm} \frac{\partial \Psi}{\partial q_{1 \beta   }^\textnormal{A}} \hspace{-0.05cm} \right)  - \left( \hspace{-0.05cm} \frac{\partial \Psi^* }{\partial q_{1 \beta}^\textnormal{A}} \hspace{-0.05cm} \right) \Biggl ( \hspace{-0.05cm} \frac{\partial \Psi}{\partial q_{1 \alpha}^\textnormal{A}} \hspace{-0.05cm} \Biggl )  \hspace{-0.05cm} \right] \hspace{-0.1cm} \right \} \hspace{-0.1cm} . \label{MMBE first version}   
\end{eqnarray}
In the equation above, in the curly brackets, the components $\Pi_{\alpha \beta}^{\textnormal{A}}(\vec q,t)$ of what is called momentum flow density tensor $\underline{\underline{\Pi}}^{\textnormal{A}}(\vec q,t)$ for the particles of the sort $\textnormal{A}$ appear:
\begin{eqnarray}
\Pi_{\alpha \beta}^{\textnormal{A}}(\vec q,t) &=& - \; N(\textnormal{A}) \int \textnormal{d} \vec Q \; \delta (\vec q - \vec q_1^{\hspace{0.05cm}\textnormal{A}}) \nonumber \\ && \times \; \frac{\hbar^2}{2 m_\textnormal{A}}  \Re  \left [ \Psi^* \frac{\partial}{\partial q_{1 \alpha}^\textnormal{A}} \left ( \frac{\partial \Psi}{\partial q_{1 \beta}^\textnormal{A}} \right)  - \left( \frac{\partial \Psi^* }{\partial q_{1 \beta}^\textnormal{A}} \right) \Biggl (\frac{\partial \Psi}{\partial q_{1 \alpha}^\textnormal{A}} \Biggl) \right]. \label{total momentum flow density tensor version 1} 
\end{eqnarray}
By applying the formula $\Re (z) = (z + z^*)/2$ on the real part appearing in Eqn.\ (\ref{total momentum flow density tensor version 1}), we recognize the symmetry 
\begin{eqnarray}
\Pi_{\alpha \beta}^{\textnormal{A}}(\vec q,t) = \Pi_{\beta \alpha}^{\textnormal{A}}(\vec q,t). \label{SymmetryPi}
\end{eqnarray}
Thus, using Eqns.\ (\ref{total momentum flow density tensor version 1}) and (\ref{SymmetryPi}), we find that Eqn.\ (\ref{MMBE first version}) can be written as: 
\begin{eqnarray}
\frac{\partial j^{\textnormal{A}}_{m,\alpha}(\vec q,t)}{\partial t} &=& f_\alpha^{\textnormal{A}}(\vec q,t) - \sum_{\beta \in K_{\textnormal{Ca}}} \frac{\partial}{\partial q_\beta} 
\Pi_{\beta \alpha}^{\textnormal{A}}(\vec q,t). \label{MMBE second version} 
\end{eqnarray}
Here, we note that the divergence $\nabla \underline{\underline{T}}(\vec q)$ of any second-rank tensor $\underline{\underline{T}}(\vec q)$ is given by: 
\begin{eqnarray}
\nabla \underline{\underline{T}}(\vec q) &=& \sum_{\alpha \in K_{\textnormal{Ca}}} \sum_{\beta \in K_{\textnormal{Ca}}} \frac{\partial T_{\alpha \beta}(\vec q)}{\partial q_\alpha} \vec e_\beta. \label{Definition Chapter 2} 
\end{eqnarray}
The equation corresponding to Eqn.\ (\ref{MMBE second version}) for the vector $\frac{\partial \vec j^{\textnormal{A}}_{m}(\vec q,t)}{\partial t}$ is the sought MPEEM for all particles of the sort $\textnormal{A}$, and by applying Eqn.\ (\ref{Definition Chapter 2}), we can write this equation in the following form:   
\begin{eqnarray}
\frac{\partial \vec j^{\textnormal{A}}_{m}(\vec q,t)}{\partial t} = \vec f^{\hspace{0.05cm} \textnormal{A}}(\vec q,t) - \nabla  \underline{\underline{\Pi}}^{\textnormal{A}}(\vec q,t). \label{MMBE}
\end{eqnarray}
Our motivation to name the equation above as the many-particle Ehrenfest equation of motion (MPEEM) is the following: In Eqn.\ (\ref{Def_operator_j}), we defined the operator $\hat{\vec j}^{\textnormal{A}}_{m}(\vec Q,\vec q)$, and we can interpret the quantity  $\vec j^{\textnormal{A}}_{m}(\vec q,t)$ 
as the expectation value for it. So, the Ehrenfest theorem (\hspace{-0.03cm}\cite{Ehrenfest_1927} and \cite{Schwabl_2005}, p.\ 28ff.) predicates that we can calculate the time derivative of the mass flux density for particles of the sort $\textnormal{A}$ by this equation: 
\begin{eqnarray}
\frac{\partial \vec j^{\textnormal{A}}_{m}(\vec q,t)}{\partial t} &=& \frac{\mathrm{i}}{\hbar} \int \textnormal{d} \vec Q \;  \Psi^*(\vec Q,t) \; \left[\hat H (\vec Q), \hat{\vec j}^{\textnormal{A}}_{m}(\vec Q,\vec q)\right] \;  \Psi(\vec Q,t).
\end{eqnarray}
Indeed, by combining Eqns.\ (\ref{Time derivative of total particle current density}) and (\ref{Ehrenfest_intermediate_result}), one can realize that the equation above is equi- valent to our result (\ref{MMBE}) for $\frac{\partial \vec j^{\textnormal{A}}_{m}(\vec q,t)}{\partial t}$.  This concept of applying the Ehrenfest theorem to derive quantum hydrodynamical equations was already discussed by Epstein in \cite{Epstein_1975}.\newline \newline 
As the next step, we discuss that Eqn.\ (\ref{total momentum flow density tensor version 1}) is not our final result for the tensor components $\Pi_{\alpha \beta}^{\textnormal{A}}(\vec q,t)$. Instead, we will show two different ways to express them. For the first of these two ways, we regard that with the definition (\ref{momentum operator}) for the canonical momentum operator $\hat {\vec p}_i^{\hspace{0.05 cm} \textnormal{A}}$ of the $(\textnormal{A},i)$-particle, Eqn.\ (\ref{total momentum flow density tensor version 1}) can be rewritten as:
\begin{eqnarray}
\Pi_{\alpha \beta}^{\textnormal{A}}(\vec q,t) &=&  N(\textnormal{A}) \int \textnormal{d} \vec Q \; \delta (\vec q - \vec q_1^{\hspace{0.05cm}\textnormal{A}}) \nonumber \\
&& \times \; \frac{1}{2 m_\textnormal{A}} \Re \left [ \Psi^* \left( \hat p_{1 \alpha}^\textnormal{A} \hat p_{1 \beta}^\textnormal{A} \Psi \right )+  \left (\hat p_{1 \beta}^\textnormal{A} \Psi \right)^* \left ( \hat p_{1 \alpha}^\textnormal{A} \Psi \right) \right] \nonumber \\
&\hspace{-1.0 cm}& \hspace{-2.5 cm} = \; \; N(\textnormal{A}) \int \textnormal{d} \vec Q \;  \delta (\vec q - \vec q_1^{\hspace{0.05 cm}\textnormal{A}} ) \; 
\frac{1}{4 m_\textnormal{A}} \times \nonumber \\
&\hspace{-1.0 cm}  &  \hspace{-2.3 cm} \; \; \;  \left [ \Psi^* \left( \hat p_{1\beta}^\textnormal{A} \hat p_{1\alpha}^\textnormal{A} \Psi  \right)  +  \left (\hat p_{1\beta}^\textnormal{A} \Psi \right)^*  \left( \hat p_{1\alpha}^\textnormal{A} \Psi \right) +  \left (\hat p_{1\alpha}^\textnormal{A} \Psi \right)^*  \left( \hat p_{1\beta}^\textnormal{A} \Psi \right) + \left( \hat p_{1\beta}^\textnormal{A} \hat p_{1\alpha}^\textnormal{A} \Psi \right)^* \Psi \right ]. \label{total momentum flow density tensor version 2}
\end{eqnarray} 
To find the second way to express the tensor components $\Pi_{\alpha \beta}^{\textnormal{A}}(\vec q,t)$, we now transform two terms appearing in Eqn.\ (\ref{total momentum flow density tensor version 1}) using Eqns.\ (\ref{Bohm representation}) and (\ref{local velocity}). Here we have the transformation of the first term:
\begin{eqnarray}
\Re \left[ \Psi^* \frac{\partial}{\partial q_{1 \alpha}^\textnormal{A}} \left( \frac{\partial \Psi} {\partial q_{1 \beta}^\textnormal{A}} \right) \right] &=& \Re \left \{ a e^{- \mathrm{i} S/\hbar} \frac{\partial}{\partial q_{1 \alpha}^\textnormal{A}} \left[  \frac{\partial }{\partial q_{1 \beta}^\textnormal{A}} \left
(    a e^{ \mathrm{i} S/\hbar}  \right) \right] \right \} \nonumber \\
&=& \Re \left [  a \frac{\partial^2 a}{\partial q_{1 \alpha}^\textnormal{A} \partial q_{1 \beta}^\textnormal{A}} - \frac{1}{\hbar^2} a^2 \frac{\partial S}{\partial q_{1 \alpha}^\textnormal{A}} \frac{\partial S}{\partial q_{1 \beta}^\textnormal{A}} \right. \nonumber \\ 
&&  \left. \; + \; \frac{\mathrm{i}}{\hbar} a \left(  \frac{\partial a}{\partial q_{1 \alpha}^\textnormal{A}}  \frac{\partial S}{\partial q_{1 \beta}^\textnormal{A}} +  \frac{\partial a}{\partial q_{1 \beta}^\textnormal{A}}  \frac{\partial S}{\partial q_{1 \alpha}^\textnormal{A}} + a \frac{\partial^2 S}{\partial q_{1 \alpha}^\textnormal{A} \partial q_{1 \beta}^\textnormal{A}} \right) \right] \nonumber \\
&=& - \; a^2 \left ( \frac{m_\textnormal{A}^2}{\hbar^2} w_{1 \alpha}^\textnormal{A} w_{1 \beta}^\textnormal{A} - \frac{1}{a} \frac{\partial^2 a}{\partial q_{1 \alpha}^\textnormal{A} \partial q_{1 \beta}^\textnormal{A} } \right). \label{Transformation 1 Pi Tensor}
\end{eqnarray}
And the transformation for the second term is:
\begin{eqnarray}
- \; \Re \left[ \left( \frac{\partial \Psi^*} {\partial q_{1 \beta}^\textnormal{A}} \right) \left ( \frac{\partial \Psi} {\partial q_{1 \alpha}^\textnormal{A}} \right) \right] &=& 
- \; \Re \left[ \left( \frac{\partial \left( a e^{- \mathrm{i} S/\hbar} \right)} {\partial q_{1 \beta}^\textnormal{A}} \right) \left ( \frac{\partial \left( a e^{ \mathrm{i} S/\hbar} \right)} {\partial q_{1 \alpha}^\textnormal{A}} \right) \right] \nonumber \\
&=& - \; \Re \left\{ \frac{\partial a}{\partial q_{1 \alpha}^\textnormal{A}}  \frac{\partial a}{\partial q_{1 \beta}^\textnormal{A}} + \frac{1}{\hbar^2} a^2 \frac{\partial S}{\partial q_{1 \alpha}^\textnormal{A}} \frac{\partial S}{\partial q_{1 \beta}^\textnormal{A}} \right. \nonumber \\
&& \left. + \; \frac{\mathrm{i}}{\hbar} a \left [ \left( \frac{\partial a}{\partial q_{1 \beta}^\textnormal{A}} \right) \left( \frac{\partial S}{\partial q_{1 \alpha}^\textnormal{A}} \right) -  \left( \frac{\partial a}{\partial q_{1 \alpha}^\textnormal{A}} \right) \left( \frac{\partial S}{\partial q_{1 \beta}^\textnormal{A}} \right) \right] \right \} \nonumber \\
&=& - \; a^2 \left ( \frac{m_\textnormal{A}^2}{\hbar^2} w_{1 \alpha}^\textnormal{A} w_{1 \beta}^\textnormal{A} + \frac{1}{a^2} \frac{\partial a}{\partial q_{1 \alpha}^\textnormal{A}} \frac{\partial a}{\partial q_{1 \beta}^\textnormal{A}} \right). \label{Transformation 2 Pi Tensor}
\end{eqnarray}
Thus, by inserting Eqns.\ (\ref{Transformation 1 Pi Tensor}) and (\ref{Transformation 2 Pi Tensor}) into Eqn.\ (\ref{total momentum flow density tensor version 1}), we get as an intermediate result for the tensor elements: 
\begin{eqnarray}
\Pi_{\alpha \beta}^{\textnormal{A}}(\vec q,t) &=& N(\textnormal{A}) \int \textnormal{d} \vec Q \; \delta (\vec q - \vec q_1^{\hspace{0.05cm}\textnormal{A}}) 
 \nonumber \\
&& \times \; a^2 \left [ m_\textnormal{A} w_{1\alpha}^\textnormal{A} w_{1\beta}^\textnormal{A} - \frac{\hbar^2}{2 m_\textnormal{A}} \left ( \frac{1}{a} \frac{\partial^2 a}{\partial q_{1 \alpha}^\textnormal{A} \partial q_{1 \beta}^\textnormal{A} } -  \frac{1}{a^2} \frac{\partial a}{\partial q_{1 \alpha}^\textnormal{A}} \frac{\partial a}{\partial q_{1 \beta}^\textnormal{A}} \right) \right]. \label{total momentum flow density tensor intermediate result}                                       
\end{eqnarray}
Regarding $a^2 = D$, it can be derived that:
\begin{eqnarray}
\frac{1}{2} \frac{\partial^2 \ln D }{\partial q_{1 \alpha}^\textnormal{A} \partial q_{1 \beta}^\textnormal{A}} &=& \frac{1}{2} \frac{\partial}{\partial q_{1 \alpha}^\textnormal{A}} \left ( \frac{1}{D}   \frac{\partial D }{\partial q_{1 \beta}^\textnormal{A}} \right) 
\; = \; \frac{1}{2} \frac{\partial}{\partial q_{1 \alpha}^\textnormal{A}} \left ( \frac{1}{a^2}   \frac{\partial a^2 }{\partial q_{1 \beta}^\textnormal{A}} \right) 
\nonumber \\
&=&      \frac{\partial}{\partial q_{1 \alpha}^\textnormal{A}} \left( \frac{1}{a} \frac{\partial a}{\partial q_{1 \beta}^\textnormal{A}} \right)  
\; = \;  \frac{1}{a} \frac{\partial^2 a}{\partial q_{1 \alpha}^\textnormal{A} \partial q_{1 \beta}^\textnormal{A} } -  \frac{1}{a^2} \frac{\partial a}{\partial q_{1 \alpha}^\textnormal{A}} \frac{\partial a}{\partial q_{1 \beta}^\textnormal{A}}. \label{Transformation ln D term}   
\end{eqnarray}
So we can write Eqn.\ (\ref{total momentum flow density tensor intermediate result}) in the following form, which is the second way to express $\Pi_{\alpha \beta}^{\textnormal{A}}(\vec q,t)$: 
\begin{eqnarray}
\Pi_{\alpha \beta}^{\textnormal{A}}(\vec q,t) &=& N(\textnormal{A}) \int \textnormal{d} \vec Q \; \delta (\vec q - \vec q_1^{\hspace{0.05cm}\textnormal{A}}) 
 \; D \left ( m_\textnormal{A} w_{1\alpha}^\textnormal{A} w_{1\beta}^\textnormal{A} - \frac{\hbar^2}{4 m_\textnormal{A}}  \frac{\partial^2 \ln D }{\partial q_{1 \alpha}^\textnormal{A} \partial q_{1 \beta}^\textnormal{A}} \right). \label{total momentum flow density tensor result 3} 
\end{eqnarray}
Having found this result for $\Pi_{\alpha \beta}^{\textnormal{A}}(\vec q,t)$,  we can define the elements $\Pi_{\alpha \beta}^{\textnormal{tot}}(\vec q,t)$ of the momentum-flow density tensor for all particles by:
\begin{eqnarray}
\Pi_{\alpha \beta}^{\textnormal{tot}}(\vec q,t) &=& \sum_{\textnormal{A}=1}^{N_S} \Pi_{\alpha \beta}^{\textnormal{A}}(\vec q,t)  \label{summing up Pi}  \\
&=& \sum_{\textnormal{A}=1}^{N_S} N(\textnormal{A}) \int \textnormal{d} \vec Q \; \delta (\vec q - \vec q_1^{\hspace{0.05cm}\textnormal{A}}) 
 \; D \left ( m_\textnormal{A} w_{1\alpha}^\textnormal{A} w_{1\beta}^\textnormal{A} - \frac{\hbar^2}{4 m_\textnormal{A}}  \frac{\partial^2 \ln D }{\partial q_{1 \alpha}^\textnormal{A} \partial q_{1 \beta}^\textnormal{A}} \right). \label{k sort momentum flow density tensor result}  
\end{eqnarray}
Moreover, both the matrix elements $\Pi_{\alpha \beta}^{\textnormal{A}}(\vec q,t)$ for a particular sort of particles $\textnormal{A}$ and the matrix elements $\Pi_{\alpha \beta}^{\textnormal{tot}}(\vec q,t)$ for all particles can be split each in a classical part (cl) and a quantum part (qu):
\begin{eqnarray}
\Pi_{\alpha \beta}^{\textnormal{A}}(\vec q,t) &=& \Pi_{\alpha \beta}^{\textnormal{A},\textnormal{cl}}(\vec q,t) + \Pi_{\alpha \beta}^{\textnormal{A},\textnormal{qu}}(\vec q,t), \label{Split Pi 1} \\
\Pi_{\alpha \beta}^{\textnormal{A},\textnormal{cl}}(\vec q,t) &=& N(\textnormal{A}) \hspace{0.075cm}  m_\textnormal{A} \int \textnormal{d} \vec Q \; \delta (\vec q - \vec q_1^{\hspace{0.05cm}\textnormal{A}}) \; D \; w_{1\alpha}^\textnormal{A} w_{1\beta}^\textnormal{A}, \label{Split Pi 2}  \\
\Pi_{\alpha \beta}^{\textnormal{A},\textnormal{qu}}(\vec q,t) &=& - \; \hbar^2 \frac{N(\textnormal{A})}{4 m_\textnormal{A}} \int \textnormal{d} \vec Q \; \delta (\vec q - \vec q_1^{\hspace{0.05cm}\textnormal{A}}) \; D \frac{\partial^2 \ln D }{\partial q_{1 \alpha}^\textnormal{A} \partial q_{1 \beta}^\textnormal{A}}, \label{Split Pi 3} \\
\Pi_{\alpha \beta}^{\textnormal{tot}}(\vec q,t) &=&  \Pi_{\alpha \beta}^{\textnormal{tot,cl}}(\vec q,t) + \Pi_{\alpha \beta}^{\textnormal{tot,qu}}(\vec q,t), \label{Split Pi 4} \\
\Pi_{\alpha \beta}^{\textnormal{tot,cl}}(\vec q,t) &=& \sum_{\textnormal{A}=1}^{N_S} \Pi_{\alpha \beta}^{\textnormal{A},\textnormal{cl}}(\vec q,t) \; = \; \sum_{\textnormal{A}=1}^{N_S} N(\textnormal{A}) \hspace{0.075cm}  m_\textnormal{A} \int \textnormal{d} \vec Q \; \delta (\vec q - \vec q_1^{\hspace{0.05cm}\textnormal{A}}) \; D \; w_{1\alpha}^\textnormal{A} w_{1\beta}^\textnormal{A}, \label{Split Pi 5}  \\
\Pi_{\alpha \beta}^{\textnormal{tot,qu}}(\vec q,t) &=& \sum_{\textnormal{A}=1}^{N_S} \Pi_{\alpha \beta}^{\textnormal{A},\textnormal{qu}}(\vec q,t) \; = \; - \; \hbar^2 \sum_{\textnormal{A}=1}^{N_S} \frac{N(\textnormal{A})}{4 m_\textnormal{A}} \int \textnormal{d} \vec Q \; \delta (\vec q - \vec q_1^{\hspace{0.05cm}\textnormal{A}}) \; D \frac{\partial^2 \ln D }{\partial q_{1 \alpha}^\textnormal{A} \partial q_{1 \beta}^\textnormal{A}}. \label{Split Pi 6} 
\end{eqnarray} 
So, the classical momentum flow density tensor $\underline{\underline{\Pi}}^{\textnormal{A},\textnormal{cl}}(\vec q,t)$ for the specific sort of particles $\textnormal{A}$ is related to a dyadic product of the velocity $\vec w_{1}^\textnormal{A}(\vec Q,t)$ of the $(\textnormal{A},1)$-particle with itself: 
\begin{eqnarray}
\underline{\underline{\Pi}}^{\textnormal{A},\textnormal{cl}}(\vec q,t) &=& N(\textnormal{A}) \hspace{0.075cm}  m_\textnormal{A} \int \textnormal{d} \vec Q \; \delta (\vec q - \vec q_1^{\hspace{0.05cm}\textnormal{A}}) \; D \; \left (\vec w_{1}^\textnormal{A} \otimes \vec w_{1}^\textnormal{A} \right). \label{Split Pi 2 tensorproduct}  
\end{eqnarray}
According to this point, the classical momentum flow density tensor $\underline{\underline{\Pi}}^{\textnormal{tot,cl}}(\vec q,t)$ for the total particle ensemble is related to dyadic products $\vec w_{1}^\textnormal{A} \otimes \vec w_{1}^\textnormal{A} $ for all sorts of particles $\textnormal{A} \in \{1,\ldots,$ \newline \noindent ${N_S}\}$:
\begin{eqnarray}
\underline{\underline{\Pi}}^{\textnormal{tot,cl}}(\vec q,t) &=& \sum_{\textnormal{A}=1}^{N_S} N(\textnormal{A}) \hspace{0.075cm}  m_\textnormal{A} \int \textnormal{d} \vec Q \; \delta (\vec q - \vec q_1^{\hspace{0.05cm}\textnormal{A}}) \; D \; \left( \vec w_{1}^\textnormal{A} \otimes \vec w_{1}^\textnormal{A} \right).  \label{Split Pi 5 tensorproduct} 
\end{eqnarray}
This relation of the classical tensors $\underline{\underline{\Pi}}^{\textnormal{A},\textnormal{cl}}(\vec q,t)$, $\underline{\underline{\Pi}}^{\textnormal{tot,cl}}(\vec q,t)$ to dyadic products of particle velocities is an analog to the calculation of the momentum flow density tensor $\underline{\underline{\Pi}}$ in classical hydrodynamics: The equation which is a classical analog to the Ehrenfest equation of motion can be found in \cite{Choudhuri_1988}, p.\ 32 and \cite{Shu_1992}, p.\ 21. Viewing this equation, one can realize that in classical hydrodynamics, the momentum flow density tensor $\underline{\underline{\Pi}}$ 
contains dyadic pro\-ducts of particle velocities. So, this is why we name $\underline{\underline{\Pi}}^{\textnormal{A,cl}}(\vec q,t)$, $\underline{ \underline{\Pi}}^{\textnormal{tot,cl}}(\vec q,t)$ classical tensors.  \newline
As a consequence of Eqns.\ (\ref{Split Pi 3}), (\ref{Split Pi 6}), compact forms also exist for the quantum tensors $\underline{\underline{\Pi}}^{\textnormal{A},\textnormal{qu}}(\vec q,t)$  and  $\underline{\underline{\Pi}}^{\textnormal{tot,qu}}(\vec q,t)$:
\begin{eqnarray}
\underline{\underline{\Pi}}^{\textnormal{A},\textnormal{qu}}(\vec q,t) &=& - \; \hbar^2 \frac{N(\textnormal{A})}{4 m_\textnormal{A}} \int \textnormal{d} \vec Q \; \delta (\vec q - \vec q_1^{\hspace{0.05cm}\textnormal{A}}) \; D \; \left( \nabla_1^\textnormal{A} \otimes \nabla_1^\textnormal{A} \right) \ln D,  \label{Split Pi 3 Tensorproduct} \\ 
\underline{\underline{\Pi}}^{\textnormal{tot},\textnormal{qu}}(\vec q,t) &=& - \;  \hbar^2 \sum_{\textnormal{A}=1}^{N_S} \frac{N(\textnormal{A})}{4 m_\textnormal{A}}
\int \textnormal{d} \vec Q \; \delta (\vec q - \vec q_1^{\hspace{0.05cm}\textnormal{A}}) \; D \;  \left( \nabla_1^\textnormal{A} \otimes \nabla_1^\textnormal{A} \right) \ln D.   \label{Split Pi 6 Tensorproduct} 
\end{eqnarray}
The term  $\nabla_1^\textnormal{A} \otimes \nabla_1^\textnormal{A}$ appearing in the two equations above is a dyadic product of the nabla operator  $\nabla_1^\textnormal{A}$ for the ($\textnormal{A},1$)-particle. In contrast to the classical parts, both the quantum tensor $\underline{\underline{\Pi}}^{\textnormal{A},\textnormal{qu}}(\vec q,t)$ for a certain sort of particles $\textnormal{A}$ and the quantum tensor $\underline{\underline{\Pi}}^{\textnormal{tot,qu}}(\vec q,t)$ for the total particle ensemble are related only to properties of $D(\vec Q,t)$, and they vanish in the limit $\hbar \rightarrow 0$. Therefore we name these tensors as quantum tensors. \newpage \noindent
Finally, we can find using Eqns.\ (\ref{Split Pi 1}), (\ref{Split Pi 4}), and  (\ref{Split Pi 2 tensorproduct}) --  (\ref{Split Pi 6 Tensorproduct}) these compact forms for the tensors $\underline{\underline{\Pi}}^{\textnormal{A}}(\vec q,t)$ and $\underline{\underline{\Pi}}^{\textnormal{tot}}(\vec q,t)$: 
\begin{eqnarray}
\underline{\underline{\Pi}}^{\textnormal{A}}(\vec q,t) &=& N(\textnormal{A}) \; \int \textnormal{d} \vec Q \; \delta (\vec q - \vec q_1^{\hspace{0.05cm}\textnormal{A}}) \; D \; \times \nonumber \\
&& \left[ m_\textnormal{A} \left ( \vec w_{1}^\textnormal{A} \otimes \vec w_{1}^\textnormal{A} \right) - \frac{\hbar^2}{4 m_\textnormal{A}} \left( \nabla_1^\textnormal{A} \otimes \nabla_1^\textnormal{A} \right) \ln D \right],  \label{Split Pi 1 Tensorproduct} \\  
\underline{\underline{\Pi}}^{\textnormal{tot}}(\vec q,t) &=& \sum_{\textnormal{A}=1}^{N_S} \underline{\underline{\Pi}}^{\textnormal{A}}(\vec q,t) \label{summing up Pi tensors} \\ \nonumber
&=& \sum_{\textnormal{A}=1}^{N_S} N(\textnormal{A}) \int \textnormal{d} \vec Q \; \delta (\vec q - \vec q_1^{\hspace{0.05cm}\textnormal{A}}) \; D \; \times \nonumber \\ \nonumber \\ 
&& \left[ m_\textnormal{A} \left ( \vec w_{1}^\textnormal{A} \otimes \vec w_{1}^\textnormal{A} \right) - \frac{\hbar^2}{4 m_\textnormal{A}} \left( \nabla_1^\textnormal{A} \otimes \nabla_1^\textnormal{A} \right) \ln D  \right].  \label{Split Pi 4 Tensorproduct}  
\end{eqnarray}
As a next task, we sum up the MPEEM for a certain sort of particles (\ref{MMBE}) over all particle sorts $\textnormal{A} \in \{1, \ldots, {N_S} \}$. Then we regard Eqns.\ (\ref{summing up flux densities}),  (\ref{summing up force density}), (\ref{summing up Pi tensors}) for the quantities $\vec j^{\textnormal{tot}}_m(\vec q,t)$, $\vec f^{\hspace{0.05cm} \textnormal{tot}}(\vec q,t)$, and $\underline{\underline{\Pi}}^{\textnormal{tot}}(\vec q,t)$, thus, finding the MPEEM for all particles: 
\begin{eqnarray}
\frac{\partial \vec j^{\textnormal{tot}}_m(\vec q,t)}{\partial t} &=& \vec f^{\hspace{0.05cm} \textnormal{tot}}(\vec q,t) - \nabla \underline{\underline{\Pi}}^{ \textnormal{tot}}(\vec q,t). \label{MMBE for all particles}
\end{eqnarray}
Therefore, there is an MPEEM both for all particles and for each sort of particles. The MPEEM for all sorts of particles (\ref{MMBE for all particles}) can be solved numerically to find $\vec j^{\textnormal{tot}}_m(\vec q,t)$ if one knows $\vec f^{\hspace{0.05cm} \textnormal{tot}}(\vec q,t)$
and $\underline{\underline{\Pi}}^{ \textnormal{tot}}(\vec q,t)$. In an analogous manner, the MPEEM for a certain sort of particles $\textnormal{A}$ (\ref{MMBE}) can be solved numerically if $\vec f^{\hspace{0.05cm} \textnormal{A}}(\vec q,t)$ and  $\underline{\underline{\Pi}}^{\textnormal{A}}(\vec q,t)$ are known. Although one might wonder why this idea is interesting because one can calculate the mass current density $\vec j_m^{\textnormal A}(\vec q,t)$ also directly with Eqn.\ (\ref{k_flux_density}) or the mass current density $\vec j_m^{\textnormal{tot}}(\vec q,t)$ also directly with Eqn.\ (\ref{Total particle current density 2}), respectively, this is an important option for the numerical application of MPQHD. The reason for this is that there are cases for molecular systems where only a wave function $\Psi^{\textnormal{BO}}(\vec Q,t)$ within the Born-Oppenheimer approximation is available and the direct calculation method for electronic mass current densities fails \cite{Barth_2009}. Therefore, one has to search for alternative approaches to calculate the electronic mass current densities, and solving the MPEEM (\ref{MMBE}) numerically for the case that the electrons are the particles of the sort $\textnormal{A}$ could be an option.\newline 
Having derived the MPEEM both for all particles and for each sort of particles, we will now derive the corresponding many-particle quantum Cauchy equations  (MPQCE). \newpage \noindent
\subsection{Derivation of the MPQCE} \label{Derivation MQNSE}
The starting point of the derivation of the MPQCE for particles of the sort $\textnormal{A}$ is the corres\-ponding MPEEM (\ref{MMBE}): \newline Taking into account the MPCE (\ref{CE particle sort k}) for particles of the sort $\textnormal{A}$, we transform the term $\frac{\partial \vec j^{\textnormal{A}}_m(\vec q,t)}{\partial t}$ appearing in Eqn.\ (\ref{MMBE}) (where $\vec e_\alpha, \alpha \in K_{\textnormal{Ca}}$ are the Cartesian unit vectors): 
\begin{eqnarray}
\frac{\partial \vec j^{\textnormal{A}}_m(\vec q,t)}{\partial t} &=& \frac{\partial}{\partial t} \left( \rho^\textnormal{A}_m \vec v^\textnormal{A} \right) =  \frac{\partial \rho^\textnormal{A}_m}{\partial t} \vec v^\textnormal{A}  + \rho^\textnormal{A}_m \frac{\partial \vec v^\textnormal{A}}{\partial t}  \nonumber \\
&=& - \left ( \nabla  \vec j^\textnormal{A}_m \right) \vec v^\textnormal{A} + \rho^\textnormal{A}_m \frac{\partial \vec v^\textnormal{A}}{\partial t} \nonumber \\
&=& - \sum_{\alpha \in K_{\textnormal{Ca}}} \vec e_\alpha \left[ v_\alpha^\textnormal{A} \left(\nabla \vec j_m^\textnormal{A}\right) \right] + \rho_m^\textnormal{A} \frac{\partial \vec v^\textnormal{A}}{\partial t} \nonumber \\ 
&=& - \sum_{\alpha \in K_{\textnormal{Ca}}}  \vec e_\alpha \left[ \nabla \left(\vec j^\textnormal{A}_m v_{\alpha}^\textnormal{A} \right) \right]  + \underbrace{\sum_{\alpha \in K_{\textnormal{Ca}}} \vec e_\alpha \left [ \vec j^\textnormal{A}_m \left (\nabla v_\alpha^\textnormal{A} \right) \right]}_{= \; \left(\rho^\textnormal{A}_m \vec v^\textnormal{A} \nabla \right) \vec v^\textnormal{A}}  + \rho^\textnormal{A}_m \frac{\partial \vec v^\textnormal{A}}{\partial t} \nonumber \\
&=& \rho^\textnormal{A}_m \left [ \frac{\partial}{\partial t} + \left( \vec v^\textnormal{A} \nabla \right) \right] \vec v^\textnormal{A} - \sum_{\alpha \in K_{\textnormal{Ca}}}  \sum_{\beta \in K_{\textnormal{Ca}}} \vec e_\alpha \left[ \frac{\partial}{\partial q_\beta} \left( \rho^\textnormal{A}_m v_\beta^\textnormal{A} v_\alpha^\textnormal{A} \right) \right] \nonumber \\
&=& \rho^\textnormal{A}_m \left [ \frac{\partial}{\partial t} + \left( \vec v^\textnormal{A} \nabla \right) \right] \vec v^\textnormal{A} - \sum_{\alpha \in K_{\textnormal{Ca}}}  \sum_{\beta \in K_{\textnormal{Ca}}} \frac{\partial \left( \rho^\textnormal{A}_m v_\alpha^\textnormal{A} v_\beta^\textnormal{A} \right)}{\partial q_\alpha}  \vec e_\beta.\label{Transformation kMBE 1}
\end{eqnarray}
Applying the definition (\ref{Definition Chapter 2}) for tensor divergences and the notation used before for dyadic products, we get: 
\begin{eqnarray}
\frac{\partial \vec j^{\textnormal{A}}_m(\vec q,t)}{\partial t} &=& \rho_m^\textnormal{A} \left[ \frac{\partial}{\partial t} + \left( \vec v^\textnormal{A} \nabla \right) \right] \vec v^\textnormal{A} - \nabla \left [ \rho_m^\textnormal{A} \left( \vec v^\textnormal{A} \otimes  \vec v^\textnormal{A} \right) \right].  \label{Transformation kMBE 2}
\end{eqnarray}
Then, we insert Eqn.\ (\ref{Transformation kMBE 2}) into Eqn.\ (\ref{MMBE}) and find:
\begin{eqnarray}
\rho^\textnormal{A}_m \left [ \frac{\partial}{\partial t} + \left( \vec v^\textnormal{A} \nabla \right) \right] \vec v^\textnormal{A} &=& \vec f^{\hspace {0.05 cm}\textnormal{A}} - \nabla \left[ \underline{\underline{\Pi}}^\textnormal{A} -  \rho_m^\textnormal{A} \left( \vec v^\textnormal{A} \otimes  \vec v^\textnormal{A} \right) \right]. \label{Transformation kMBE 3}
\end{eqnarray}
Now, we introduce a new quantity called pressure tensor $\underline{\underline {p}}^\textnormal{A}(\vec q,t)$ for the sort of particles $\textnormal{A}$:
\begin{eqnarray}
\underline{\underline {p}}^\textnormal{A}(\vec q,t) &=& \underline{\underline {\Pi}}^\textnormal{A}(\vec q,t) - \rho_m^\textnormal{A}(\vec q,t) \left[ \vec v^\textnormal{A} (\vec q,t) \otimes  \vec v^\textnormal{A} (\vec q,t) \right]. \label{first def p sort A}
\end{eqnarray}
More properties of this tensor are discussed in Sec.\ \ref{pressure tensor}. \newline
So, we get the MPQCE for the sort of particles $\textnormal{A}$ by combining Eqns.\ (\ref{Transformation kMBE 3}) and (\ref{first def p sort A}):    
\begin{eqnarray}
\rho^\textnormal{A}_m(\vec q,t) \left [ \frac{\partial}{\partial t} + \left( \vec v^\textnormal{A}(\vec q,t) \nabla \right) \right] \vec v^\textnormal{A}(\vec q,t) &=& \vec f^{\hspace{0.05 cm \textnormal{A}}}(\vec q,t)  - \nabla \underline{\underline{p}}^\textnormal{A}(\vec q,t).  \label{MQNSE for particle sort k}
\end{eqnarray}
Now, we discuss why we name the equation above many-particle quantum Cauchy equation (MPQCE): \newline 
In classical hydrodyamics, there is a differential equation named Cauchy's equation of motion, which is related to the momentum balance in a liquid. It is given by \cite{Acheson_2005}, p. 205, \cite{Stokes_1966}:  
\begin{eqnarray}
\rho_m(\vec q,t) \frac{d \vec v(\vec q,t)}{d t}  &=&  \vec f(\vec q,t)  + \nabla \underline{\underline{\sigma}}(\vec q,t), \label{3a}
\end{eqnarray}
In the equation above, the quantity $\underline{\underline{\sigma}} (\vec q,t)$ is the stress tensor. Moreover, the term $\frac{d \vec v(\vec q,t)}{d t}$ is the total rate of change of the velocity and it is given by \cite{Acheson_2005}, p. 4-6: 
\begin{eqnarray}
\frac{d \vec v(\vec q,t)}{d t} &=&  \left [ \frac{\partial}{\partial t} + \left( \vec v(\vec q,t) \nabla \right) \right] \vec v(\vec q,t).  \label{3b}
\end{eqnarray}
This equation shows that the total rate of change of the velocity $\frac{d \vec v(\vec q,t)}{d t}$ is given by the sum of two terms: The first term $\frac{\partial \vec v(\vec q,t)}{\partial t}$ is the local rate of change of the velocity at a fixed position $\vec q$, and the second term $\left( \vec v(\vec q,t) \nabla \right) \vec v(\vec q,t)$ is related to the effect that the flow transports the fluid elements to other positions where the velocity of the streaming can differ. \newline 
If we now identify   
\begin{eqnarray}
\underline{\underline{\sigma}} (\vec q,t) = - \underline{\underline{p}} (\vec q,t), \label{3}
\end{eqnarray}
and insert Eqns.\ (\ref{3b}) and (\ref{3}) into Cauchy's equation of motion (\ref{3a}), we realize that Cauchy's equation of motion takes indeed the form of the MPQCE (\ref{MQNSE for particle sort k}). So, the MPQCE, which we derived with basic quantum mechanics, is a quantum analog to Cauchy's equation of motion known in classical hydrodynamics.\newline
We mention that in classical hydrodynamics, Cauchy's equation of motion becomes the Navier-Stokes equation by applying the approximation:   
\begin{eqnarray}
\nabla \underline{\underline{\sigma}}(\vec q,t) &\approx& - \nabla  P(\vec q,t) + \eta \laplace \vec v(\vec q,t) + \left(\zeta + \frac{\eta}{3} \right) \nabla \left ( \nabla \vec v(\vec q,t) \right), 
\end{eqnarray} 
where  $P(\vec q,t)$ is the scalar pressure, and $\zeta$ and $\eta$ are called coefficients of viscosity. So, the Navier-Stokes equation has the following form (\hspace{-0.03 cm}\cite{Landau_2000}, p.\ 44f. and \cite{Discussion}):  
\begin{eqnarray}
&& \rho_m(\vec q,t)  \left [ \frac{\partial}{\partial t} + \left( \vec v(\vec q,t) \nabla \right) \right] \vec v(\vec q,t) \; = \nonumber \\ 
&& \hspace{1.0cm} \vec f(\vec q,t)  - \nabla  P(\vec q,t) + \eta \laplace \vec v(\vec q,t) + \left(\zeta + \frac{\eta}{3} \right) \nabla \left ( \nabla \vec v(\vec q,t) \right). \label{Navier-Stokes-equation}
\end{eqnarray}
In \cite{Harvey_1966}, Harvey called the MPQCE (90) for the case of a quantum system for a single particle ``quantum-mechanical Navier-Stokes equation''. However, we think that this analogy is less precise than the analogy of the MPQCE (\ref{MQNSE for particle sort k}) to Cauchy's equation of motion (\ref{3a}). The reason for this is that the analogy of the tensor gradient term $-\nabla \underline{\underline{p}}^{\textnormal A}(\vec q,t)$ appearing in the MPQCE (\ref{MQNSE for particle sort k}) to the tensor gradient term $\nabla  \underline{\underline{\sigma}}(\vec q,t) = - \nabla  \underline{\underline{p}}(\vec q,t)$ in Cauchy's equation of motion is much closer than the analogy of the mentioned tensor gradient term $-\nabla  \underline{\underline{p}}^{\textnormal A}(\vec q,t)$ to the complicated term $- \nabla  P(\vec q,t) + \eta \laplace \vec v(\vec q,t) + (\zeta + \frac{\eta}{3}) \nabla(\nabla \vec v(\vec q,t))$ in the Navier-Stokes equation (\ref{Navier-Stokes-equation}).
\newline \newline
As the next step, we derive the MPQCE for all particles. The MPQCEs which are specific for a certain sort of particles are non-linear differential equations because of the non-linear term $\rho^\textnormal{A}_m \left ( \vec v^\textnormal{A}  \nabla \right) \vec v^\textnormal{A}$ in Eqn.\ (\ref{MQNSE for particle sort k}). Therefore, we cannot derive the MPQCE for all particles just by summing up all the MPQCEs specific for a certain sort of particles. \newline 
Here, one realizes a contrast to the derivation of the MPCE and the MPEEM for the total particle ensemble (see Eqns.\ (\ref{CE all particles}), (\ref{MMBE for all particles})), which could be derived by summing up all the corresponding equations for the particular sorts of particles (see Eqns.\ (\ref{CE particle sort k}), (\ref{MMBE})). This context is related to the point that both the MPCE and the MPEEM for a particular sort of particles and the corresponding equations for the total particle ensemble are linear differential equations for which the superposition principle is true, which says that linear combinations of their solutions form new solutions of these equations. \newline 
However, the following derivation for the MPQCE for all particles is still quite similar to the derivation of Eqn.\ (\ref{MQNSE for particle sort k}) because the MPQCE for all particles can be derived from the MPEEM (\ref{MMBE for all particles}) for all particles in an analogous manner like the MPQCE (\ref{MQNSE for particle sort k}) specific for a certain sort of particles can be derived from the MPEEM (\ref{MMBE}) for a certain sort of particles. \newline \newline 
Therefore, we find a new expression for the time derivation term  $\frac{\partial \vec j^{\textnormal{tot}}_m(\vec q,t)}{\partial t}$ in Eqn.\ (\ref{MMBE for all particles}), and doing so, we insert the MPCE (\ref{CE all particles}) for all particles: 
\begin{eqnarray}
\frac{\partial \vec j^{\textnormal{tot}}_m(\vec q,t)}{\partial t} &=& \frac{\partial}{\partial t} \left( \rho^{\textnormal{tot}}_m \vec v^{\hspace{0.05cm} \textnormal{tot}} \right) =  \frac{\partial \rho^{\textnormal{tot}}_m}{\partial t} \vec v^{\hspace{0.05cm} \textnormal{tot}}  + \rho^{\textnormal{tot}}_m \frac{\partial \vec v^{\hspace{0.05cm} \textnormal{tot}}}{\partial t}  \nonumber \\
&=& - \left ( \nabla  \vec j^{\textnormal{tot}}_m \right) \vec v^{\hspace{0.05cm} \textnormal{tot}} + \rho^{\textnormal{tot}}_m \frac{\partial \vec v^{\hspace{0.05cm} \textnormal{tot}}}{\partial t} \nonumber \\
&=& -  \sum_{\alpha \in K_{\textnormal{Ca}}} \vec e_\alpha \left[ v_\alpha^\textnormal{tot} \left(\nabla \vec j_m^\textnormal{tot} \right) \right] + \rho_m^\textnormal{tot} \frac{\partial \vec v^{\hspace{0.05cm} \textnormal{tot}}}{\partial t} \nonumber \\ 
&=& - \sum_{\alpha \in K_{\textnormal{Ca}}} \vec e_\alpha \left[ \nabla \left(\vec j^{\textnormal{tot}}_m v_{\alpha}^{\textnormal{tot}} \right) \right]  + \underbrace{\sum_{\alpha \in K_{\textnormal{Ca}}} \vec e_\alpha \left [ \vec j^{\textnormal{tot}}_m \left (\nabla v_\alpha^{\textnormal{tot}} \right) \right]}_{= \; \left(\rho^{\textnormal{tot}}_m \vec v^{\hspace{0.05cm} \textnormal{tot}} \nabla \right) \vec v^{\hspace{0.05cm} \textnormal{tot}}}  + \rho^{\textnormal{tot}}_m \frac{\partial \vec v^{\hspace{0.05cm} \textnormal{tot}}}{\partial t} \nonumber \\ 
&=& \rho^{\textnormal{tot}}_m \left [ \frac{\partial}{\partial t} + \left( \vec v^{\hspace{0.05cm} \textnormal{tot}} \nabla \right) \right] \vec v^{\hspace{0.05cm} \textnormal{tot}} - \sum_{\alpha \in K_{\textnormal{Ca}}}  \sum_{\beta \in K_{\textnormal{Ca}}} \vec e_\alpha \left[ \frac{\partial}{\partial q_\beta} \left( \rho^{\textnormal{tot}}_m v_\beta^{\textnormal{tot}} v_\alpha^{\textnormal{tot}} \right) \right] \nonumber \\
&=& \rho^{\textnormal{tot}}_m \left [ \frac{\partial}{\partial t} + \left( \vec v^{\hspace{0.05cm} \textnormal{tot}} \nabla \right) \right] \vec v^{\hspace{0.05cm} \textnormal{tot}} - \sum_{\alpha \in K_{\textnormal{Ca}}} \sum_{\beta \in K_{\textnormal{Ca}}}  \frac{\partial \left( \rho^{\textnormal{tot}}_m v_\alpha^{\textnormal{tot}} v_\beta^{\textnormal{tot}} \right)}{\partial q_\alpha}  \vec e_\beta. \label{Transformation kMBE 1 tot}
\end{eqnarray}
Regarding the tensor divergence definition (\ref{Definition Chapter 2}), we now write the last term in the equation above as a tensor divergence: 
\begin{eqnarray}
\frac{\partial \vec j^{\textnormal{tot}}_m(\vec q,t)}{\partial t} &=& \rho_m^\textnormal{tot} \left[ \frac{\partial}{\partial t} + \left( \vec v^{\hspace{0.05cm} \textnormal{tot}} \nabla \right) \right] \vec v^{\hspace{0.05cm} \textnormal{tot}} - \nabla \left [ \rho_m^\textnormal{tot} \left( \vec v^{\hspace{0.05cm} \textnormal{tot}} \otimes  \vec v^{\hspace{0.05cm} \textnormal{tot}} \right) \right].  \label{Transformation kMBE 2 tot}
\end{eqnarray}
After that, we use Eqn.\ (\ref{Transformation kMBE 2 tot}) for a transformation of Eqn.\ (\ref{MMBE for all particles}) and find:  
\begin{eqnarray}
\rho^{\textnormal{tot}}_m \left [ \frac{\partial}{\partial t} + \left( \vec v^{\hspace{0.05cm} \textnormal{tot}} \nabla \right) \right] \vec v^{\hspace{0.05cm} \textnormal{tot}} &=& \vec f^{\hspace {0.05 cm} \textnormal{tot}} - \nabla \left[\underline{\underline{\Pi}}^{\textnormal{tot}} - \rho_m^{\textnormal{tot}} \left( \vec v^{\hspace{0.05cm} \textnormal{tot}} \otimes  \vec v^{\hspace{0.05cm} \textnormal{tot}}  \right) \right]. \label{Transformation kMBE 3 tot}
\end{eqnarray}
Here, a new quantity is introduced called pressure tensor $\underline{\underline {p}}^\textnormal{tot}(\vec q,t)$ for the total particle ensemble:
\begin{eqnarray}
\underline{\underline {p}}^\textnormal{tot}(\vec q,t) &=& \underline{\underline {\Pi}}^\textnormal{tot}(\vec q,t) - \rho_m^\textnormal{tot}(\vec q,t) \left[ \vec v^{\hspace{0.05cm} \textnormal{tot}}(\vec q,t) \otimes  \vec v^{\hspace{0.05cm} \textnormal{tot}}(\vec q,t) \right]. \label{first def p all sorts}
\end{eqnarray}
We will discuss this tensor in more detail in Sec.\ \ref{pressure tensor}. \newline
Finally, we insert Eqn.\ (\ref{first def p all sorts}) for $\underline{\underline {p}}^\textnormal{tot}(\vec q,t)$ into Eqn.\ (\ref{Transformation kMBE 3 tot}) and get the MPQCE for the total ensemble of particles:
\begin{eqnarray}
\rho^{\textnormal{tot}}_m(\vec q,t) \left [ \frac{\partial}{\partial t} + \left( \vec v^{\hspace{0.05cm} \textnormal{tot}}(\vec q,t) \nabla \right) \right] \vec v^{\hspace{0.05cm} \textnormal{tot}}(\vec q,t) = \vec f^{\hspace{0.05 cm} \textnormal{tot}}(\vec q,t)  - \nabla \underline{\underline{p}}^{\textnormal{tot}}(\vec q,t).  \label{MQNSE for all particles}
\end{eqnarray}
As the next issue, we investigate the properties of the pressure tensors $\underline{\underline {p}}^\textnormal{A}(\vec q,t)$, $\underline{\underline {p}}^\textnormal{tot}(\vec q,t)$. 
\subsection{Pressure tensor} \label{pressure tensor}
We define the pressure tensor $\underline{\underline {p}}^\textnormal{A}(\vec q,t)$ for the sort of particles $\textnormal{A}$ as:
\begin{eqnarray}
\underline{\underline {p}}^\textnormal{A}(\vec q,t) &:=& N(\textnormal{A}) \int \textnormal{d} \vec Q \; \delta (\vec q - \vec q_1^{\hspace{0.05cm}\textnormal{A}}) \; D \; \times \nonumber \\ && 
\left[ m_\textnormal{A} \left ( \vec u_{1}^\textnormal{A} \otimes \vec u_{1}^\textnormal{A} \right) - \frac{\hbar^2}{4 m_\textnormal{A}} \left( \nabla_1^\textnormal{A} \otimes \nabla_1^\textnormal{A} \right) \ln D  \right],  \label{Definiton p sort A}
\end{eqnarray}
so that its components are given by: 
\begin{eqnarray}
p_{\alpha \beta}^\textnormal{A}(\vec q,t) &=& N(\textnormal{A}) \int \textnormal{d} \vec Q \; \delta (\vec q - \vec q_1^{\hspace{0.05cm}\textnormal{A}}) 
 \; D \left ( m_\textnormal{A} u_{1\alpha}^\textnormal{A} u_{1\beta}^\textnormal{A} - \frac{\hbar^2}{4 m_\textnormal{A}}  \frac{\partial^2 \ln D }{\partial q_{1 \alpha}^\textnormal{A} \partial q_{1 \beta}^\textnormal{A}} \right). \label{k sort pressure tensor result}  
\end{eqnarray}
$\left. \right.$ \newpage \noindent Moreover, we define the pressure tensor $\underline{\underline {p}}^\textnormal{tot}(\vec q,t)$ for the total particle ensemble as: 
\begin{eqnarray}
\underline{\underline {p}}^\textnormal{tot}(\vec q,t) &:=& \sum_{\textnormal{A}=1}^{N_S} N(\textnormal{A}) \int \textnormal{d} \vec Q \; \delta (\vec q - \vec q_1^{\hspace{0.05cm}\textnormal{A}}) \; D \; \times \nonumber \\ && 
\left[ m_\textnormal{A} \left ( \vec {\mathfrak{u}}_{1}^\textnormal{A} \otimes \vec {\mathfrak{u}}_{1}^\textnormal{A} \right) - \frac{\hbar^2}{4 m_\textnormal{A}} \left( \nabla_1^\textnormal{A} \otimes \nabla_1^\textnormal{A} \right) \ln D  \right],  \label{Definiton p all sorts}
\end{eqnarray}
and we can write its components as follows: 
\begin{eqnarray}
p_{\alpha \beta}^\textnormal{tot}(\vec q,t) &=&  \sum_{\textnormal{A}=1}^{N_S}  N(\textnormal{A}) \int \textnormal{d} \vec Q \; \delta (\vec q - \vec q_1^{\hspace{0.05cm}\textnormal{A}}) 
 \; D \left ( m_\textnormal{A} {\mathfrak{u}}_{1\alpha}^\textnormal{A} {\mathfrak{u}}_{1\beta}^\textnormal{A} - \frac{\hbar^2}{4 m_\textnormal{A}}  \frac{\partial^2 \ln D }{\partial q_{1 \alpha}^\textnormal{A} \partial q_{1 \beta}^\textnormal{A}} \right). \label{total pressure tensor result}  
\end{eqnarray} 
At the end of this Sec.\ \ref{pressure tensor}, we will prove that the definition (\ref{Definiton p sort A}) for $\underline{\underline {p}}^\textnormal{A}(\vec q,t)$ 
is equivalent to Eqn.\ (\ref{first def p sort A}), and that the definition (\ref{Definiton p all sorts}) for  $\underline{\underline {p}}^\textnormal{tot}(\vec q,t)$
is equivalent to Eqn.\ (\ref{first def p all sorts}). \newline \newline
But before we do that, we will first discuss that we can split up both the tensor components $p_{\alpha \beta}^\textnormal{A}(\vec q,t)$ and $p_{\alpha \beta}^{\textnormal{tot}}(\vec q,t)$ in a classical part and a quantum part in a similar manner like the tensor components $\Pi_{\alpha \beta}^\textnormal{A}(\vec q,t)$ and $\Pi_{\alpha \beta}^{\textnormal{tot}}(\vec q,t)$. We mention here that a splitting of the pressure tensor components in a classical part and a quantum part was already described by Wong \cite{Wong_1976}: 
\begin{eqnarray} 
p_{\alpha \beta}^{\textnormal{A}}(\vec q,t) &=& p_{\alpha \beta}^{\textnormal{A},\textnormal{cl}}(\vec q,t) + p_{\alpha \beta}^{\textnormal{A},\textnormal{qu}}(\vec q,t), \label{Split p 1} \\
p_{\alpha \beta}^{\textnormal{A},\textnormal{cl}}(\vec q,t) &=& N(\textnormal{A}) \hspace{0.075cm}  m_\textnormal{A}  \hspace{-0.05 cm} \int  \hspace{-0.05 cm} \textnormal{d} \vec Q \; \delta (\vec q - \vec q_1^{\hspace{0.05cm}\textnormal{A}}) \; D \; u_{1\alpha}^\textnormal{A} u_{1\beta}^\textnormal{A}, \label{Split p 2}  \\
p_{\alpha \beta}^{\textnormal{A},\textnormal{qu}}(\vec q,t) &=& - \; \hbar^2 \frac{N(\textnormal{A})}{4 m_\textnormal{A}}  \hspace{-0.05 cm} \int  \hspace{-0.05 cm} \textnormal{d} \vec Q \; \delta (\vec q - \vec q_1^{\hspace{0.05cm}\textnormal{A}}) \; D \frac{\partial^2 \ln D }{\partial q_{1 \alpha}^\textnormal{A} \partial q_{1 \beta}^\textnormal{A}}, \label{Split p 3} \\
p_{\alpha \beta}^{\textnormal{tot}}(\vec q,t) &=&  p_{\alpha \beta}^{\textnormal{tot,cl}}(\vec q,t) + p_{\alpha \beta}^{\textnormal{tot,qu}}(\vec q,t), \label{Split p 4} \\
p_{\alpha \beta}^{\textnormal{tot,cl}}(\vec q,t) &=& \sum_{\textnormal{A}=1}^{N_S} N(\textnormal{A}) \hspace{0.075cm}  m_\textnormal{A}  \hspace{-0.05 cm} \int  \hspace{-0.05 cm} \textnormal{d} \vec Q \; \delta (\vec q - \vec q_1^{\hspace{0.05cm}\textnormal{A}}) \; D \; \mathfrak{u}_{1\alpha}^\textnormal{A} \mathfrak{u}_{1\beta}^\textnormal{A}, \label{Split p 5}  \\
p_{\alpha \beta}^{\textnormal{tot,qu}}(\vec q,t) &=& \sum_{\textnormal{A}=1}^{N_S} p_{\alpha \beta}^{\textnormal{A},\textnormal{qu}}(\vec q,t) \; \hspace{-0.05 cm} = \; \hspace{-0.05 cm}  - \; \hbar^2 \sum_{\textnormal{A}=1}^{N_S} \frac{N(\textnormal{A})}{4 m_\textnormal{A}}  \hspace{-0.05 cm} \int  \hspace{-0.05 cm} \textnormal{d} \vec Q \; \delta (\vec q - \vec q_1^{\hspace{0.05cm}\textnormal{A}}) \; D \frac{\partial^2 \ln D }{\partial q_{1 \alpha}^\textnormal{A} \partial q_{1 \beta}^\textnormal{A}}. \label{Split p 6} 
\end{eqnarray}
Note that  
\begin{eqnarray}
p_{\alpha \beta}^{\textnormal{tot,cl}}(\vec q,t) &\neq& \sum_{\textnormal{A}=1}^{N_S} p_{\alpha \beta}^{\textnormal{A},\textnormal{cl}}(\vec q,t) \label{not additive 1}  \\
\Longrightarrow \; \; \; \; p_{\alpha \beta}^{\textnormal{tot}}(\vec q,t) &\neq& \sum_{\textnormal{A}=1}^{N_S} p_{\alpha \beta}^{\textnormal{A}}(\vec q,t), \label{not additive}
\end{eqnarray}
because in Eqn.\ (\ref{k sort pressure tensor result}), components of the velocity $\vec u_{1}^\textnormal{A}(\vec Q,t)$ appear which are defined by the velocity of the $(\textnormal{A},1)$-particle relative to the mean particle velocity $\vec v^\textnormal{A}(\vec q_1^{\hspace{0.05 cm} \textnormal{A}},t)$ of particles of the sort $\textnormal{A}$, while in Eqn.\ (\ref{total pressure tensor result}), components of the velocity $\vec{\mathfrak{u}}_{1}^\textnormal{A}(\vec Q,t)$ appear which are defined by the velocity of the $(\textnormal{A},1)$-particle relative to the mean particle velocity $\vec v^{\hspace{0.05cm}\textnormal{tot}}(\vec q_1^{\hspace{0.05 cm} \textnormal{A}},t)$ for the total particle ensemble. \newline
The inequation (\ref{not additive}) can be related to the fact mentioned above that the MPQCE (\ref{MQNSE for particle sort k}) for a certain sort of particles is a non-linear differential equation. So, the sum over this equation for all different sorts of particles does not yield the MPQCE (\ref{MQNSE for all particles}) for the total particle ensemble. Thus, it is reasonable that the sum over all sorts of particles for the pressure tensors $p_{\alpha \beta}^{\textnormal{A}}(\vec q,t)$ on the right side of Eqn.\ (\ref{not additive}) does not yield the pressure tensor $p_{\alpha \beta}^{\textnormal{tot}}(\vec q,t)$ for the total particle ensemble.   \newline \newline 
Moreover, we can write the classical pressure tensor $\underline{\underline{p}}^{\textnormal{A},\textnormal{cl}}(\vec q,t)$ for a specific sort of particles $\textnormal{A}$ in a compact tensor notation, where a dyadic product of the relative velocity $\vec {u}_{1}^\textnormal{A}(\vec Q,t)$ of the $(\textnormal{A},1)$-particle appears: 
\begin{eqnarray}
\underline{\underline{{p}}}^{\textnormal{A},\textnormal{cl}}(\vec q,t) &=& N(\textnormal{A}) \hspace{0.075cm}  m_\textnormal{A} \int \textnormal{d} \vec Q \; \delta (\vec q - \vec q_1^{\hspace{0.05cm}\textnormal{A}}) \; D \; \left (\vec u_{1}^\textnormal{A} \otimes \vec u_{1}^\textnormal{A} \right). \label{Split p 2 tensorproduct}  
\end{eqnarray}
In an analogous manner, the classical pressure tensor $\underline{\underline{p}}^{\textnormal{tot,cl}}(\vec q,t)$ for the total particle ensemble is related to dyadic products $\vec {\mathfrak{u}}_{1}^\textnormal{A}  \otimes \vec {\mathfrak{u}}_{1}^\textnormal{A}$ for all sorts of particles $\textnormal{A} \in \{1,\ldots, {N_S}\}$:
\begin{eqnarray}
\underline{\underline{{p}}}^{\textnormal{tot,cl}}(\vec q,t) &=& \sum_{\textnormal{A}=1}^{N_S} N(\textnormal{A}) \hspace{0.075cm}  m_\textnormal{A} \int \textnormal{d} \vec Q \; \delta (\vec q - \vec q_1^{\hspace{0.05cm}\textnormal{A}}) \; D \; \left( \vec {\mathfrak{u}}_{1}^\textnormal{A} \otimes \vec {\mathfrak{u}}_{1}^\textnormal{A} \right).  \label{Split p 5 tensorproduct} 
\end{eqnarray}
This relation of the classical pressure tensors $\underline{\underline{p}}^{\textnormal{A},\textnormal{cl}}(\vec q,t)$, $\underline{\underline{p}}^{\textnormal{tot,cl}}(\vec q,t)$ to dyadic products of relative particle velocities is an analog to the calculation of the pressure tensor $\underline{\underline{p}}$ in classical hydrodynamics (\hspace{-0.15 cm} \cite{Choudhuri_1988}, p.\ 32 and \cite{Shu_1992}, p.\ 21). So, this is the motivation to call  $\underline{\underline{p}}^{\textnormal{A},\textnormal{cl}}(\vec q,t)$ and  $\underline{\underline{p}}^{\textnormal{tot,cl}}(\vec q,t)$ classical tensors. \newline 
As a remark, we note that for the special case of a one-particle system, the associated classical pressure tensor  $\underline{\underline{p}}^{\textnormal{cl}}$ vanishes. The reason for this is that for this system, the mean particle velocity $\vec v$ and the velocity $\vec w$ of the single particle are obviously the same, so, the relative velocity $\vec u$ of this particle vanishes. Since the classical pressure tensor  $\underline{\underline{p}}^{\textnormal{cl}}$ depends on the dyadic product 
$\vec u \otimes \vec u$, thus, the classical pressure tensor  $\underline{\underline{p}}^{\textnormal{cl}}$ vanishes, too. Therefore, the classical pressure tensor does not appear in the analysis of one-particle systems in these references \cite{Sonego_1991,Lopreore_1999,Lopreore_2000,Wyatt_2002}, and \cite{Wyatt_2005}, p.\ 56f. Please note that for a one-particle system the classical momentum flow density tensor $\underline{\underline{\Pi}}^{\textnormal{cl}}$ does not vanish generally because the particle velocity $\vec w$ does not vanish for some one-particle systems (see for such a case the analysis in chapter 13 of \cite{Wyatt_2005}). \newpage \noindent    
Now we turn our focus back to many-particle systems with different sorts of particles. For these systems, both the quantum pressure tensor elements $p_{\alpha \beta}^{\textnormal{A},\textnormal{qu}}(\vec q,t)$ for a certain sort of particles $\textnormal{A}$ and the quantum pressure tensor elements $p_{\alpha \beta}^{\textnormal{tot,qu}}(\vec q,t)$ for the total particle ensemble are just equal to the corresponding quantum momentum-flow density tensor elements: \vspace{0.3 cm}
\begin{eqnarray}
 p_{\alpha \beta}^{\textnormal{A},\textnormal{qu}}(\vec q,t) &=& \Pi_{\alpha \beta}^{\textnormal{A},\textnormal{qu}}(\vec q,t), \label{eqn for quantum part} \\
 p_{\alpha \beta}^{\textnormal{tot,qu}}(\vec q,t) &=& \Pi_{\alpha \beta}^{\textnormal{tot,qu}}(\vec q,t),  \label{quantum parts equal p}  
\end{eqnarray}
\vspace{-0.1 cm} \newline \noindent
so that we can write for the corresponding tensors 
\vspace{0.3 cm}
\begin{eqnarray}
\hspace{-0.5cm} \underline{\underline{p}}^{\textnormal{A},\textnormal{qu}}(\vec q,t) \hspace{-0.15 cm}  &=& \hspace{-0.15 cm}   \underline{\underline{\Pi}}^{\textnormal{A},\textnormal{qu}}(\vec q,t)  \hspace{0.05 cm} = \hspace{0.05 cm} - \hspace{0.1 cm} \hbar^2 \frac{N(\textnormal{A})}{4 m_\textnormal{A}} \hspace{-0.1 cm}
\int  \hspace{-0.1 cm} \textnormal{d} \vec Q \; \delta (\vec q - \vec q_1^{\hspace{0.05cm}\textnormal{A}}) \hspace{-0.1 cm}  \; D \; \hspace{-0.15 cm} \left( \nabla_1^\textnormal{A} \otimes \nabla_1^\textnormal{A} \right) \ln D, \label{eqn for quantum part tensor}  \\ 
\hspace{-0.5cm} \underline{\underline{p}}^{\textnormal{tot,qu}}(\vec q,t) \hspace{-0.15 cm} &=& \hspace{-0.15 cm} \underline{\underline{\Pi}}^{\textnormal{tot,qu}}(\vec q,t) \hspace{0.05 cm} = \hspace{0.05 cm} - \hspace{0.1 cm} \hbar^2 \hspace{-0.05 cm} \sum_{\textnormal{A}=1}^{N_S} \hspace{-0.05 cm} \frac{N(\textnormal{A})}{4 m_\textnormal{A}} \hspace{-0.1 cm}
\int \hspace{-0.1 cm} \textnormal{d} \vec Q \; \delta (\vec q - \vec q_1^{\hspace{0.05cm}\textnormal{A}})  \hspace{-0.1 cm}  \; D \; \hspace{-0.15 cm} \left( \nabla_1^\textnormal{A} \otimes \nabla_1^\textnormal{A} \right) \ln D. \label{quantum parts equal p tensor}
\end{eqnarray}
\vspace{-0.1 cm} \newline \noindent
Thus, like the quantum momentum-flow density tensors $\underline{\underline{\Pi}}^{\textnormal{A},\textnormal{qu}}(\vec q,t)$ and $\underline{\underline{\Pi}}^{\textnormal{tot,qu}}(\vec q,t)$, both the quantum pressure tensor $\underline{\underline{p}}^{\textnormal{A},\textnormal{qu}}(\vec q,t)$ specific for a given sort $\textnormal{A}$ of particles and the quantum pressure tensor $\underline{\underline{p}}^{\textnormal{tot,qu}}(\vec q,t)$ for the total particle ensemble are related only to properties of $D(\vec Q,t)$, and they vanish in the limit $\hbar \rightarrow 0$. Now, it becomes clear why we named 
$\underline{\underline{p}}^{\textnormal{A},\textnormal{qu}}(\vec q,t)$ and $\underline{\underline{p}}^{\textnormal{tot,qu}}(\vec q,t)$ as quantum tensors. \newline \newline 
At the end of this chapter, here we prove that the definition (\ref{Definiton p sort A}) for $\underline{\underline {p}}^\textnormal{A}(\vec q,t)$ 
is equi\- valent to Eqn.\ (\ref{first def p sort A}), and the definition (\ref{Definiton p all sorts}) for  $\underline{\underline {p}}^\textnormal{tot}(\vec q,t)$
is equivalent to Eqn.\ (\ref{first def p all sorts}). \newline \newline 
In order to prove the equivalence of Eqns.\ (\ref{Definiton p sort A}) and (\ref{first def p sort A}), we show that the quantity $p_{\alpha \beta}^{\textnormal{A},\textnormal{cl}}(\vec q,t)$ can be expressed in the following way by applying Eqns.\ (\ref{One-particle density of sort k}), (\ref{k_flux_density_bohm_representation}), (\ref{definition relative velocity}), (\ref{Split Pi 2}), and (\ref{Split p 2}): \hspace{0.1 cm}
\begin{eqnarray}
p_{\alpha \beta}^{\textnormal{A},\textnormal{cl}}(\vec q,t) &=& N(\textnormal{A}) \hspace{0.075cm}  m_\textnormal{A} \int \textnormal{d} \vec Q \; \delta (\vec q - \vec q_1^{\hspace{0.05cm}\textnormal{A}}) \; D(\vec Q,t)  \;  u_{1\alpha}^\textnormal{A}(\vec Q,t) \;  u_{1\beta}^\textnormal{A}(\vec Q,t)  \nonumber \\
&=& N(\textnormal{A}) \hspace{0.075cm}  m_\textnormal{A} \int \textnormal{d} \vec Q \; \delta (\vec q - \vec q_1^{\hspace{0.05cm}\textnormal{A}}) \; D(\vec Q,t)  \; \times \nonumber \\   
&& \left [  w_{1 \alpha}^\textnormal{A}(\vec Q,t) - v_{\alpha}^\textnormal{A}(\vec q_1^{\hspace{0.05cm}\textnormal{A}},t)  \right] \; \left [ w_{1 \beta}^\textnormal{A}(\vec Q,t) -  v_{\beta}^\textnormal{A}(\vec q_1^{\hspace{0.05cm}\textnormal{A}},t) \right] \nonumber \\ \nonumber \\  \nonumber \\  \nonumber \\  
&=& \underbrace{ N(\textnormal{A}) \hspace{0.075cm}  m_\textnormal{A}  \int \textnormal{d} \vec Q_1^\textnormal{A} \; D(\vec Q_1^\textnormal{A}(\vec q),t)  \;  w_{1\alpha}^\textnormal{A}(\vec Q_1^\textnormal{A}(\vec q),t) \; w_{1\beta}^\textnormal{A}(\vec Q_1^\textnormal{A}(\vec q),t)}_{= \; \Pi_{\alpha \beta}^{\textnormal{A},\textnormal{cl}}(\vec q,t)} \nonumber \\  
&&  - \; v_{\alpha}^\textnormal{A}(\vec q,t) \; \underbrace{N(\textnormal{A}) \hspace{0.075cm}  m_\textnormal{A} \int \textnormal{d} \vec Q_1^\textnormal{A} \; D(\vec Q_1^\textnormal{A}(\vec q),t) \; w_{1\beta}^\textnormal{A}(\vec Q_1^\textnormal{A}(\vec q),t)}_{= \; j^\textnormal{A}_{m,\beta} (\vec q,t) \; = \; \rho_m^\textnormal{A}(\vec q,t) \; v_{\beta}^\textnormal{A} (\vec q,t) } \nonumber \\
&&  - \; v_{\beta}^\textnormal{A}(\vec q,t)  \; \underbrace{N(\textnormal{A}) \hspace{0.075cm}  m_\textnormal{A} \int \textnormal{d} \vec Q_1^\textnormal{A} \; D(\vec Q_1^\textnormal{A}(\vec q),t) \; w_{1\alpha}^\textnormal{A}(\vec Q_1^\textnormal{A}(\vec q),t)}_{= \; 
j^\textnormal{A}_{m,\alpha}(\vec q,t) \; = \; \rho_m^\textnormal{A}(\vec q,t) \; v_{\alpha}^\textnormal{A} (\vec q,t) } \nonumber \\    
&& + \; v_{\alpha}^\textnormal{A}(\vec q,t) \; v_{\beta}^\textnormal{A}(\vec q,t) \; \underbrace{N(\textnormal{A}) \hspace{0.075cm}  m_\textnormal{A} \int \textnormal{d} \vec Q_1^\textnormal{A} \; D(\vec Q_1^\textnormal{A}(\vec q),t)}_{= \; \rho_m^\textnormal{A}(\vec q,t)} \nonumber \\
&=& \Pi_{\alpha \beta}^{\textnormal{A},\textnormal{cl}}(\vec q,t) -  \rho_m^\textnormal{A}(\vec q,t) \; v_{\alpha}^\textnormal{A}(\vec q,t) \; v_{\beta}^\textnormal{A}(\vec q,t). \label{eqn for classical part} 
\end{eqnarray}
\vspace{-0.1 cm} \newline \noindent Then, we find a formula that relates the pressure tensor elements $p_{\alpha \beta}^\textnormal{A}(\vec q,t)$ for the sort of particles $\textnormal{A}$ with the corresponding momentum flow density tensor elements $\Pi_{\alpha \beta}^\textnormal{A}(\vec q,t)$ by adding Eqn.\ (\ref{eqn for quantum part}) and Eqn.\ (\ref{eqn for classical part}):
\vspace{0.3 cm}
\begin{align}
p_{\alpha \beta}^{\textnormal{A}}(\vec q,t) &= \Pi_{\alpha \beta}^{\textnormal{A}}(\vec q,t) -  \rho_m^\textnormal{A}(\vec q,t) \;  v_{\alpha}^\textnormal{A}(\vec q,t) \; v_{\beta}^\textnormal{A}(\vec q,t). \label{eqn for classical and quantum part} 
\end{align}
\vspace{-0.3 cm} \newline \noindent One can find similar equations in classical hydrodynamics (\hspace {-0.15 cm} \cite{Landau_2000}, p.\ 11 and p.\ 44). Rewriting the equation above as an equation for tensors instead of their components, we find just Eqn.\ (\ref{first def p sort A}) -- so we have shown the equivalence of Eqns.\ (\ref{first def p sort A}) and  (\ref{Definiton p sort A}). \newline \newline 
Finally, it remains to show that Eqns.\ (\ref{first def p all sorts}) and  (\ref{Definiton p all sorts})  are equivalent equations for the calculation of $\underline{\underline {p}}^\textnormal{tot}(\vec q,t)$:
For this derivation, the quantity $p_{\alpha \beta}^{\textnormal{tot,cl}}(\vec q,t)$ is transformed by Eqns.\ (\ref{Total one-particle density}), (\ref{tot_flux_density_bohm_representation}), (\ref{alternative definition relative velocity}), (\ref{Split Pi 5}), and (\ref{Split p 5}) analogously to how Eqn.\ (\ref{eqn for classical part}) was derived: 
\vspace{0.1 cm}
\begin{eqnarray}
p_{\alpha \beta}^{\textnormal{tot,cl}}(\vec q,t) &=& \sum_{\textnormal{A}=1}^{N_S} N(\textnormal{A}) \hspace{0.075cm}  m_\textnormal{A} \int \textnormal{d} \vec Q \; \delta (\vec q - \vec q_1^{\hspace{0.05cm}\textnormal{A}}) \; D(\vec Q,t)  \;  \mathfrak{u}_{1\alpha}^\textnormal{A}(\vec Q,t) \;  \mathfrak{u}_{1\beta}^\textnormal{A}(\vec Q,t)  \nonumber \\
&=&  \sum_{\textnormal{A}=1}^{N_S} N(\textnormal{A}) \hspace{0.075cm}  m_\textnormal{A} \int \textnormal{d} \vec Q \; \delta (\vec q - \vec q_1^{\hspace{0.05cm}\textnormal{A}}) \; D(\vec Q,t)  \; \times \nonumber \\
&& \left [ w_{1 \alpha}^\textnormal{A}(\vec Q,t) - v_{\alpha}^{\textnormal{tot}}(\vec q_1^{\hspace{0.05cm}\textnormal{A}},t)  \right] \; \left [  w_{1 \beta}^\textnormal{A}(\vec Q,t) - v_{\beta}^{\textnormal{tot}}(\vec q_1^{\hspace{0.05cm}\textnormal{A}},t)   \right] \nonumber \\  \nonumber \\ 
&=& \underbrace{\sum_{\textnormal{A}=1}^{N_S} N(\textnormal{A}) \hspace{0.075cm}  m_\textnormal{A}  \int \textnormal{d} \vec Q_1^\textnormal{A} \; D(\vec Q_1^\textnormal{A}(\vec q),t)  \;  w_{1\alpha}^\textnormal{A}(\vec Q_1^\textnormal{A}(\vec q),t) \; w_{1\beta}^\textnormal{A}(\vec Q_1^\textnormal{A}(\vec q),t)}_{= \; \Pi_{\alpha \beta}^{\textnormal{tot,cl}}(\vec q,t)} \nonumber \\  
&&  - \; v_{\alpha}^{\textnormal{tot}}(\vec q,t) \; \underbrace{\sum_{\textnormal{A}=1}^{N_S} N(\textnormal{A}) \hspace{0.075cm}  m_\textnormal{A} \int \textnormal{d} \vec Q_1^\textnormal{A} \; D(\vec Q_1^\textnormal{A}(\vec q),t) \; w_{1\beta}^\textnormal{A}(\vec Q_1^\textnormal{A}(\vec q),t)}_{= \; j^{\textnormal{tot}}_{m,\beta} (\vec q,t) \; = \; \rho_m^{\textnormal{tot}}(\vec q,t) \; v_{\beta}^{\textnormal{tot}} (\vec q,t) } \nonumber \\
&&  - \; v_{\beta}^{\textnormal{tot}}(\vec q,t)  \; \underbrace{\sum_{\textnormal{A}=1}^{N_S} N(\textnormal{A}) \hspace{0.075cm}  m_\textnormal{A} \int \textnormal{d} \vec Q_1^\textnormal{A} \; D(\vec Q_1^\textnormal{A}(\vec q),t) \; w_{1\alpha}^\textnormal{A}(\vec Q_1^\textnormal{A}(\vec q),t)}_{= \; j^{\textnormal{tot}}_{m,\alpha}(\vec q,t) \; = \; \rho_m^{\textnormal{tot}}(\vec q,t) \; v_{\alpha}^{\textnormal{tot}} (\vec q,t) } \nonumber \\    
&& + \; v_{\alpha}^{\textnormal{tot}}(\vec q,t) \; v_{\beta}^{\textnormal{tot}}(\vec q,t) \; \underbrace{\sum_{\textnormal{A}=1}^{N_S} N(\textnormal{A}) \hspace{0.075cm}  m_\textnormal{A} \int \textnormal{d} \vec Q_1^\textnormal{A} \; D(\vec Q_1^\textnormal{A}(\vec q),t)}_{= \; \rho_m^{\textnormal{tot}}(\vec q,t)}  \nonumber \\
&=& \Pi_{\alpha \beta}^{\textnormal{tot,cl}}(\vec q,t) -  \rho_m^{\textnormal{tot}}(\vec q,t) \; v_{\alpha}^{\textnormal{tot}}(\vec q,t) \; v_{\beta}^{\textnormal{tot}}(\vec q,t). \label{eqn for classical part total ensemble}
\end{eqnarray}
Summing up Eqns.\ (\ref{quantum parts equal p}) and (\ref{eqn for classical part total ensemble}), we obtain a formula that relates the tensor elements  $p_{\alpha \beta}^{\textnormal{tot}}(\vec q,t)$ and $\Pi_{\alpha \beta}^{\textnormal{tot}}(\vec q,t)$ to each other:
\begin{eqnarray}
p_{\alpha \beta}^{\textnormal{tot}}(\vec q,t) &=& \Pi_{\alpha \beta}^{\textnormal{tot}}(\vec q,t) - \rho_m^{\textnormal{tot}}(\vec q,t) \; v_{\alpha}^{\textnormal{tot}}(\vec q,t) \; v_{\beta}^{\textnormal{tot}}(\vec q,t). \label{eqn for classical and quantum part tot}
\end{eqnarray}
By writing the equation above in a representation with tensors instead of tensor components, we find Eqn.\ (\ref{first def p all sorts}). So, the proof of the equivalence of Eqns.\ (\ref{first def p all sorts}) and (\ref{Definiton p all sorts}) is provided. 
\subsection{External fields}
In this chapter, it is now briefly discussed which basic formulas and main results of the derivations above change if external electric and magnetic fields $\vec {\mathcal{E}}(\vec q,t)$, $\vec {\mathcal{B}}(\vec q,t)$ are present. Here, we mention that the following results are similar to the results in \cite{Kuzmenkov_1999}, where analogous equations for MPQHD were derived like here -- but, in \cite{Kuzmenkov_1999}, first, the presence of different sorts was not discussed, and second, external fields were taken into account. These fields are described by a vector potential $\vec {\mathcal{A}}(\vec q,t)$ and a scalar potential $\Phi(\vec q,t)$ by 
\begin{eqnarray}
\vec {\mathcal{B}}(\vec q,t) &=& \nabla \times \vec {\mathcal{A}}(\vec q, t), \label{rotA} \\
\vec {\mathcal{E}}(\vec q,t) &=&  - \nabla \Phi(\vec q,t) -  \frac{\partial \vec{\mathcal{A}}(\vec q, t)}{\partial t}.
\end{eqnarray}
Moreover, we introduce the kinematic momentum operator $\hat {\vec {\mathcal{D}}}_i^{\hspace{0.05 cm}\textnormal{A}}$; it is given by
\begin{eqnarray}
\hat {\vec {\mathcal{D}}}_i^{\hspace{0.05 cm}\textnormal{A}} &=& \hat {\vec p}_i^{\hspace{0.05 cm}\textnormal{A}} - e_\textnormal{A} \vec {\mathcal{A}}(\vec q_i^{\hspace{0.05 cm}\textnormal{A}}, t). 
\end{eqnarray}
In our analysis above without external fields held $\vec {\mathcal{A}}(\vec q, t) = \vec 0$, so, there was no need to distinguish between the kinematic momentum operator $\hat {\vec {\mathcal{D}}}_i^{\hspace{0.05 cm}\textnormal{A}}$ and the canonical momentum operator $\hat {\vec p}_i^{\hspace{0.05 cm}\textnormal{A}}$. But now we analyze situations where, in general, this is not true anymore, and we have to distinguish these operators. As a rule, in all the equations we previously derived for the field-free case, where the canonical momentum operator $\hat {\vec p}_i^{\hspace{0.05 cm}\textnormal{A}}$ appears, it has to be substituted in these equations by the kinematic momentum operator $\hat {\vec {\mathcal{D}}}_i^{\hspace{0.05 cm}\textnormal{A}}$ for the presence of external fields. \newline  
Now, the Hamilton operator has this time-dependent form instead of Eqn.\ (\ref{Hamiltion operator with momentum operator}): 
\begin{eqnarray}
\hat H (\vec Q,t) &=&  \sum_{\textnormal{A} = 1}^{{N_S}} \sum_{i = 1}^{N(\textnormal{A})} \left[ \frac { (\hat {\vec {\mathcal{D}}}_i^{\hspace{0.05 cm} \textnormal{A}})^2} {2 m_\textnormal{A}} + e_\textnormal{A} \Phi(\vec q_i^{\hspace{0.05 cm}\textnormal{A}} ,t)\right] + \frac{1}{2} \sum_{\textnormal{A} = 1}^{{N_S}} \sum_{i = 1}^{N(\textnormal{A})} \sum_{\textnormal{B} = 1}^{{N_S}} \sum_{j=1}^{N(\textnormal{B})} V_{ij}^{\textnormal{AB}}. \label{Hamiltion operator with kinematic momentum operator}   
\end{eqnarray}
For the total particle mass current density $\vec j_m^{\textnormal{tot}}(\vec q,t)$, we find instead of Eqns.\ (\ref{Total particle current density with momentum operator}) and (\ref{Total particle current density 2}): 
\begin{eqnarray}
\vec j^{\textnormal{tot}}_m(\vec q,t) &=& \sum_{\textnormal{A}=1}^{N_S} \sum_{i=1}^{N(\textnormal{A})} \int \textnormal{d} \vec Q \;  \delta (\vec q - \vec q_i^{\hspace{0.05 cm}\textnormal{A}} ) \; \Re
\left [ \Psi^*(\vec Q,t) \; \hat {\vec {\mathcal{D}}}_i^{\hspace{0.05 cm}\textnormal{A}} \; \Psi(\vec Q,t) \right] \label{Total particle current density with kinematic momentum operator} \\
&=& \sum_{\textnormal{A}=1}^{N_S} N(\textnormal{A}) \int \textnormal{d} \vec Q \;  \delta (\vec q - \vec q_1^{\hspace{0.05 cm}\textnormal{A}} ) \; \Re
\left [ \Psi^*(\vec Q,t) \; \hat {\vec {\mathcal{D}}}_1^{\hspace{0.05 cm}\textnormal{A}} \; \Psi(\vec Q,t) \right], 
\end{eqnarray}
so the particle mass current density $\vec j_m^{\textnormal{A}}(\vec q,t)$ for the sort of particles $\textnormal{A}$ is now described by
\begin{eqnarray}
\vec j^{\textnormal{A}}_m(\vec q,t) &=& \sum_{i=1}^{N(\textnormal{A})}  \int \textnormal{d} \vec Q \;  \delta (\vec q - \vec q_i^{\hspace{0.05 cm}\textnormal{A}} ) \; \Re
\left [ \Psi^*(\vec Q,t) \; \hat {\vec {\mathcal{D}}}_i^{\hspace{0.05 cm}\textnormal{A}} \; \Psi(\vec Q,t) \right]  \\ &=&
N(\textnormal{A}) \int \textnormal{d} \vec Q \;  \delta (\vec q - \vec q_1^{\hspace{0.05 cm}\textnormal{A}} ) \; \Re
\left [ \Psi^*(\vec Q,t) \; \hat {\vec {\mathcal{D}}}_1^{\hspace{0.05 cm}\textnormal{A}} \; \Psi(\vec Q,t) \right] \label{k_flux_density with kinematic momentum operator} 
\end{eqnarray}
instead by Eqns.\ (\ref{k_flux_density with momentum operator}), (\ref{k_flux_density}). \newline 
Moreover, the definition of the velocity $\vec w_i^\textnormal{A}(\vec Q,t)$ for the $(\textnormal{A},i)$-particle shown in Eqn.\ (\ref{local velocity}) changes:
\begin{eqnarray}
\vec w_i^\textnormal{A}(\vec Q,t) &=& \frac{1}{m_\textnormal{A}} \left( \nabla_i^\textnormal{A} S(\vec Q,t) - e_\textnormal{A} \vec {\mathcal{A}}(\vec q_i^{\hspace{0.05 cm}\textnormal{A}}, t) \right). \label{local velocity for A}
\end{eqnarray}
While the rotation of the velocity $\vec w_i^{\textnormal{A}}(\vec Q,t)$ described by Eqn.\ (\ref{local velocity}) always vanishes, this is not true anymore for the more general Eqn.\ (\ref{local velocity for A}) for $\vec w_i^{\textnormal{A}}(\vec Q,t)$: \newline
As a consequence of Eqns.\ (\ref{rotA}) and (\ref{local velocity for A}), we find that 
\begin{eqnarray}
\nabla_i^{\textnormal{A}} \times \vec w_i^{\textnormal{A}}(\vec Q,t) = - \frac{e_\textnormal{A}}{m_\textnormal{A}} \vec {\mathcal{B}}(\vec q,t).
\end{eqnarray}
In addition, we note that the old formulas (\ref{tot_flux_density_bohm_representation}) and (\ref{k_flux_density_bohm_representation}) for $\vec j^{\textnormal{tot}}_m(\vec q,t)$ or $\vec j^{\textnormal{A}}_m(\vec q,t)$, respectively, do not change explicitly for the case that external fields are present. However, an implicit change occurs due to the changed definition for the velocity $\vec w_i^\textnormal{A}(\vec Q,t)$. Applying the same argumentation, 
 Eqn.\ (\ref{mean particle velocity def}) for $\vec v^{\textnormal{A}}(\vec q,t)$, Eqn.\ (\ref{definition relative velocity}) for $\vec u_i^{\textnormal{A}}(\vec Q,t)$, Eqn.\ (\ref{total mean particle velocity}) for $\vec v^{\hspace{0.05cm}\textnormal{tot}}(\vec q,t)$, and Eqn.\ (\ref{alternative definition relative velocity}) for $\vec {\mathfrak{u}}_i^{\textnormal{A}}(\vec Q,t)$ do not change explicitly, but they do change implicitly for the presence of external fields. \newline 
Moreover, the force density $\vec f^{\hspace{0.05 cm} \textnormal{A}}(\vec q,t)$ for the particles of the sort $\textnormal{A}$ is now described by the following equation, which replaces Eqn.\ (\ref{Total force density 2}):
\begin{eqnarray}
\vec f^{\hspace{0.05 cm} \textnormal{A}}(\vec q,t) &=& - \; N(\textnormal{A}) \left[ \sum_{\textnormal{B}=1}^{N_S} \left (N(\textnormal{B}) - \delta_{\textnormal{AB}} \right) \int \textnormal{d} \vec Q_{1}^\textnormal{A} \; D(\vec Q_1^\textnormal{A}(\vec q),t) \nabla V^{\textnormal{AB}}(|\vec q - \vec q_{N(\textnormal{B})}^{\hspace{0.05cm} \textnormal{B}}|) \right] \nonumber \\
&& + \; \frac{e_\textnormal{A}}{m_\textnormal{A}} \rho_m^\textnormal{A} (\vec q,t) \left [ \vec {\mathcal{E}}(\vec q,t) + \vec v^\textnormal{A}(\vec q,t) \times \vec {\mathcal{B}}(\vec q,t) \right].
\label{Total force density with external fields}                              
\end{eqnarray}
In the second line of Eqn.\ (\ref{Total force density with external fields}), extra terms relative to Eqn.\ (\ref{Total force density 2}) appear because of the external fields. We did not derive Eqn.\ (\ref{Total force density with external fields}) here in detail but the extra field terms in this equation are intuitively clear. \newline 
So, the force density $\vec f^{\hspace{0.05 cm} \textnormal{tot}}(\vec q,t)$ for all particles is given for the presence of external fields by
\begin{eqnarray}
\vec f^{\hspace{0.05 cm} \textnormal{tot}}(\vec q,t) &=& - \sum_{\textnormal{A}=1}^{N_S} \left \{ N(\textnormal{A}) \left[ \sum_{\textnormal{B}=1}^{N_S} \left (N(\textnormal{B}) - \delta_{\textnormal{AB}} \right) \int \textnormal{d} \vec Q_{1}^\textnormal{A} \; D(\vec Q_1^\textnormal{A}(\vec q),t) \nabla V^{\textnormal{AB}}(|\vec q - \vec q_{N(\textnormal{B})}^{\hspace{0.05cm}\textnormal{B}}|) \right] \right. \nonumber \\
&& \left. + \; \frac{e_\textnormal{A}}{m_\textnormal{A}} \rho_m^\textnormal{A} (\vec q,t) \left [ \vec {\mathcal{E}}(\vec q,t) + \vec v^\textnormal{A}(\vec q,t) \times \vec {\mathcal{B}}(\vec q,t) \right] \right\},
\label{Total force density with external fields 2}                              
\end{eqnarray}
which replaces Eqn.\ (\ref{Force density for all k}). \newline  
In addition, in the old calculations we found two different representations for the momentum flow density tensor elements $\Pi_{\alpha \beta}^{\textnormal{A}}(\vec q,t)$ of the sort $\textnormal{A}$. The first one is Eqn.\ (\ref{total momentum flow density tensor version 2}), which contains components of the canonical momentum flow operator $\hat {\vec p}_1^{\hspace{0.05 cm}\textnormal{A}}$. For the presence of external fields, these components must be exchanged because of the rule mentioned above by the components of the corresponding kinematic momentum operator $\hat {\vec {\mathcal{D}}}_1^{\hspace{0.05 cm}\textnormal{A}}$:
\begin{eqnarray}
\Pi_{\alpha \beta}^{\textnormal{A}}(\vec q,t) &=&  N(\textnormal{A}) \int \textnormal{d} \vec Q \;  \delta (\vec q - \vec q_1^{\hspace{0.05 cm}\textnormal{A}} ) \; 
\frac{1}{4 m_\textnormal{A}} \times \nonumber \\
&&  \hspace{-2.5 cm} \; \; \;  \left [ \Psi^* \hspace{-0.05cm} \left( \hspace{-0.05cm} \hat {\mathcal{D}}_{1\beta}^\textnormal{A} \hat {\mathcal{D}}_{1\alpha}^\textnormal{A} \Psi \hspace{-0.05cm} \right)  +  \left ( \hspace{-0.05cm} \hat {\mathcal{D}}_{1\beta}^\textnormal{A} \Psi \hspace{-0.05cm} \right)^*  \left( \hspace{-0.05cm} \hat {\mathcal{D}}_{1\alpha}^\textnormal{A} \Psi \hspace{-0.05cm} \right) +  \left ( \hspace{-0.05cm} \hat {\mathcal{D}}_{1\alpha}^\textnormal{A} \Psi \hspace{-0.05cm} \right)^*  \left( \hspace{-0.05cm} \hat {\mathcal{D}}_{1\beta}^\textnormal{A} \Psi \hspace{-0.05cm} \right) + \left( \hspace{-0.05cm} \hat {\mathcal{D}}_{1\beta}^\textnormal{A} \hat {\mathcal{D}}_{1\alpha}^\textnormal{A} \Psi \hspace{-0.05cm} \right)^* \Psi \right ]\hspace{-0.1cm}. \label{total kinematic momentum flow density tensor version 2} 
\end{eqnarray} 
The second one is Eqn.\ (\ref{total momentum flow density tensor result 3}), which contains components of the vector $\vec w_{1}^\textnormal{A}(\vec Q,t)$.  This representation for the tensor elements $\Pi_{\alpha \beta}^{\textnormal{A}}(\vec q,t)$ does not change explicitly but it changes implicitly because for the presence of external fields, the new formula (\ref{local velocity for A}) holds for the vector $\vec w_1^\textnormal{A}(\vec Q,t)$. Applying an analogous argumentation, it can be found that the formula (\ref{k sort momentum flow density tensor result}) for the momentum flow density tensor elements $\Pi_{\alpha \beta}^{\textnormal{tot}}(\vec q,t)$ for the total particle ensemble does not change explicitly but implicitly, too.  \newline
It can be found for the pressure tensor elements $p_{\alpha \beta}^\textnormal{A}(\vec q,t)$ and $p_{\alpha \beta}^{\textnormal{tot}}(\vec q,t)$ that the 
corres\-ponding Eqns.\ (\ref{k sort pressure tensor result}) and (\ref{total pressure tensor result}) remain valid explicitly. However, implicit changes occur due to 
the components of the velocities $\vec u_1^{\textnormal{A}}(\vec Q,t)$ and $\vec {\mathfrak{u}}_1^{\textnormal{A}}(\vec Q,t)$ appearing in Eqn.\ (\ref{k sort pressure tensor result}) or Eqn.\ (\ref{total pressure tensor result}), respectively. \newline \newline 
Taking all these changes into account for the different quantities discussed above, we eventually find that both for a certain sort of particles $\textnormal{A}$ and the total particle ensemble the corresponding MPCEs, MPEEMs, and the MPQCEs, given in Eqns.\ (\ref{CE particle sort k}), (\ref{CE all particles}), (\ref{MMBE}), (\ref{MMBE for all particles}), (\ref{MQNSE for particle sort k}), and (\ref{MQNSE for all particles}), remain valid explicitly for the presence of external fields. \newline
For all the following considerations  we assume that no external fields are present.
\section{Transformations of the $\underline{\underline{\Pi}}$ and $\underline{\underline{p}}$ tensors} \label{Tensor transformations}
The following analysis is done for quantities for a particular sort of particles denoted with a corresponding index $\textnormal{A}$. It can be made in an analogous way for the corresponding quantities for the total particle ensemble denoted with an index tot. Since we focus in our following analysis on quantities for a particular sort of particles $\textnormal{A}$, we will only indicate by the index $\textnormal{A}$ when quantities are related to this sort of particles, but we will not mention this extra verbally anymore. 
\subsection{Kuzmenkov tensors $\underline{\underline{\Pi}}^{K\textnormal{A}}$ and $\underline{\underline{p}}^{K\textnormal{A}}$}
In the calculations above we found the formula (\ref{total momentum flow density tensor result 3}) for the elements of the momentum flow density tensor $\underline{\underline{\Pi}}^\textnormal{A}(\vec q,t)$, and the formula (\ref{k sort pressure tensor result}) for the elements of the pressure tensor $\underline{\underline{p}}^\textnormal{A}(\vec q,t)$. Since these equations are similar to results stated in \cite{Kuzmenkov_1999}, from now on, we call these tensors, and their corresponding quantum parts and classical parts, Kusmenkov tensors and denote them with a superscript $K$.  \newline
Due to Eqns.\ (\ref{Split Pi 2 tensorproduct}) and (\ref{Split p 2 tensorproduct}), the classical Kuzmenkov tensors $\underline{\underline{\Pi}}^{K\textnormal{A},\textnormal{cl}}(\vec q,t)$ and $\; \underline{\underline{p}}^{K\textnormal{A},\textnormal{cl}}(\vec q,t)$ are related to dyadic products $\vec w_1^\textnormal{A} \otimes \vec w_1^\textnormal{A}$ or $\vec u_1^\textnormal{A} \otimes \vec u_1^\textnormal{A}$, respectively. So  -- as mentioned above -- these tensors are related to the momentum-flow density tensor $\underline{\underline{\Pi}}$ or to the pressure tensor $\underline{\underline{p}}$, respectively, in classical hydrodynamics, and their interpretation is clear. \newline 
However, a clear interpretation for the quantum quantities $\underline{\underline{\Pi}}^{K\textnormal{A},\textnormal{qu}}(\vec q,t)$, $\underline{\underline{p}}^{K\textnormal{A},\textnormal{qu}}(\vec q,t)$ is missing except for the aspect that they are related to quantum effects. This problem occurs because the term  $D \left(\nabla_1^\textnormal{A} \otimes \nabla_1^\textnormal{A}\right) \ln D$ appearing in Eqn.\ (\ref{eqn for quantum part tensor}) for these quantities is difficultly to understand. In order to close this gap, an alternative to the Kuzmenkov versions $\underline{\underline{\Pi}}^{K \textnormal{A}}(\vec q,t)$ and $\; \underline{\underline{p}}^{K \textnormal{A}}(\vec q,t)$ of the momentum flow density tensor and the pressure tensor is analyzed in the following Sec.\ \ref{W_and_P_tensors}. 
\subsection{Wyatt tensors $\underline{\underline{\Pi}}^{W\textnormal{A}}$ and $\underline{\underline{p}}^{W\textnormal{A}}$ \label{W_and_P_tensors}}
The formula (1.57) in R. E. Wyatts book \cite{Wyatt_2005}, p.\ 31, implies that the momentum flow density tensor $\underline{\underline{\Pi}}^\textnormal{A}(\vec q,t)$ can be calculated in the following manner (here, $\underline{\underline {1}}$ is the unit matrix): 
\begin{eqnarray}
\underline{\underline{\Pi}}^{W\textnormal{A}}(\vec q,t) &=&  \underline{\underline{1}} P_\textnormal{A}  + \sum_{i=1}^{N(\textnormal{A})} m_{\textnormal{A}}  \int \textnormal{d} \vec Q \; \delta (\vec q - \vec q_i^{\hspace{0.05 cm} \textnormal{A} } ) \; D \; \hspace{-0.1cm} \left[ \left(\vec w_{i}^\textnormal{A} \otimes \vec w_{i}^\textnormal{A} \right)  + \left(\vec d_{i}^{\hspace{0.05 cm} \textnormal{A}}  \otimes \vec d_{i}^{\hspace{0.05 cm} \textnormal{A}}  \right) \right] \hspace{-0.1cm}, \label{momentum_flow_densitytensor no N_k tensor version} 
\end{eqnarray}
so that its elements are given for Cartesian coordinates by: 
\begin{eqnarray}
\Pi_{\alpha \beta}^{W\textnormal{A}}(\vec q,t) &=&  P_\textnormal{A} \delta_{\alpha \beta} + \sum_{i=1}^{N(\textnormal{A})} m_\textnormal{A} \int \textnormal{d} \vec Q \; \delta (\vec q - \vec q_i^{\hspace{0.05 cm} \textnormal{A} } ) \; D \; \left (w_{i \alpha}^\textnormal{A} w_{i \beta}^\textnormal{A} + d_{i \alpha}^\textnormal{A} d_{i \beta}^\textnormal{A} \right). \label{momentum_flow_densitytensor no N_k} 
\end{eqnarray}
The upper extra superscript $W$ in the equations above for the tensor $\underline{\underline{\Pi}}^{W\textnormal{A}}(\vec q,t)$ and its elements $\Pi_{\alpha \beta}^{W\textnormal{A}}(\vec q,t)$ refers to the fact that this is a version of the tensor $\underline{\underline{\Pi}}^{\textnormal{A}}(\vec q,t)$ related to \cite{Wyatt_2005}. \newline
The quantity $P_\textnormal{A}$ appearing in Eqns.\ (\ref{momentum_flow_densitytensor no N_k tensor version}) and (\ref{momentum_flow_densitytensor no N_k}) is the scalar quantum pressure given by:
\begin{eqnarray}
P_\textnormal{A}(\vec q,t) =  - \sum_{i=1}^{N(\textnormal{A})} \frac{\hbar^2}{4 m_\textnormal{A}} \int \textnormal{d} \vec Q \; \delta (\vec q - \vec q_i^{\hspace{0.05 cm}\textnormal{A}} ) \laplace_i^\textnormal{A} D. \label{scalar_pressure no N_k}
\end{eqnarray}
The naming of the scalar quantum pressure comes from the dependence of $P_\textnormal{A}(\vec q,t)$ on the probability density $D(\vec Q,t)$, which is a pure quantum density. \newline
In addition, in Eqn.\ (\ref{momentum_flow_densitytensor no N_k tensor version}), the dyadic product of a vector $\vec d_i^{\hspace{0.05 cm}\textnormal{A}}$ appears; this vector is defined by
\begin{eqnarray}
\vec d_i^{\hspace{0.05 cm} \textnormal{A}}(\vec Q,t)  = - \frac{\hbar}{2 m_\textnormal{A}} \frac{\nabla_i^\textnormal{A} {D}}{D}. \label{definition d}
\end{eqnarray}
This vector $\vec d_i^{\hspace{0.05 cm} \textnormal{A}}(\vec Q,t)$ is named osmotic velocity of the $(\textnormal{A},i)$-particle corresponding to the nomenclature in \cite{Wyatt_2005}, p.\ 327. It is the quantum analog to the particle velocity $\vec w_i^\textnormal{A}(\vec Q,t)$, and it is related to the shape of $D(\vec Q,t)$. \newline
It can be shown in a straightforward calculation that the rotation of the osmotical velocity  $\vec d_i^{\hspace{0.05 cm} \textnormal A}(\vec Q,t)$ relative to the coordinate $\vec q_i^{\hspace{0.05 cm} \textnormal A}$ vanishes:
\begin{align}
\nabla_i^{\textnormal A} \times \vec d_i^{\hspace{0.05 cm} \textnormal A}(\vec Q,t) &= - \frac{\hbar}{2m_\textnormal{A}} \nabla_i^{\textnormal A} \times \left( \frac{1}{D} \nabla_i^{\textnormal A} D \right) \nonumber  \\
&= - \frac{\hbar}{2m_\textnormal{A}} \left \lbrace \frac{1}{D} \underbrace{\nabla_i^{\textnormal A} \times \left( \nabla_i^{\textnormal A} D \right)}_{= \; \vec 0} +
\left[ \nabla_i^{\textnormal A} \left( \frac{1}{D} \right) \right]  \times \left( \nabla_i^{\textnormal A} D \right) \right \rbrace \nonumber \\
&= \frac{\hbar}{2m_\textnormal{A}} \frac{1}{D^2} \underbrace{\left[ \left( \nabla_i^{\textnormal A} D \right)  \times \left( \nabla_i^{\textnormal A} D \right) \right]}_{= \; \vec 0}.
\end{align}
Due to the indistinguishability of the particles of the sort $\textnormal{A}$, we can also write Eqns.\ (\ref{momentum_flow_densitytensor no N_k tensor version}),   (\ref{momentum_flow_densitytensor no N_k}) and (\ref{scalar_pressure no N_k}) in the form
\begin{eqnarray}
\underline{\underline{\Pi}}^{W\textnormal{A}}(\vec q,t) &=& \underline{\underline{1}} P_\textnormal{A} + N(\textnormal{A}) \hspace{0.075cm}  m_{\textnormal{A}} \hspace{-0.05 cm} \int \hspace{-0.05 cm}    \textnormal{d} \vec Q \; \delta (\vec q - \vec q_1^{\hspace{0.05 cm} \textnormal{A} } ) \hspace{-0.05 cm} \; D \; \hspace{-0.1 cm} \left[ \left(\vec w_{1}^\textnormal{A} \otimes \vec w_{1}^\textnormal{A} \right)  + \left(\vec d_{1}^{\hspace{0.05 cm} \textnormal{A}}  \otimes \vec d_{1}^{\hspace{0.05 cm} \textnormal{A}}  \right) \right] \hspace{-0.1cm}, \label{momentum_flow_densitytensor tensor version} \\ 
\Pi_{\alpha \beta}^{W\textnormal{A}}(\vec q,t)
&=& P_\textnormal{A} \delta_{\alpha \beta} + N(\textnormal{A}) \hspace{0.075cm}  m_\textnormal{A} \hspace{-0.05 cm} \int  \hspace{-0.05 cm}  \textnormal{d} \vec Q \; \delta (\vec q - \vec q_1^{\hspace{0.05 cm}\textnormal{A}} ) \; \hspace{-0.05 cm}  D \; \hspace{-0.05 cm} \left (w_{1 \alpha}^\textnormal{A} w_{1 \beta}^\textnormal{A} + d_{1 \alpha}^\textnormal{A} d_{1 \beta}^\textnormal{A} \right), \label{momentum_flow_densitytensor} \\
P_\textnormal{A}(\vec q,t) &=&  - N(\textnormal{A}) \frac{\hbar^2}{4 m_\textnormal{A}} \int \textnormal{d} \vec Q \; \delta (\vec q - \vec q_1^{\hspace{0.05 cm}\textnormal{A}} ) \laplace_1^\textnormal{A} D. \label{scalar_pressure} 
\end{eqnarray}
In order to achieve a better understanding for the different meanings of the particle velocities $\vec d_i^{\hspace{0.05 cm}\textnormal{A}}(\vec Q,t)$ and $\vec w_i^\textnormal{A}(\vec Q,t)$,  we calculate, as a small excursion, the velocities $w$ and $d$ for an one-dimensional free Gaussian wave packet for a single particle at the start time $t=0$. In literature (\hspace{-0.01cm}\cite{Wyatt_2000_a, Wyatt_2000_b, Sonego_1991}, and \cite{Wyatt_2005}, p.\ 327), this system is popular for the explanation of hydrodynamical quantities. The wave function $\Psi(x,t=0)$ of this Gaussian wave packet is given by:
\begin{eqnarray}
\Psi(x,0) &=&  \left( \frac{1}{2 \pi \sigma^2} \right)^{\frac{1}{4}} e^{-{x^2}/({4 \sigma^2})} e^{i k_0 x}.
\end{eqnarray}
Here, $\sigma$ is related to the width of the wave packet and $k_0$ is a space-independent and time-independent wave number. Then, we find:
\begin{eqnarray}
S(x,0) &=& \hbar k_0 x, \\
D(x,0) &=& \left( \frac{1}{2 \pi \sigma^2} \right)^{\frac{1}{2}} e^{-{x^2}/({2 \sigma^2})},
\end{eqnarray}
and the velocities $w$ and $d$ are given by:
\begin{eqnarray}
w(x,0) &=& \frac{1}{m} \frac {\partial S}{\partial x} \; = \; \frac{\hbar k_0}{m}, \\
d(x,0) &=& - \frac{\hbar}{2m} \frac{1}{D} \frac{\partial D}{\partial x} \; = \; \frac{\hbar}{2 m \sigma^2} x.
\end{eqnarray}
Thus, at $t=0$, the whole wave packet $\Psi(x,0)$ moves like a classical particle with a corres\-ponding velocity $w = \hbar k_0/m$, independent of the position $x$. \newline 
As a supplement to this result, it can be shown in a straightforward calculation that the expectation value $\langle \hat p \rangle$ of the momentum operator $\hat p = \frac{\hbar}{\mathrm{i}} \frac{\partial}{\partial x}$ for the wave function $\Psi(x,0)$ is given by 
\begin{equation}
\langle \hat p \rangle = \langle \Psi(x,0) \hspace{0.05cm}  | \hspace{0.05cm}  \hat p \hspace{0.05cm} | \hspace{0.05cm}  \Psi(x,0) \rangle = \hbar k_0 = m w.
\end{equation} 
Moreover, the wave packet disperses due to the shape of $D(x,0)$, and this dispersion can be explained with additional movements of local parts of the wave packet $\Psi(x,0)$. These dispersion movements vary depending on what part of the wave packet is considered, and they are described by the osmotic velocity $d(x,0)$. In particular, $d(x,0)$ is proportional to the position $x$, so for $x>0$, this velocity is positive and is related to a forward movement of the front wave packet shoulder, and for $x<0$, it is negative and is related to a rear movement of the backward wave packet shoulder  (see for this dispersion discussion also  \cite{Sonego_1991} and \cite{Wyatt_2005}, p.\ 327). \newline \newline
Resuming our general analysis, as can be realized by Eqns.\ (\ref{Split Pi 1 Tensorproduct}) and (\ref{Definiton p sort A}), one can get the pressure tensor by substituting the particle velocities $ \vec w_{1}^\textnormal{A}(\vec Q,t)$ in the equation for the momentum flow density tensor by the corresponding relative velocities $\vec u_{1}^\textnormal{A}(\vec Q,t)$. So, using Eqn.\ (\ref{momentum_flow_densitytensor tensor version}) for $\underline{\underline{\Pi}}^{W\textnormal{A}}(\vec q,t)$ as a starting point, we get a ``Wyatt version'' $\underline{\underline{p}}^{W\textnormal{A}}(\vec q,t)$ of the pressure tensor. This tensor and its elements are: 
\begin{eqnarray}
\underline{\underline{p}}^{W\textnormal{A}}(\vec q,t) &=& \underline{\underline{1}} P_\textnormal{A} + N(\textnormal{A}) \hspace{0.075cm}  m_{\textnormal{A}} \hspace{-0.05 cm} \int \hspace{-0.05 cm}    \textnormal{d} \vec Q \; \delta (\vec q - \vec q_1^{\hspace{0.05 cm} \textnormal{A} } ) \hspace{-0.05 cm} \; D \; \hspace{-0.1 cm} \left[ \left(\vec u_{1}^\textnormal{A} \otimes \vec u_{1}^\textnormal{A} \right)  + \left(\vec d_{1}^{\hspace{0.05 cm} \textnormal{A}}  \otimes \vec d_{1}^{\hspace{0.05 cm} \textnormal{A}}  \right) \right] \hspace{-0.1cm}, \label{pressure_tensor Wyatt tensor version} \\ 
p_{\alpha \beta}^{W\textnormal{A}}(\vec q,t) &=& P_\textnormal{A} \delta_{\alpha \beta} + N(\textnormal{A}) \hspace{0.075cm}  m_\textnormal{A} \int \textnormal{d} \vec Q \; \delta (\vec q - \vec q_1^{\hspace{0.05 cm}\textnormal{A}} ) \; D \;  \left (u_{1 \alpha}^\textnormal{A} u_{1 \beta}^\textnormal{A} + d_{1 \alpha}^\textnormal{A} d_{1 \beta}^\textnormal{A} \right). \label{pressure_tensor_Wyatt}
\end{eqnarray} \hspace{-0.1 cm}
So, due to Eqns.\ (\ref{Split Pi 2 tensorproduct}), (\ref{Split p 2 tensorproduct}), 
(\ref{momentum_flow_densitytensor tensor version}), and (\ref{pressure_tensor Wyatt tensor version}), we can split the Wyatt tensors $\underline{\underline{\Pi}}^{W\textnormal{A}}(\vec q,t)$ and $\underline{\underline{p}}^{W\textnormal{A}}(\vec q,t)$ in the following form each into a classical part and a quantum part:
\begin{eqnarray}
\underline{\underline{\Pi}}^{W\textnormal{A}}(\vec q,t) &=& \underline{\underline{\Pi}}^{W\textnormal{A},\textnormal{cl}}(\vec q,t) \; + \; \underline{\underline{\Pi}}^{W\textnormal{A},\textnormal{qu}}(\vec q,t), \\
\underline{\underline{\Pi}}^{W\textnormal{A},\textnormal{cl}}(\vec q,t) &=& \underline{\underline{\Pi}}^{K\textnormal{A},\textnormal{cl}}(\vec q,t) \; \hspace{-0.05 cm} = \; \hspace{-0.05 cm} N(\textnormal{A}) \hspace{0.075cm}  m_\textnormal{A}  \hspace{-0.1 cm} \int  \hspace{-0.1 cm} \textnormal{d} \vec Q \; \delta (\vec q - \vec q_1^{\hspace{0.05cm}\textnormal{A}}  \hspace{-0.05 cm} ) \; D \;  \hspace{-0.1 cm} \left (\vec w_{1}^\textnormal{A} \otimes \vec w_{1}^\textnormal{A}    \right) \hspace{-0.1 cm}  ,  \\
\underline{\underline{\Pi}}^{W\textnormal{A},\textnormal{qu}}(\vec q,t) &=& \underline{\underline{1}} P_\textnormal{A} \hspace{-0.05 cm} + \hspace{-0.05 cm} N(\textnormal{A}) \hspace{0.075cm}  m_\textnormal{A} \hspace{-0.1 cm} \int \hspace{-0.1 cm} \textnormal{d} \vec Q \; \delta (\vec q - \vec q_1^{\hspace{0.05cm}\textnormal{A}}  \hspace{-0.05 cm} ) \; D \;  \hspace{-0.1 cm} \left ( \hspace{-0.05 cm}   \vec d_{1}^{\hspace{0.05cm}\textnormal{A}} \otimes \vec d_{1}^{\hspace{0.05cm}\textnormal{A}} \hspace{-0.05 cm} \right)  \hspace{-0.1 cm} , \\ && \nonumber \\  && \nonumber \\
\underline{\underline{p}}^{W\textnormal{A}}(\vec q,t) &=& \underline{\underline{p}}^{W\textnormal{A},\textnormal{cl}}(\vec q,t) \; + \; \underline{\underline{p}}^{W\textnormal{A},\textnormal{qu}}(\vec q,t), \\
\underline{\underline{p}}^{W\textnormal{A},\textnormal{cl}}(\vec q,t) &=& \underline{\underline{p}}^{K\textnormal{A},\textnormal{cl}}(\vec q,t) \;\hspace{-0.05 cm}  = \; \hspace{-0.05 cm}  N(\textnormal{A}) \hspace{0.075cm}  m_\textnormal{A} \hspace{-0.1 cm} \int \hspace{-0.1 cm} \textnormal{d} \vec Q \; \delta (\vec q - \vec q_1^{\hspace{0.05cm}\textnormal{A}}  \hspace{-0.05 cm} ) \; D \; \hspace{-0.1 cm}  \left (   \vec u_{1}^\textnormal{A} \otimes \vec u_{1}^\textnormal{A} \right)  \hspace{-0.1 cm}  ,  \\
\underline{\underline{p}}^{W\textnormal{A},\textnormal{qu}}(\vec q,t) &=& \underline{\underline{\Pi}}^{W\textnormal{A},\textnormal{qu}}(\vec q,t) \;  \hspace{-0.05 cm} = \; \hspace{-0.05 cm}  \underline{\underline{1}} P_\textnormal{A} \hspace{-0.05 cm} + \hspace{-0.05 cm} N(\textnormal{A}) \hspace{0.075cm}  m_\textnormal{A} \hspace{-0.1 cm} \int \hspace{-0.1 cm} \textnormal{d} \vec Q \;  \delta (\vec q - \vec q_1^{\hspace{0.05cm}\textnormal{A}}  \hspace{-0.05 cm} ) \; D \; \hspace{-0.1 cm} \left ( \hspace{-0.05 cm} \vec d_{1}^{\hspace{0.05cm}\textnormal{A}} \otimes \vec d_{1}^{\hspace{0.05cm}\textnormal{A}}  \hspace{-0.05 cm} \right) \hspace{-0.1 cm}  .
\end{eqnarray}
By the equations above we realize that for the Wyatt tensors $\underline{\underline{\Pi}}^{W\textnormal{A}}(\vec q,t)$ and $\underline{\underline{p}}^{W\textnormal{A}}(\vec q,t)$, their quantum parts are related to dyadic products $\vec d_{1}^{\hspace{0.05cm}\textnormal{A}} \otimes \vec d_{1}^{\hspace{0.05cm}\textnormal{A}}$ of the osmotic velocity $\vec d_{1}^{\hspace{0.05cm}\textnormal{A}}(\vec Q,t)$ and to the scalar quantum pressure $P_\textnormal{A}(\vec q,t)$. So, the advantage of the Wyatt tensor versions $\underline{\underline{\Pi}}^{W\textnormal{A}}(\vec q,t)$ and $\underline{\underline{p}}^{W\textnormal{A}}(\vec q,t)$
relative to the corresponding Kuzmenkov tensors is that their quantum parts $\underline{\underline{\Pi}}^{W\textnormal{A},\textnormal{qu}}(\vec q,t)$ and $\underline{\underline{p}}^{W\textnormal{A},\textnormal{qu}}(\vec q,t)$, which are identical, are more clearly to interpret.
\newline \newline 
Now, it remains to show that the Wyatt tensors $\underline{\underline{\Pi}}^{W\textnormal{A}}(\vec q,t)$ and $\underline{\underline{p}}^{W\textnormal{A}}(\vec q,t)$ are physically equivalent to the corresponding Kuzmenkov tensors $\underline{\underline{\Pi}}^{K\textnormal{A}}(\vec q,t)$ or $\underline{\underline{p}}^{K\textnormal{A}}(\vec q,t)$, respectively. First, we explain this proof for the pressure tensors $\underline{\underline{p}}^{W\textnormal{A}}(\vec q,t)$ and  $\underline{\underline{p}}^{K\textnormal{A}}(\vec q,t)$: \newline
As will be shown below, the Wyatt pressure tensor $\underline{\underline{p}}^{W\textnormal{A}}(\vec q,t)$ is in general not equal to the Kuzmenkov pressure tensor $\underline{\underline{p}}^{K\textnormal{A}}(\vec q,t)$:
\begin{eqnarray}
\underline{\underline{p}}^{W \textnormal{A}}(\vec q,t) &\neq& \underline{\underline{p}}^{K \textnormal{A}}(\vec q,t). \label{pressure tensors not equal}
\end{eqnarray}
However, it will be shown below, too, that
\begin{eqnarray}
\nabla  \underline{\underline{p}}^{W\textnormal{A}}(\vec q,t) &=& \nabla \underline{\underline{p}}^{K\textnormal{A}}(\vec q,t) \label{Equal_divergences}
\end{eqnarray}
holds, so that both tensors lead to an equivalent input in the MPQCE (\ref{MQNSE for particle sort k}) -- and this is the property which makes them physically equivalent. \newline \newline
So, as a general statement, for a pressure tensor $\underline{\underline p}(\vec q,t)$ only its divergnce  $\nabla \underline{\underline p}(\vec q,t)$ is physically important, and in this sense, it behaves like a scalar potential $\phi(\vec q,t)$ for which only the gradient $\nabla \phi(\vec q,t)$ is physically important. Note here that both $\nabla \underline{\underline p}(\vec q,t)$ and $\nabla \phi(\vec q,t)$ are vectors. Thus, for pressure tensors $\underline{\underline p}(\vec q,t)$ and for the scalar potential $\phi(\vec q,t)$ mentioned above, we can apply both transformations that keep  $\nabla \underline{\underline p}(\vec q,t)$ or $\nabla \phi(\vec q,t)$, respectively, constant. For the scalar potential $\nabla \phi(\vec q,t)$, the only degree of freedom for such a transformation is an additive constant independent of the position $\vec q$. However, since in Cartesian coordinates, the divergence of the tensor $\nabla \underline{\underline p}(\vec q,t)$ is calculated by 
\begin{eqnarray}
\nabla \underline{\underline{p}}(\vec q,t)  &=&  \left( \begin{array} {ccc} \frac{\partial p_{xx}}{\partial x}  + \frac{\partial p_{yx}}{\partial y} + \frac{\partial p_{zx}}{\partial z}  \\ 
\frac{\partial p_{xy}}{\partial x} + \frac{\partial p_{yy}}{\partial y} +  \frac{\partial p_{zy}}{\partial z} \\  
\frac{\partial p_{xz}}{\partial x} + \frac{\partial p_{yz}}{\partial y} + \frac{\partial p_{zz}}{\partial z} 
\end{array} \right),   \label{pressure_tensor_divergence}   
\end{eqnarray}
there are transformations for the different pressure tensor elements $p_{\alpha \beta}(\vec q,t)$ that make the coordinate derivations of these tensor elements vary but keep $\nabla \underline{\underline{p}}(\vec q,t)$ constant (e.\hspace{-0.2 cm} g. $p_{xy} \rightarrow  p_{xy} + Cx$ and  $p_{yy} \rightarrow p_{yy} - Cy$, all other $p_{\alpha \beta}$ remain unmodified).  \newline \newline
In this sense, we can understand the definition  (\ref{pressure_tensor_Wyatt}) for the Wyatt pressure tensor $\underline{\underline{p}}^{W\textnormal{A}}(\vec q,t)$ as a tensor version where the physical interpretation of all quantities appearing in this defi\-nition are clear. But because of the transformation freedom for the pressure tensor elements $p^{\textnormal{A}}_{\alpha \beta}(\vec q,t)$ explained above, there are other versions for the pressure tensor $\underline{\underline{p}}^{\textnormal{A}}(\vec q,t)$ where this physical interpretation is not so clear -- and the Kuzmenkov pressure tensor $\underline{\underline{p}}^{K\textnormal{A}}(\vec q,t)$ is one of these other versions. \newline \newline
At this point, we notice that it was already mentioned in \cite{Wong_1976, Deb_1979_1} that there exist seve\-ral versions of the pressure tensor $\underline{\underline{p}}(\vec q,t)$. In addition, Sonego discussed already in \cite{Sonego_1991} the reason mentioned above for the ambiguity of the pressure tensor  $\underline{\underline{p}}(\vec q,t)$. In contrast to our discussion, Sonego restricts the allowed transformations of this tensor to transformations which keep a pressure tensor with symmetric matrix elements ($p_{\alpha \beta}(\vec q,t) = p_{\beta \alpha} (\vec q,t)$) symmetric -- but we think that this condition is not mandatory because only the conservation of the tensor divergence $\nabla \underline{\underline{p}}(\vec q,t)$ is required physically. Moreover, in the same reference, Sonego presented two versions of the pressure tensor, which we would call in our nomenclature the Kuzmenkov pressure tensor $\underline{\underline{p}}^K(\vec q,t)$ and the Wyatt pressure tensor $\underline{\underline{p}}^W(\vec q,t)$. But in contrast to our work, Sonego prefers using the Kuzmenkov pressure tensor  $\underline{\underline{p}}^K(\vec q,t)$ to using the Wyatt pressure tensor $\underline{\underline{p}}^W(\vec q,t)$. His reason for this is that he uses a function called Wigner function to describe the system in the phase space which yields as a result the Kuzmenkov pressure tensor $\underline{\underline{p}}^K(\vec q,t)$. However, Sonego himself stated that ``we do not claim at all that the Wigner function is the correct phase space distribution, nor that such a distribution exists: (...)'' \cite{Sonego_1991}, p.\ 1166. In this sense, we think that it is still reasonable to prefer the Wyatt pressure tensor $\underline{\underline{p}}^W(\vec q,t)$. \newline 
We now finish our discussion about the ambiguity of the pressure tensor $\underline{\underline{p}}(\vec q,t)$ with the remark that similar ambiguities of quantities also appear in other fields of physics. One example for this context is the energy-momentum tensor $\underline{\underline{T}}(\vec q,t)$ in relativistic physics, which has to fulfill the condition that its four-divergence vanishes, but this condition does not determine the tensor uniquely -- a discussion about this context can be found in \cite{Blaschke_2016}. Another example is the gauge ambiguity of the vector potential $\vec {\mathcal{A}}(\vec q,t)$ and the scalar potential $\Phi(\vec q,t)$ in electrodynamics: \newline 
There is the Lorenz gauge  
\begin{equation}
\nabla \vec {\mathcal{A}}(\vec q,t) + \frac{1}{c^2} \frac{\partial}{\partial t}  \Phi(\vec q,t) = 0, 
\end{equation}
which has the advantage that the description of electrodynamics in relativistic physics becomes very elegant if one applies this gauge. This elegance is a good reason to prefer the Lorenz gauge to other gauges \cite{Jackson_1962}, p. 179-181, 377-380.
\newline 
Nevertheless, the Lorenz gauge is not the only gauge for $\vec {\mathcal{A}}(\vec q,t)$ and $\Phi(\vec q,t)$ that one can find in literature; there is the Coulomb gauge  
\begin{equation}
\nabla \vec {\mathcal{A}}(\vec q,t) = 0  
\end{equation}
as well, where the divergence of the vector potential $\vec {\mathcal{A}}(\vec q,t)$ vanishes. The Coulomb gauge can be advantageous for applications where no charges are present. For more details see \cite{Jackson_1962}, p. 181-183. \newline \newline 
In order to prove now Eqns.\ (\ref{pressure tensors not equal}) and (\ref{Equal_divergences}), we first transform the term $- \frac{\hbar^2}{4 m_\textnormal{A}} \frac {\partial^2 \ln D}{\partial q_{1 \alpha}^\textnormal{A} \partial q_{1 \beta}^\textnormal{A}}$ appearing in Eqn.\ (\ref{k sort pressure tensor result}) for the matrix elements $p_{\alpha \beta}^{K\textnormal{A}}(\vec q,t)$ using Eqn.\ (\ref{definition d}): 
\begin{eqnarray}
- \frac{\hbar^2}{4 m_\textnormal{A}} \frac {\partial^2 \ln D}{\partial q_{1 \alpha}^\textnormal{A} \partial q_{1 \beta}^\textnormal{A}} &=& - \frac{\hbar^2}{4 m_\textnormal{A}} \frac{\partial}{\partial q_{1 \alpha}^\textnormal{A}} \left ( \frac{1}{D} \frac{\partial D}{\partial q_{1 \beta}^\textnormal{A}} \right) \nonumber \\ 
&=& \frac{\hbar^2}{4m_\textnormal{A}} \frac{1}{D^2} \frac{\partial D}{\partial  q_{1 \alpha}^\textnormal{A}} \frac{\partial D}{\partial q_{1 \beta}^\textnormal{A}} - \frac{\hbar^2}{4m_\textnormal{A}} \frac{1}{D} \frac{\partial^2 D} {\partial q_{1 \alpha}^\textnormal{A} \partial q_{1 \beta}^\textnormal{A}} \nonumber \\
 &=& m_\textnormal{A} d_{1 \alpha}^\textnormal{A} d_{1 \beta}^\textnormal{A} - \frac{\hbar^2}{4m_\textnormal{A}} \frac{1}{D} \frac{\partial^2 D} {\partial q_{1 \alpha}^\textnormal{A} \partial q_{1 \beta}^\textnormal{A}}. \label{intermediate_result_1}
\end{eqnarray}
As a next step, we insert the intermediate result (\ref{intermediate_result_1}) into Eqn.\ (\ref{k sort pressure tensor result}) for $p_{\alpha \beta}^{K\textnormal{A}}(\vec q,t)$. 
Now, $p_{\alpha \beta}^{K\textnormal{A}}(\vec q,t)$ can be splitted in a sum 
\begin{eqnarray}
p_{\alpha \beta}^{K\textnormal{A}}(\vec q,t) &=& p_{\alpha \beta}^{K\textnormal{A},1}(\vec q,t) + p_{\alpha \beta}^{K\textnormal{A},2}(\vec q,t), 
\end{eqnarray}
where
\begin{eqnarray}
p_{\alpha \beta}^{K\textnormal{A},1}(\vec q,t) &=& N(\textnormal{A}) \hspace{0.075cm}  m_\textnormal{A} \int \textnormal{d} \vec Q \; \delta (\vec q - \vec q_1^{ \hspace{0.05 cm} \textnormal{A}}) \;  D  \; \left (u_{1 \alpha}^\textnormal{A} u_{1 \beta}^\textnormal{A} + d_{1 \alpha}^\textnormal{A} d_{1 \beta}^\textnormal{A} \right),  \label{Kusmenkov_K1} \\
p_{\alpha \beta}^{K\textnormal{A},2}(\vec q,t) &=& - N(\textnormal{A}) \frac{\hbar^2}{4 m_\textnormal{A}} \int \textnormal{d} \vec Q \; \delta (\vec q - \vec q_1^{\hspace{0.05 cm} \textnormal{A}}) \;  \frac{\partial^2 D} {\partial q_{1 \alpha}^\textnormal{A} \partial q_{1 \beta}^\textnormal{A}}. \label{pKk1}
\end{eqnarray}
Here, we make the following remark: The naming of the terms $p_{\alpha \beta}^{K\textnormal{A},1}(\vec q,t)$ and $p_{\alpha \beta}^{K\textnormal{A},2}(\vec q,t)$ is not just a simple numbering, but there is a deeper meaning: The term $p_{\alpha \beta}^{K\textnormal{A},1}(\vec q,t)$ contains products of two factors being 
first-order Cartesian coordinate derivations of $S(\vec Q,t)$ or $D(\vec Q,t)$, and the term $p_{\alpha \beta}^{K\textnormal{A},2}(\vec q,t)$ contains second-order Cartesian coordinate derivations of $D(\vec Q,t)$. \newline
In an analogous manner, we can also split the corresponding Wyatt matrix element $p_{\alpha \beta}^{W\textnormal{A}}(\vec q,t)$ in two summands using Eqn.\ (\ref{pressure_tensor_Wyatt}):
\begin{eqnarray}
p_{\alpha \beta}^{W\textnormal{A}}(\vec q,t) &=& p_{\alpha \beta}^{W\textnormal{A},1}(\vec q,t) + p_{\alpha \beta}^{W\textnormal{A},2}(\vec q,t). \label{Zerlegung_Wyatt_pressure_tensor}
\end{eqnarray}
Here, the summands $p_{\alpha \beta}^{W\textnormal{A},1}(\vec q,t)$ and $p_{\alpha \beta}^{W\textnormal{A},2}(\vec q,t)$ are given by: 
\begin{eqnarray}
p_{\alpha \beta}^{W\textnormal{A},1}(\vec q,t) &=& N(\textnormal{A}) \hspace{0.075cm}  m_\textnormal{A} \hspace{-0.05 cm} \int  \hspace{-0.05 cm}  \textnormal{d} \vec Q \; \delta (\vec q - \vec q_1^{ \hspace{0.05 cm} \textnormal{A}}) \; \hspace{-0.1 cm} D  \; \hspace{-0.15 cm}  \left (u_{1 \alpha}^\textnormal{A} u_{1 \beta}^\textnormal{A} + d_{1 \alpha}^\textnormal{A} d_{1 \beta}^\textnormal{A} \right)  \hspace{-0.05 cm}  \; = \;  p_{\alpha \beta}^{K\textnormal{A},1}(\vec q,t), \label{pWk1} \\
p_{\alpha \beta}^{W\textnormal{A},2}(\vec q,t) &=&  P_\textnormal{A}(\vec q,t) \; \delta_{\alpha \beta}. \label{pWK2}
\end{eqnarray}
So, the summands $p_{\alpha \beta}^{W\textnormal{A},1}(\vec q,t)$ and $p_{\alpha \beta}^{K\textnormal{A},1}(\vec q,t)$ are equal. But in general, $p_{\alpha \beta}^{K\textnormal{A},2}(\vec q,t)$ and $p_{\alpha \beta}^{W\textnormal{A},2}(\vec q,t)$ are not equal -- in particular, $p_{\alpha \beta}^{W\textnormal{A},2}(\vec q,t)$ is always diagonal, and $p_{\alpha \beta}^{K\textnormal{A},2}(\vec q,t)$ is in general non-diagonal.  \newline 
Thus, first, we haven proven the inequation (\ref{pressure tensors not equal}) that in general $\underline{\underline{p}}^{W \textnormal{A}}(\vec q,t)$ and $\underline{\underline{p}}^{K \textnormal{A}}(\vec q,t)$ are not equal. 
\newline Second, to prove Eqn.\ (\ref{Equal_divergences}), the remaining task is to show that the following equation is true: 
\begin{eqnarray}
\nabla \underline{\underline{p}}^{K\textnormal{A},2}(\vec q,t) = \nabla \underline{\underline{p}}^{W\textnormal{A},2}(\vec q,t). \label{Second_order_equal}
\end{eqnarray}
The proof for this equation can be done with the following straightforward calculation:
We analyze the $\beta$-component of the tensor divergence $\nabla \underline{\underline{p}}^{K\textnormal{A},2}(\vec q,t)$ in Cartesian coordinates:
\begin{eqnarray}
\left[\nabla \underline{\underline{p}}^{K\textnormal{A},2}(\vec q,t)\right]_\beta &=& \sum_{\alpha \in K_{\textnormal{Ca}}}  \frac{\partial p_{\alpha \beta}^{K\textnormal{A},2}(\vec q,t)}{\partial q_{\alpha}} \nonumber \\
                                                             &=& \sum_{\alpha \in K_{\textnormal{Ca}}}  \frac{\partial}{\partial q_\alpha} \left [ - N(\textnormal{A}) \frac{\hbar^2}{4 m_\textnormal{A}} \int \textnormal{d} \vec Q \; \delta (\vec q - \vec q_1^{\hspace{0.05cm} \textnormal{A}}) \; \frac{\partial^2 D} {\partial q_{1 \alpha}^\textnormal{A} \partial q_{1 \beta}^\textnormal{A}} \right] \nonumber  \\   
                                                             &=&  -  N(\textnormal{A}) \frac{\hbar^2}{4 m_\textnormal{A}} \sum_{\alpha \in K_{\textnormal{Ca}}}  \int \textnormal{d} \vec Q \; \delta (\vec q - \vec q_1^{\hspace{0.05cm} \textnormal{A}}) \; \frac{\partial}{\partial q_{1 \alpha}^\textnormal{A}}  \frac{\partial^2 D} {\partial q_{1 \alpha}^\textnormal{A} \partial q_{1 \beta}^\textnormal{A}} \label{intermediate_result_3a} \\
                                                             &=& - N(\textnormal{A}) \frac{\hbar^2}{4 m_\textnormal{A}}  \int \textnormal{d} \vec Q \; \delta (\vec q - \vec q_1^{\hspace{0.05cm} \textnormal{A}}) \; \frac{\partial}{\partial q_{1 \beta}^\textnormal{A}} \sum_{\alpha \in K_{\textnormal{Ca}}}  \frac{\partial^2 D} {\partial q_{1 \alpha}^\textnormal{A} \partial q_{1 \alpha}^\textnormal{A}} \nonumber \\
                                                             &=& \frac{\partial}{\partial q_{\beta}} \left [ - N(\textnormal{A}) \frac{\hbar^2}{4 m_\textnormal{A}}  \int \textnormal{d} \vec Q \; \delta (\vec q - \vec q_1^{\hspace{0.05cm} \textnormal{A}}) \; \laplace_1^\textnormal{A} D \right] \nonumber \\
                                                             &=& \frac{\partial}{\partial q_{\beta}} P_\textnormal{A} (\vec q,t) \nonumber \\
                                                             &=& \sum_{\alpha \in K_{\textnormal{Ca}}} \frac{\partial}{\partial q_{\alpha}} \left [P_\textnormal{A} (\vec q,t) \; \delta_{\alpha \beta} \right]  \label{intermediate_result_4a}\\
                                                             &=& \sum_{\alpha \in K_{\textnormal{Ca}}}   \frac{\partial  p^{W\textnormal{A},2}_{\alpha \beta} (\vec q,t)}{\partial q_{\alpha}} \nonumber \\
                                                             &=& \left[\nabla \underline{\underline{p}}^{W\textnormal{A},2}(\vec q,t)\right]_\beta \nonumber \\
\Longrightarrow \nabla \underline{\underline{p}}^{K\textnormal{A},2}(\vec q,t) &=& \nabla \underline{\underline{p}}^{W\textnormal{A},2}(\vec q,t). \label{Second_order_equal_proof}
\end{eqnarray}
The crucial steps of the proof shown above occur between Eqns.\ (\ref{intermediate_result_3a}) and (\ref{intermediate_result_4a}), where a rearrangement of the spatial derivations is done. This rearrangement is possible due to the sum $\sum_{\alpha \in K_{\textnormal{Ca}}}$ appearing both in Eqn.\ (\ref{intermediate_result_3a}) and Eqn.\ (\ref{intermediate_result_4a}) because this sum corresponds to the fact that, in Cartesian coordinates, for each vector component of a tensor divergence there is a sum with three summands, where each of these three summands depends on spatial derivatives of different tensor matrix elements -- Eqn.\ (\ref{pressure_tensor_divergence}) is an illustration of this fact. The rearrangement above changes each of the three summands in this sum but doing this, the value of the total sum remains unchanged.  \newline 
Finally, with the proof of Eqn.\ (\ref{Second_order_equal}), we have the evidence that Eqn.\ (\ref{Equal_divergences}) is true -- thus, both the Kuzmenkov pressure tensor $\underline{\underline{p}}^{K\textnormal{A}}(\vec q,t)$ and the Wyatt pressure tensor $\underline{\underline{p}}^{W\textnormal{A}}(\vec q,t)$ lead to an equivalent input in the MPQCE (\ref{MQNSE for particle sort k}), and therefore, they are physically equivalent. \newline \newline 
Now, it remains to prove that the momentum-flow density tensors $\underline{\underline{\Pi}}^{W\textnormal{A}}(\vec q,t)$ and $\underline{\underline{\Pi}}^{K\textnormal{A}}(\vec q,t)$ are physically equivalent. Analogously to our analysis for the pressure tensors, we will show below that in general $\underline{\underline{\Pi}}^{W\textnormal{A}}(\vec q,t)$ and $\underline{\underline{\Pi}}^{K\textnormal{A}}(\vec q,t)$ are not equal:
\begin{eqnarray}
\underline{\underline{\Pi}}^{W\textnormal{A}}(\vec q,t) &\neq& \underline{\underline{\Pi}}^{K\textnormal{A}}(\vec q,t). \label{pi tensors not equal}
\end{eqnarray} 
But, we will also show below that these tensors have the property: 
\begin{eqnarray}
\nabla \underline{\underline{\Pi}}^{W\textnormal{A}}(\vec q,t) &=& \nabla \underline{\underline{\Pi}}^{K\textnormal{A}}(\vec q,t). \label{Equal_divergences Pi}
\end{eqnarray}
Due to this property, both tensors lead to an equivalent input in the MPEEM (\ref{MMBE}) making these tensors physically equivalent. \newline 
The first step to prove Eqns.\ (\ref{pi tensors not equal}) and (\ref{Equal_divergences Pi}) is inserting Eqn.\ (\ref{intermediate_result_1}) into Eqn.\ (\ref{total momentum flow density tensor result 3}) for the Kuzmenkov tensor elements $\Pi_{\alpha \beta}^{K\textnormal{A}}(\vec q,t)$, and to split each of them into a term $\Pi_{\alpha \beta}^{K\textnormal{A},1}(\vec q,t)$ containing products of two factors being first-order Cartesian derivations of $S(\vec Q,t)$ or $D(\vec Q,t)$, and a term $\Pi_{\alpha \beta}^{K\textnormal{A},2}(\vec q,t)$ containing products of second-order Cartesian derivations of $D(\vec Q,t)$.  Thus, we get: 
\begin{eqnarray}
\Pi_{\alpha \beta}^{K\textnormal{A}}(\vec q,t)   &=& \Pi_{\alpha \beta}^{K\textnormal{A},1}(\vec q,t) + \Pi_{\alpha \beta}^{K\textnormal{A},2}(\vec q,t), \\
\Pi_{\alpha \beta}^{K\textnormal{A},1}(\vec q,t) &=& N(\textnormal{A}) \hspace{0.075cm}  m_\textnormal{A} \int \textnormal{d} \vec Q \; \delta (\vec q - \vec q_1^{ \hspace{0.05 cm} \textnormal{A}}) \;  D  \; \left (w_{1 \alpha}^\textnormal{A} w_{1 \beta}^\textnormal{A} + d_{1 \alpha}^\textnormal{A} d_{1 \beta}^\textnormal{A} \right), \label{Kusmenkov_K2} \\
\Pi_{\alpha \beta}^{K\textnormal{A},2}(\vec q,t) &=& - N(\textnormal{A}) \frac{\hbar^2}{4 m_\textnormal{A}} \int \textnormal{d} \vec Q \; \delta (\vec q - \vec q_1^{\hspace{0.05 cm} \textnormal{A}}) \;  \frac{\partial^2 D} {\partial q_{1 \alpha}^\textnormal{A} \partial q_{1 \beta}^\textnormal{A}} \; = \; p_{\alpha \beta}^{K\textnormal{A},2}(\vec q,t). \label{Pi_part_2_Kusmenkov}
\end{eqnarray}
An analogous splitting can be done for the Wyatt tensor elements $\Pi_{\alpha \beta}^{W\textnormal{A}}(\vec q,t)$ given in Eqn.\ (\ref{momentum_flow_densitytensor}): 
\begin{eqnarray}
\Pi_{\alpha \beta}^{W\textnormal{A}}(\vec q,t)   &=& \Pi_{\alpha \beta}^{W\textnormal{A},1}(\vec q,t) + \Pi_{\alpha \beta}^{W\textnormal{A},2}(\vec q,t), \\
\Pi_{\alpha \beta}^{W\textnormal{A},1}(\vec q,t) &=& \Pi_{\alpha \beta}^{K\textnormal{A},1}(\vec q,t), \label{Kusmenkov_K3} \\
\Pi_{\alpha \beta}^{W\textnormal{A},2}(\vec q,t) &=& P_\textnormal{A}(\vec q,t) \delta_{\alpha \beta} \; = \;  p_{\alpha \beta}^{W\textnormal{A},2}(\vec q,t). \label{Pi_part_2_Wyatt}
\end{eqnarray}
We realize that the terms $\Pi_{\alpha \beta}^{W\textnormal{A},1}(\vec q,t)$ and $\Pi_{\alpha \beta}^{K\textnormal{A},1}(\vec q,t)$ are equal, but in general 
$\Pi_{\alpha \beta}^{W \textnormal{A},2}(\vec q,t)$ and $\Pi_{\alpha \beta}^{K \textnormal{A},2}(\vec q,t)$ are not equal.\newline 
So, the inequation (\ref{pi tensors not equal}) is proven that in general $\underline{\underline{\Pi}}^{W \textnormal{A}}(\vec q,t)$ and $\underline{\underline{\Pi}}^{K \textnormal{A}}(\vec q,t)$ are not equal. \newline 
For the proof of Eqn.\ (\ref{Equal_divergences Pi}), which we need to show the physical equivalence of the tensors $\underline{\underline{\Pi}}^{W \textnormal{A}}(\vec q,t)$ and $\underline{\underline{\Pi}}^{K \textnormal{A}}(\vec q,t)$, it remains to show that 
\begin{eqnarray}
\nabla \underline{\underline{\Pi}}^{W \textnormal{A},2}(\vec q,t) &=& \nabla \underline{\underline{\Pi}}^{K \textnormal{A},2}(\vec q,t). \label{Second_order_equal Pi} 
\end{eqnarray}
Therefore, we take into account that $\underline{\underline{\Pi}}^{K\textnormal{A},2}(\vec q,t)$ is just equal to  $\underline{\underline{p}}^{K\textnormal{A},2}(\vec q,t)$, and $\underline{\underline{\Pi}}^{W\textnormal{A},2}(\vec q,t)$ is just equal to $\underline{\underline{p}}^{W\textnormal{A},2}(\vec q,t)$.  As we proved Eqn.\ (\ref{Second_order_equal}) already, Eqn.\ (\ref{Second_order_equal Pi}) must also be true. Thus, we proved Eqn.\ (\ref{Equal_divergences Pi}), and finally, we showed that the Wyatt and Kuzmenkov momentum flow density tensors $\underline{\underline{\Pi}}^{W\textnormal{A}}(\vec q,t)$ and $\underline{\underline{\Pi}}^{K\textnormal{A}}(\vec q,t)$ are physically equivalent. \newline \newline 
As an intermediate conclusion, we found that the Wyatt and the Kuzmenkov pressure tensors  $\underline{\underline{p}}^{W\textnormal{A}}(\vec q,t)$,  $\underline{\underline{p}}^{K\textnormal{A}}(\vec q,t)$ are physically equivalent, and the same holds for the Wyatt and Kuzmenkov momentum flow density tensors $\underline{\underline{\Pi}}^{W\textnormal{A}}(\vec q,t)$, $\underline{\underline{\Pi}}^{K\textnormal{A}}(\vec q,t)$. Moreover, the quantum parts of the Wyatt tensors are more easily to interpret than the quantum parts of the Kuzmenkov tensors. 
\subsection{System with cylindrical symmetry}
As an additional task, we will now show for an example system that the divergence of the Wyatt pressure tensor  $\underline{\underline{p}}^{W\textnormal{A}}(\vec q,t)$ is more easily to calculate than the divergence of the Kuzmenkov pressure tensor  $\underline{\underline{p}}^{K\textnormal{A}}(\vec q,t)$, so, the clearer interpretation is not the only advantage of the Wyatt pressure tensor  $\underline{\underline{p}}^{W\textnormal{A}}(\vec q,t)$. \newline \newline 
For the analyzed example system with cylindrical symmetry, the use of cylindrical coordinates is advantageous, and it means that we represent the position vector $\vec q$ by
\begin{eqnarray}
\vec q &=& q_\rho \vec e_{\rho} + q_\varphi \vec e_{\varphi } + q_z \vec e_{z},
\end{eqnarray}
with the cylindrical basis vectors $\vec e_{\rho}$, $\vec e_{\varphi}$, $\vec e_{z}$ instead of the Cartesian representation
\begin{eqnarray}
\vec q &=& q_x \vec e_{x} + q_y \vec e_{y} + q_z \vec e_{z}.
\end{eqnarray}
Now, we introduce the radius $\rho$, the phase $\varphi$, and the coordinates $x$, $y$, and $z$, depending on $q_x$, $q_y$, and $q_z$ by
\begin{eqnarray}
q_x & \equiv & x \; = \; \rho \cos \varphi, \label{Transformationx} \\
q_y & \equiv & y \; = \; \rho \sin \varphi, \label{Transformationy} \\
q_z & \equiv & z.                           \label{Transformationz} 
\end{eqnarray}
We will show how the cylindrical vector components $q_\rho$, $q_\varphi$, and $q_z$ depend on $\rho$, $\varphi$, and $z$ -- in particular we will find that  $q_\varphi$ vanishes. \newline     
The transformation between the basis vectors $\vec e_{\rho}$, $\vec e_{\varphi}$,  $\vec e_{z}$ in cylindrical coordinates and the basis vectors in Cartesian coordinates is described by what is called rotation matrix $\underline{\underline{\Lambda}}(\varphi)$. \newline
This rotation matrix $\underline{\underline{\Lambda}}(\varphi)$ depends on the geometrical orientation of the position vector $\vec q$ via the phase $\varphi$, and its matrix elements $\Lambda_{\alpha' \alpha}(\varphi)$ have the following form  (\hspace{-0.01cm}\cite{Shapiro_2013}, p.\ 231):
\begin{eqnarray} 
\left( \begin{array} {ccc}  \vec e_{\rho}  \\ \vec e_{\varphi } \\ \vec e_{z } \end{array} \right ) &=& \left( \begin{array}{ccc} 
\Lambda_{\rho x}    & \Lambda_{\rho y}    & \Lambda_{\rho z}            \\
\Lambda_{\varphi x} & \Lambda_{\varphi y} & \Lambda_{\varphi z}         \\
\Lambda_{z x}       & \Lambda_{z y}       & \Lambda_{z z} \end{array} \right ) \hspace{-0.15 cm} \left( \begin{array} {ccc}  \vec e_{x}  \\ \vec e_{y} \\ \vec e_{z} \end{array} \right ) = \left( \hspace{-0.1 cm}  \begin{array}{ccc} 
 \cos \varphi & \sin \varphi & 0 \\
-\sin \varphi & \cos \varphi & 0 \\
      0       &      0       & 1 \end{array} \right ) \hspace{-0.15 cm} \left( \begin{array} {ccc}  \vec e_{x}  \\ \vec e_{y} \\ \vec e_{z} \end{array} \right ) \hspace{-0.1 cm}.  \label{Definition_Coordinate_Trafos} 
\end{eqnarray} 
As a convention for notation, we write in the following matrix elements of a tensor field $\underline{\underline{T}}(\vec q)$ or components of a vector field $\vec b (\vec q)$ in Cartesian coordinates as $T_{\alpha \beta} (\vec q)$ or as $b_{\alpha} (\vec q)$, respectively, but in cylindrical coordinates as $T_{\alpha' \beta'}(\vec q)$ or as $b_{\alpha'}(\vec q)$, respectively, if it is not explicitly specified what components are meant. Here, the Cartesian indices $\alpha$, $\beta$ are elements of the set $K_{\textnormal{Ca}} = \{x,y,z\}$, and the cylindrical indices $\alpha'$, $\beta'$ are elements of the set $K_{\textnormal{cy}} = \{\rho,\varphi,z\}$. \newline 
As a consequence of Eqn.\ (\ref{Definition_Coordinate_Trafos}), vector components $b_{\alpha}(\vec q)$ and tensor elements $T_{\alpha\beta}(\vec q)$ are transformed via (\hspace{-0.1 cm} \cite{Phan-Thien_2013}, p.\ 4f.):
\begin{eqnarray} 
b_{\alpha'} (\vec q)        &=& \sum_{\alpha \in K_{\textnormal{Ca}}} \Lambda_{\alpha' \alpha} (\varphi) \hspace{0.1cm} b_{\alpha} (\vec q),  \label{Transformationvector}  \\ 
T_{\alpha'\beta'} (\vec q)  &=& \sum_{\alpha \in K_{\textnormal{Ca}}} \sum_{\beta \in K_{\textnormal{Ca}}} \Lambda_{\alpha' \alpha} (\varphi) \hspace{0.1cm} \Lambda_{\beta' \beta} (\varphi) \hspace{0.1cm} T_{\alpha \beta} (\vec q). \label{Transformationtensor}
\end{eqnarray}
Applying Eqn.\ (\ref{Transformationvector}), we find for the particular case that the vector field $\vec b(\vec q)$ is the position vector $\vec q$ itself:
\begin{eqnarray}
\vec q &=& \rho \vec e_{\rho} + z \vec e_{z},
\end{eqnarray}
so $q_\rho \equiv \rho$,  and  $q_\varphi$ vanishes. However, for vectors $\vec b(\vec q)$ which are not equal to the position vector $\vec q$ itself the  component $b_\varphi(\vec q)$ does not need to vanish.  \newline   
For the coordinate transformation of the tensor elements $p^{K\textnormal{A}}_{\alpha \beta}(\vec q,t)$ and $p^{W\textnormal{A}}_{\alpha \beta}(\vec q,t)$, we take into account the cylindrical symmetry of the system mentioned above. Because of this symmetry, we assume that the wave function $\Psi$ describing this system has the following properties: \newline
The wave function describes a system for ${N_S}$ different sorts of particles like in our previous analysis, so $\Psi = \Psi(\vec Q,t)$. Moreover, as an additional symmetry property, we assume that the wave function $\Psi(\vec Q,t)$ does not depend on the polar angles $\varphi_{i\textnormal{A}}$ of all the $(\textnormal{A},i)$-particles for a certain sort of particles $\textnormal{A}$. \newline 
An example for a system with such a wave function is a $\textnormal{H}_2^+$ molecule in its electronic and rotational ground state because for fixed nuclei we can choose the coordinate system in a manner that the wave function is independent of the polar angle $\varphi_\textnormal{e}$ of the electron. \newline 
Thus, for the $S(\vec Q,t)$ and $D(\vec Q,t)$ functions related to a wave function $\Psi(\vec Q,t)$  of such a system it holds that the following equations are true for any natural number $n = 1,2,\ldots$ and any particle index $i = 1,2,\ldots, N(\textnormal{A})$:  
\begin{eqnarray}
\frac{\partial^n S}{\partial \varphi_{i\textnormal{A}}^n} &=& 0 \label{SymmetryS}, \\
\frac{\partial^n D}{\partial \varphi_{i\textnormal{A}}^n} &=& 0.  \label{SymmetryD} 
\end{eqnarray}
Moreover, using Eqns.\ (\ref{k sort pressure tensor result}) and (\ref{pressure_tensor_Wyatt}) it can be easily realized that the matrices for $\underline{\underline{p}}^{X\textnormal{A}}(\vec q,t)$, where $X$ stands both for the Kuzmenkov and the Wyatt pressure tensor, are symmetric for Cartesian coordinates:
\begin{eqnarray}
p^{X\textnormal{A}}_{\alpha \beta}(\vec q,t) &=& p^{X\textnormal{A}}_{\beta \alpha}(\vec q,t).  \label{symmetry_cartesian} 
\end{eqnarray}
Regarding this symmetry $p^{X\textnormal{A}}_{\alpha \beta}(\vec q,t)  = p^{X\textnormal{A}}_{\beta \alpha}(\vec q,t)$ for Cartesian coordinates, we can prove using Eqn.\ (\ref{Transformationtensor}) for matrix element transformations that this symmetry is true for cylindrical coordinates, too: 
\begin{eqnarray}
p^{X\textnormal{A}}_{\beta' \alpha'}(\vec q,t) &=& \sum_{\alpha \in K_{\textnormal{Ca}}} \sum_{\beta \in K_{\textnormal{Ca}}}  \Lambda_{\beta' \alpha} (\varphi) \hspace{0.1cm} \Lambda_{\alpha' \beta} (\varphi) \hspace{0.1cm} p^{X\textnormal{A}}_{\alpha \beta}  (\vec q,t) \nonumber \\ 
& \underset{\text{Commutate } \alpha, \beta}{=} & \sum_{\beta \in K_{\textnormal{Ca}}} \sum_{\alpha \in K_{\textnormal{Ca}}}  \Lambda_{\alpha' \alpha} (\varphi) \hspace{0.1cm} \Lambda_{\beta' \beta} (\varphi) \hspace{0.1cm} \underbrace{p^{X\textnormal{A}}_{\beta \alpha}}_{= \; p^{X\textnormal{A}}_{\alpha \beta}} (\vec q,t) \nonumber \\
                                  &=&  p^{X\textnormal{A}}_{\alpha' \beta'}  (\vec q,t) \; \; \square.  \label{symmetry_cylindric}  
\end{eqnarray}
For the following analysis, it is advantageous to split the pressure tensor elements $p^{X\textnormal{A}}_{\alpha \beta}(\vec q,t)$ 
into two parts $p^{X\textnormal{A},1}_{\alpha \beta}(\vec q,t)$  and $p^{X\textnormal{A},2}_{\alpha \beta}(\vec q,t)$, analogously to the discussions above. Then, we transform each part separately into corresponding cylindrical coordinate matrix elements  $p^{X\textnormal{A},1}_{\alpha' \beta'}(\vec q,t)$  or $p^{X\textnormal{A},2}_{\alpha' \beta'}(\vec q,t)$, respectively. \newline 
Because of Eqns.\ (\ref{Kusmenkov_K1}),  (\ref{pKk1}), (\ref{pWk1}), and (\ref{pWK2}), the Cartesian coordinate matrix elements  $p^{X\textnormal{A},1}_{\alpha \beta}(\vec q,t)$ and $p^{X\textnormal{A},2}_{\alpha \beta}(\vec q,t)$ are symmetric: 
\begin{eqnarray}
p^{X\textnormal{A},1}_{\beta \alpha}(\vec q,t) &=& p^{X\textnormal{A},1}_{\alpha \beta}(\vec q,t),      \label{symmetry_cartesian_part_1}  \\                                    
p^{X\textnormal{A},2}_{\beta \alpha}(\vec q,t) &=& p^{X\textnormal{A},2}_{\alpha \beta}(\vec q,t).      \label{symmetry_cartesian_part_2}                                 
\end{eqnarray}
We find that the cylindrical coordinate matrix elements  $p^{X\textnormal{A},1}_{\alpha' \beta'}(\vec q,t)$ and $p^{X\textnormal{A},2}_{\alpha' \beta'}(\vec q,t)$ are symmetric, too, by applying a calculation similar to the derivation of Eqn.\ (\ref{symmetry_cylindric}): 
\begin{eqnarray}
p^{X\textnormal{A},1}_{\beta' \alpha'}(\vec q,t) &=& p^{X\textnormal{A},1}_{\alpha' \beta'}(\vec q,t), \label{symmetry_cylindric_part_1}  \\
p^{X\textnormal{A},2}_{\beta' \alpha'}(\vec q,t) &=& p^{X\textnormal{A},2}_{\alpha' \beta'}(\vec q,t).  \label{symmetry_cylindric_part_2} 
\end{eqnarray}
Now, we regard these four points to make the transformations $p^{X\textnormal{A},1}_{\alpha \beta}(\vec q,t) \rightarrow p^{X\textnormal{A},1}_{\alpha' \beta'}(\vec q,t)$ and $p^{X\textnormal{A},2}_{\alpha \beta}(\vec q,t) \rightarrow p^{X\textnormal{A},2}_{\alpha' \beta'}(\vec q,t)$: \newline \newline 
First, for the calculation of the matrix elements $p^{X\textnormal{A},1}_{\alpha' \beta'}(\vec q,t)$ and $p^{X\textnormal{A},2}_{\alpha' \beta'}(\vec q,t)$, 
one has to eva\-luate the tensor transformation law  (\ref{Transformationtensor}), which leads to sums over corresponding Cartesian matrix elements $p^{X\textnormal{A},1}_{\alpha \beta}(\vec q,t)$ or $p^{X\textnormal{A},2}_{\alpha \beta}(\vec q,t)$: 
\begin{eqnarray}
p^{X\textnormal{A},1}_{\alpha' \beta'} (\vec q,t) &=& \sum_{\alpha \in K_{\textnormal{Ca}}}  \sum_{\beta \in K_{\textnormal{Ca}}} \Lambda_{\alpha' \alpha} (\varphi)   \hspace{0.1 cm} \Lambda_{\beta' \beta} (\varphi) \hspace{0.1 cm} p^{X\textnormal{A},1}_{\alpha \beta} (\vec q,t), \label{Help Transform1}  \\
p^{X\textnormal{A},2}_{\alpha' \beta'} (\vec q,t) &=& \sum_{\alpha \in K_{\textnormal{Ca}}} \sum_{\beta \in K_{\textnormal{Ca}}} \Lambda_{\alpha' \alpha} (\varphi) \hspace{0.1 cm}  \Lambda_{\beta' \beta} (\varphi) \hspace{0.1 cm} p^{X\textnormal{A},2}_{\alpha \beta} (\vec q,t). \label{Help Transform2}
\end{eqnarray}
Second, the Cartesian matrix elements $p^{X\textnormal{A},1}_{\alpha \beta}(\vec q,t)$ depend on the Cartesian vector components $u_{1\alpha}^\textnormal{A}(\vec Q,t)$, $d_{1\alpha}^{\textnormal{A}}(\vec Q,t)$ (see Eqn.\ (\ref{pWk1})). So,
when one calculates the matrix elements $p^{X\textnormal{A},1}_{\alpha' \beta'} (\vec q,t)$ using Eqn.\ (\ref{Help Transform1}), one has to transform the Cartesian vector components $u_{1\alpha}^\textnormal{A}(\vec Q,t)$, $d_{1\alpha}^{\textnormal{A}}(\vec Q,t)$ using Eqn.\ (\ref{Transformationvector}) into the vector components  $u_{1\alpha'}^\textnormal{A}(\vec Q,t)$, $d_{1\alpha'}^{\textnormal{A}}(\vec Q,t)$ for each of the Cartesian matrix elements $p^{X\textnormal{A},1}_{\alpha \beta}(\vec q,t)$, which appear on the right side of Eqn.\ (\ref{Help Transform1}). These velocity components $u_{1 \alpha'}^\textnormal{A}(\vec Q,t)$ and $d_{1 \alpha'}^\textnormal{A}(\vec Q,t) $ can be calculated from the quantities $S(\vec Q,t)$ and $D(\vec Q,t)$ by using Eqns.\ (\ref{One-particle density of sort k}), (\ref{mean particle velocity def}), (\ref{local velocity}),  (\ref{k_flux_density_bohm_representation}),  (\ref{definition relative velocity}), and (\ref{definition d}). We regard for this calculation that the divergence $\nabla_i^\textnormal{A} S(\vec Q,t)$ appears in  Eqn.\ (\ref{local velocity}), and that the divergence  $\nabla_i^\textnormal{A} D(\vec Q,t)$ appears in Eqn.\ (\ref{definition d}) -- we calculate these divergences in cylindrical coordinates by applying 
that the divergence of any scalar function $\Phi(\vec Q,t)$ related to the coordinate $\vec q_i^{\hspace{0.05 cm} \textnormal{A}}$ is given in cylindrical coordinates by:
\begin{eqnarray}
\nabla_i^{\textnormal{A}} \Phi(\vec Q,t) &=& \frac{\partial \Phi}{\partial \rho_{i\textnormal{A}}} \vec e_\rho + \frac{1}{\rho_{i\textnormal{A}}} \frac{\partial \Phi}{\partial \varphi_{i\textnormal{A}}} \vec e_\varphi + \frac{\partial \Phi}{\partial z_{i\textnormal{A}}} \vec e_z.  \label{Cylindrical_Divergence_vector} 
\end{eqnarray}
Third, the Cartesian coordinate derivations $\frac{\partial}{\partial q_{1x}^\textnormal{A}} \equiv \frac{\partial}{\partial x_{1 \textnormal{A}}} $ and $\frac{\partial}{\partial q_{1y}^\textnormal{A}} \equiv \frac{\partial}{\partial y_{1\textnormal{A}}}$ are present in Eqn.\ (\ref{pKk1}) for all of the Cartesian matrix elements $p^{K\textnormal{A},2}_{\alpha \beta}(\vec q,t)$ (besides the $zz$-element). Thus, when one calculates the matrix elements $p^{K\textnormal{A},2}_{\alpha' \beta'} (\vec q,t)$, one has to transform the Cartesian coordinate derivations $\frac{\partial}{\partial q_{1x}^\textnormal{A}} \equiv \frac{\partial}{\partial x_{1\textnormal{A}}} $ and $\frac{\partial}{\partial q_{1y}^\textnormal{A}} \equiv \frac{\partial}{\partial y_{1\textnormal{A}}}$ for the Cartesian matrix elements $p^{K\textnormal{A},2}_{\alpha \beta}(\vec q,t)$, which appear on the right side of Eqn.\ (\ref{Help Transform2}) for $X=K$. Hereby, one has to regard:  
\begin{eqnarray}
\frac{\partial}{\partial x_{1\textnormal{A}}} &=& \underbrace{\frac{\partial \rho_{1\textnormal{A}}}   {\partial x_{1\textnormal{A}}}}_{= \; \cos \varphi_{1\textnormal{A}}}               \frac{\partial}{\partial \rho_{1\textnormal{A}}} + 
                                     \underbrace{\frac{\partial \varphi_{1\textnormal{A}}}{\partial x_{1\textnormal{A}}}}_{= \; -\frac{\sin \varphi_{1\textnormal{A}}}{\rho_{1\textnormal{A}}}} \frac{\partial}{\partial \varphi_{1\textnormal{A}}} \; = \; \cos \varphi_{1\textnormal{A}} \frac{\partial}{\partial \rho_{1\textnormal{A}}} - \frac{\sin \varphi_{1\textnormal{A}}}{\rho_{1\textnormal{A}}} \frac{\partial}{\partial \varphi_{1\textnormal{A}}}, \label{trafo1} \\
\frac{\partial}{\partial y_{1\textnormal{A}}} &=& \underbrace{\frac{\partial \rho_{1\textnormal{A}}}   {\partial y_{1\textnormal{A}}}}_{= \;\sin \varphi_{1\textnormal{A}}}              \frac{\partial}{\partial \rho_{1\textnormal{A}}} + 
                                     \underbrace{\frac{\partial \varphi_{1\textnormal{A}}}{\partial y_{1\textnormal{A}}}}_{= \; \frac{\cos \varphi_{1\textnormal{A}}}{\rho_{1\textnormal{A}}}} \frac{\partial}{\partial \varphi_{1\textnormal{A}}} \; = \; \sin \varphi_{1\textnormal{A}} \frac{\partial}{\partial \rho_{1\textnormal{A}}} + \frac{\cos \varphi_{1\textnormal{A}}}{\rho_{1\textnormal{A}}} \frac{\partial}{\partial \varphi_{1\textnormal{A}}}.   \label{trafo2}                                           
\end{eqnarray}
Fourth, one can simplify the transformation calculations by taking into account the symmetry properties (\ref{SymmetryS}) and (\ref{SymmetryD}) for the $S(\vec Q,t)$ and $D(\vec Q,t)$ functions. However, we point out that in spite of these symmetry properties one cannot always omit the derivation relative to the $\varphi_{1\textnormal{A}}$-coordinate in Eqns.\ (\ref{trafo1}) and (\ref{trafo2}) -- this is important for the calculation of the matrix element $p_{\varphi  \varphi }^{K\textnormal{A},2} (\vec q,t)$. 
\newline \newline
Then, we find for the first-order tensor components $p^{X\textnormal{A},1}_{\alpha' \beta'}(\vec q,t)$: 
\begin{eqnarray}
p_{\alpha' \beta'}^{K\textnormal{A},1}(\vec q,t) &=& p_{\alpha' \beta'}^{W\textnormal{A},1}(\vec q,t) \nonumber \\
&=& N(\textnormal{A}) \int \textnormal{d}Q \; \delta (\vec q - \vec q_1^{\hspace{0.05 cm} \textnormal{A}}) \; D \; m_\textnormal{A} \left ( u_{1 \alpha'}^\textnormal{A} u_{1 \beta'}^\textnormal{A} + d_{1 \alpha'}^\textnormal{A} d_{1 \beta'}^\textnormal{A}  \right).     \label{W1_cylindrical}
\end{eqnarray}
That $p_{\alpha' \beta'}^{K\textnormal{A},1}(\vec q,t)$ and $p_{\alpha' \beta'}^{W\textnormal{A},1}(\vec q,t)$ are equal is trivial since the corresponding Cartesian matrix elements are equal (see Eqn.\ (\ref{pWk1})). Here, we note that the velocity components $u_{1 \varphi}^\textnormal{A}$ and $d_{1 \varphi}^\textnormal{A}$ vanish because of Eqn.\ (\ref{Cylindrical_Divergence_vector}) and the symmetry properties described by Eqns.\ (\ref{SymmetryS}) and (\ref{SymmetryD}): 
\begin{eqnarray}
w_{1 \varphi}^\textnormal{A} &=& \frac{1}{m_\textnormal{A}} \frac{1}{\rho_{1 \textnormal A}} \underbrace{\frac {\partial S}{\partial \varphi_{1 \textnormal A}}}_{= \; 0} \; = \; 0 \; \; \Longrightarrow \; \;  u_{1 \varphi}^\textnormal{A} \;= \;  0, \\
d_{1 \varphi}^\textnormal{A} &=& - \frac{\hbar}{2 m_\textnormal{A}} \frac{1}{D} \frac{1}{\rho_{1 \textnormal A}} \underbrace{\frac {\partial D}{\partial \varphi_{1 \textnormal A}}}_{= \; 0} \; = \; 0. 
\end{eqnarray}
Therefore, any tensor element $p_{\alpha' \beta'}^{X\textnormal{A},1}(\vec q,t)$ vanishes where $\alpha'$ or $\beta'$ is $\varphi$. \newline \newline 
Moreover, the calculation of the Kuzmenkov second-order tensor elements $p_{\alpha' \beta'}^{K\textnormal{A},2}(\vec q,t)$ yields these results:  \newline \newline
For the $\rho \rho $-matrix element: 
\begin{eqnarray}
p_{\rho  \rho }^{K\textnormal{A},2} (\vec q,t) &=& - N(\textnormal{A}) \frac{\hbar^2}{4 m_\textnormal{A}}  \int \textnormal{d} \vec Q \; \delta (\vec q - \vec q_1^{\hspace{0.05cm}\textnormal{A}}) \; \frac{\partial^2 D}{\partial \rho_{1\textnormal{A}}^2}. \label{Kusmenkov rhorho}
\end{eqnarray}
For the $\rho  \varphi $-matrix element:
\begin{eqnarray}
p_{\rho  \varphi }^{K\textnormal{A},2}  &=& 0. \label{Kusmenkov rhovarphi}
\end{eqnarray}
For the $\rho  z $-matrix element:
\begin{eqnarray}
p_{\rho  z }^{K\textnormal{A},2} (\vec q,t) &=&  - N(\textnormal{A}) \frac{\hbar^2}{4 m_\textnormal{A}} \int \textnormal{d} \vec Q \; \delta (\vec q - \vec q_1^{\hspace{0.05cm}\textnormal{A}}) \frac{\partial^2 D}{\partial \rho_{1\textnormal{A}} \partial z_{1\textnormal{A}}}. \label{Kusmenkov rhoz}
\end{eqnarray}
For the $\varphi \varphi $-matrix element:
\begin{eqnarray}
p_{\varphi  \varphi }^{K\textnormal{A},2} (\vec q,t) &=&  N(\textnormal{A}) \frac{\hbar}{2} \int \textnormal{d} \vec Q \; \delta (\vec q - \vec q_1^{\hspace{0.05cm}\textnormal{A}}) \; \frac{D  \hspace{0.05 cm}  d_{1 \rho }^\textnormal{A}}{\rho_{1\textnormal{A}}}. \label{Kusmenkov varphivarphi}
\end{eqnarray}
For the $\varphi  z $-matrix element:
\begin{eqnarray}
p_{\varphi  z }^{K\textnormal{A},2} &=& 0.  \label{Kusmenkov varphiz}
\end{eqnarray}
For the $z  z $-matrix element:
\begin{eqnarray}
p_{z  z }^{K\textnormal{A},2} (\vec q,t) &=& - N(\textnormal{A}) \frac{\hbar^2}{4 m_\textnormal{A}}  \int \textnormal{d} \vec Q \; \delta (\vec q - \vec q_1^{\hspace{0.05cm}\textnormal{A}}) \; \frac{\partial^2 D}{\partial z_{1\textnormal{A}}^2}. \label{Kusmenkov zz}
\end{eqnarray}
Hereby, we do not need to state separate results for the remaining $\varphi \rho $, $z  \varphi $, and $z \rho $-matrix elements because of the symmetry described by Eqn.\ (\ref{symmetry_cylindric_part_2}). \newline \newline 
As the next step, we find that the transformation of the second-order Wyatt Cartesian matrix elements $p_{\alpha \beta}^{W\textnormal{A},2}(\vec q,t)$ into the corresponding cylindrical matrix elements $p_{\alpha' \beta'}^{W\textnormal{A},2}(\vec q,t)$ is trivial: Because of Eqn.\ (\ref{pWK2}),  $\underline{\underline{p}}^{W\textnormal{A},2}(\vec q,t)  = P_\textnormal{A} (\vec q,t) \hspace{0.05 cm} \underline {\underline{1}}$ holds. The unity tensor matrix elements --  in Cartesian coordinates being equal to the Kronecker symbol -- remain equal to this symbol under the transformation into cylindrical coordinates done by Eqn.\ (\ref{Transformationtensor}):
\begin{eqnarray}
1_{\alpha \beta} = \delta_{\alpha \beta} \Longrightarrow 1_{\alpha' \beta'} = \delta_{\alpha' \beta'}.
\end{eqnarray}
Combining this with the context that the scalar quantum pressure $P_\textnormal{A}(\vec q,t)$ does not change in a coordinate transformation yields this result for the matrix elements $p^{W\textnormal{A},2}_{\alpha' \beta'}(\vec q,t)$: 
\begin{eqnarray}
p^{W\textnormal{A},2}_{\alpha' \beta'}(\vec q,t)  &=& P_\textnormal{A} (\vec q,t)  \; \delta_{\alpha' \beta'}. \label{W2_cylindrical}
\end{eqnarray} 
We mention that the quantity $P_\textnormal{A}(\vec q,t)$ does not change itself under a coordinate transformation because it is a scalar field. However, the coordinate transformation from Cartesian to cylindrical coordinates changes as follows how $P_\textnormal{A}(\vec q,t)$ is calculated: \newline
We evaluate $P_\textnormal{A}(\vec q,t)$ from the total particle density $D(\vec Q,t)$ using Eqn.\ (\ref{scalar_pressure}), where the Laplace operator $\laplace_1^\textnormal{A}$ relative to the coordinate $\vec q_1^{\hspace{0.05cm} \textnormal{A}}$ appears. So, we have to regard that in cylindrical coordinates this operator is given by:
\begin{eqnarray}
\laplace_1^\textnormal{A} &=& \frac{\partial^2} {\partial \rho_{1\textnormal{A}}^2} + \frac{1}{\rho_{1\textnormal{A}}} \frac{\partial }{\partial \rho_{1\textnormal{A}}} + \frac{1}{\rho_{1\textnormal{A}}^2} \frac{\partial^2} {\partial \varphi_{1\textnormal{A}}^2} + \frac{\partial^2 } {\partial z_{1\textnormal{A}}^2}. \label{Laplacian_cylindrical_coordinates} 
\end{eqnarray}
$\left. \right.$ \newline  After having calculated the first- and second-order cylindrical elements $p^{X\textnormal{A},1}(\vec q,t)$ and $p^{X\textnormal{A},2}(\vec q,t)$, we calculate the corresponding tensor divergences. For this objective, the gene\-ral equation for calculating the divergence of a tensor field $\nabla \underline{ \underline {T}}(\vec q)$ in cylindrical coordinates has to be evaluated  (\hspace{0.01cm}\cite{Lai_2010}, p.\ 60):
\begin{eqnarray} 
\nabla \underline{\underline{T}}(\vec q) &=& \left[\frac {\partial T_{\rho  \rho }}{\partial \rho}  +  \frac{1}{\rho} \left(\frac{\partial T_{\varphi  \rho }}{\partial \varphi}   + T_{\rho  \rho } - T_{\varphi  \varphi } \right) + \frac{\partial T_{z \rho }}{\partial z} \right] \vec e_{\rho }  \nonumber  \\
&& + \left [  \frac {\partial T_{\rho  \varphi }}{\partial \rho} + \frac{1}{\rho} \left(\frac{\partial T_{\varphi  \varphi }}{\partial \varphi}   + T_{\rho  \varphi } + T_{\varphi  \rho } \right) +  \frac{\partial T_{z \varphi }}{\partial z} \right ] \vec e_{\varphi } \nonumber \\
&& + \left [  \frac {\partial T_{\rho  z }}{\partial \rho} + \frac{1}{\rho} \left(\frac{\partial T_{\varphi  z }}{\partial \varphi}   + T_{\rho  z } \right) +  \frac{\partial T_{z z }}{\partial z} \right ]  \vec e_{z }. \label{tensor_divergences_cylindrical_coordinates} 
\end{eqnarray}
Here, we mention that Andreev and Kuzmenkov analyze in \cite{Andreev_2014a} QHD in cylindrical coordinates, too. However, in their approach, they calculate a tensor divergence of the momentum flow density tensor $\nabla \underline{\underline{\Pi}}$ in cylindrical coordinates by applying the $\nabla$-operator on a tensor component set $\{ \Pi_{\alpha \rho}, \Pi_{\alpha \varphi}, \Pi_{\alpha z} \}$ as if these three components were components of a vector with a parameter $\alpha \in K_{\textnormal{cy}}$, and then they treat the result of this calculation as if it were the $\alpha$-component of the tensor divergence $\nabla \underline{\underline{\Pi}}$. Andreev and Kuzmenkov compensate their error by this approach by introducing in their QHD equations an additonal inertia force. But we think that if one applies Eqn.\  (\ref{tensor_divergences_cylindrical_coordinates}) for calculating tensor divergences instead, it is not necessary to introduce this inertia force. 
\newline \newline 
Using Eqn.\ (\ref{tensor_divergences_cylindrical_coordinates}) and the symmetry $p^{K\textnormal{A},1}_{\beta' \alpha'}(\vec q,t) = p^{K\textnormal{A},1}_{\alpha' \beta'}(\vec q,t)$  for calculating the divergence of the first-order tensors $\underline{\underline{p}}^{X\textnormal{A},1}(\vec q,t)$, we find: 
\begin{eqnarray}
\nabla \underline{\underline{p}}^{K\textnormal{A},1}(\vec q,t) &=& \nabla \underline{\underline{p}}^{W\textnormal{A},1} (\vec q,t) \nonumber \\
                                                  &=& \left(\frac {\partial p_{\rho  \rho }^{W\textnormal{A},1}}{\partial \rho}  +  \frac{1}{\rho} p_{\rho  \rho }^{W\textnormal{A},1} + \frac{\partial p_{\rho z}^{W\textnormal{A},1}}{\partial z}  \right) \vec e_{\rho }  \nonumber \\
                                                  && + \left (  \frac {\partial p_{\rho  z }^{W\textnormal{A},1}}{\partial \rho} + \frac{1}{\rho} p_{\rho  z }^{W\textnormal{A},1}  +  \frac{\partial p_{z z }^{W\textnormal{A},1}}{\partial z} \right )  \vec e_{z }.  \label{tensor_divergences_cylindrical_coordinates_W} 
\end{eqnarray}
In addition, for the divergence of the second-order Kuzmenkov tensor $\underline{\underline{p}}^{K\textnormal{A},2}(\vec q,t)$, we find: 
\begin{eqnarray}
\nabla \underline{\underline{p}}^{K\textnormal{A},2}(\vec q,t) &=&  \left[ \frac {\partial p_{\rho \rho }^{K\textnormal{A},2}}{\partial \rho}  +  \frac{1}{\rho} \left( p_{\rho  \rho }^{K\textnormal{A},2} -  p_{\varphi \varphi }^{K\textnormal{A},2} \right) + \frac{\partial p_{\rho z}^{K\textnormal{A},2}}{\partial z} \right] \vec e_{\rho }  \nonumber \\ 
&& + \left (  \frac {\partial p_{\rho  z }^{K\textnormal{A},2}}{\partial \rho} + \frac{1}{\rho} p_{\rho  z }^{K\textnormal{A},2}  +  \frac{\partial p_{z z }^{K\textnormal{A},2}}{\partial z} \right )  \vec e_{z }.  \label{tensor_divergences_cylindrical_coordinates_W 2}
\end{eqnarray}
When we calculated  $\nabla \underline{\underline{p}}^{K\textnormal{A},2}(\vec q,t)$ using Eqn.\ (\ref{tensor_divergences_cylindrical_coordinates}) and  $p^{K\textnormal{A},2}_{\beta' \alpha'}(\vec q,t)$ $= p^{K\textnormal{A},2}_{\alpha' \beta'}(\vec q,t)$, we regarded
that the derivation $\frac{\partial p_{\varphi  \varphi }^{K\textnormal{A},2} (\vec q,t) }{\partial \varphi}$ vanishes because of the symmetry property (\ref{SymmetryD}). Moreover, for the divergence of the second-order Wyatt tensor $\underline{\underline{p}}^{W\textnormal{A},2}(\vec q,t)$, we initially find this intermediate result: 
\begin{eqnarray}
\nabla \underline{\underline{p}}^{W\textnormal{A},2}(\vec q,t) = \frac{\partial P_\textnormal{A}}{\partial \rho} \vec e_{\rho } + \frac{1}{\rho} \frac{\partial P_\textnormal{A}}{\partial \varphi} \vec e_{\varphi } + \frac{\partial P_\textnormal{A}}{\partial z} \vec e_{z }.  \label{tensor_divergences_scalar_coordinates_P}
\end{eqnarray}
As a next step, we regard that the differential operators $\laplace_1^\textnormal{A}$ and $\frac{\partial}{\partial \varphi_{1\textnormal{A}}}$ commutate -- 
this can be proven trivially by Eqn.\ (\ref{Laplacian_cylindrical_coordinates}) for the Laplace operator $\laplace_1^\textnormal{A}$ in cylindrical coordinates.  
From this context, we conclude that the derivative $\frac{\partial P_\textnormal{A}(\vec q,t)}{\partial \varphi}$ vanishes due to the symmetry property (\ref{SymmetryD}): 
\begin{eqnarray}
\frac{\partial P_\textnormal{A}(\vec q,t)}{\partial \varphi} &=& - \frac{\partial}{\partial \varphi} N(\textnormal{A}) \frac{\hbar^2}{4 m_\textnormal{A}} \int  \textnormal{d} \vec Q \;  \delta (\vec q - \vec q_1^{\hspace{0.05 cm} \textnormal{A}}) \;    \laplace_1^\textnormal{A} D \nonumber \\
                                    &=& - N(\textnormal{A}) \frac{\hbar^2}{4 m_\textnormal{A}} \int  \textnormal{d} \vec Q \;  \delta (\vec q - \vec q_1^{\hspace{0.05 cm} \textnormal{A}}) \; \frac{\partial}{\partial \varphi_{1\textnormal{A}}} \laplace_1^\textnormal{A} D \nonumber \\ 
                                    &=& - N(\textnormal{A}) \frac{\hbar^2}{4 m_\textnormal{A}} \int  \textnormal{d} \vec Q \;  \delta (\vec q - \vec q_1^{\hspace{0.05 cm} \textnormal{A}}) \; \laplace_1^\textnormal{A}  \underbrace{\frac{\partial D}{\partial \varphi_{1\textnormal{A}}}}_{=0}  \nonumber \\
                                    &=& 
0. \label{Laplace_D_derivated} 
\end{eqnarray}
Then, we get this simplified result for $\nabla \underline{\underline{p}}^{W\textnormal{A},2}(\vec q,t)$:
\begin{eqnarray}
\nabla \underline{\underline{p}}^{W\textnormal{A},2}(\vec q,t) &=& \frac{\partial P_\textnormal{A}}{\partial \rho} \vec e_{\rho } + \frac{\partial P_\textnormal{A}}{\partial z} \vec e_{z }.  \label{tensor_divergences_scalar_coordinates_P 2}
\end{eqnarray}
Finally, by adding Eqns.\ (\ref{tensor_divergences_cylindrical_coordinates_W}) and (\ref{tensor_divergences_cylindrical_coordinates_W 2}), we find for the divergence of the Kuzmenkov pressure tensor $\nabla \underline{\underline{p}}^{K\textnormal{A}}(\vec q,t)$: 
\begin{eqnarray}
\nabla \underline{\underline{p}}^{K\textnormal{A}}(\vec q,t) &=&  \nabla \underline{\underline{p}}^{K\textnormal{A},1}(\vec q,t) + \nabla \underline{\underline{p}}^{K\textnormal{A},2}(\vec q,t) \nonumber \\ 
&=& \left[ \frac {\partial p_{\rho  \rho }^{W\textnormal{A},1}}{\partial \rho}  +  \frac{1}{\rho} p_{\rho  \rho }^{W\textnormal{A},1} + \frac{\partial p_{\rho z }^{W\textnormal{A},1}}{\partial z}  \right. \nonumber  \\
&& \left. + \; \frac {\partial p_{\rho  \rho }^{K\textnormal{A},2}}{\partial \rho} + \frac{1}{\rho} \left(  p_{\rho  \rho }^{K\textnormal{A},2} - p_{\varphi \varphi }^{K\textnormal{A},2} \right) + \frac {\partial p_{\rho  z }^{K\textnormal{A},2}}{\partial z} \right] \vec e_{\rho }  \nonumber \\
                                               && + \left (  \frac {\partial p_{\rho  z }^{W\textnormal{A},1}}{\partial \rho} + \frac{1}{\rho} p_{\rho  z }^{W\textnormal{A},1} + \frac{\partial p_{z z }^{W\textnormal{A},1}}{\partial z} \right. \nonumber \\
                                               && \left. \; \; \; +  \; \frac {\partial p_{\rho  z }^{K\textnormal{A},2}}{\partial \rho} + \frac{1}{\rho} p_{\rho  z }^{K\textnormal{A},2} +  \frac {\partial p_{zz}^{K\textnormal{A},2}}{\partial z} \right ) \vec e_{z}, \label{Result Nabla p Kusmenkov}
\end{eqnarray}
and by adding Eqns.\ (\ref{tensor_divergences_cylindrical_coordinates_W}) and (\ref{tensor_divergences_scalar_coordinates_P 2}), we find for the divergence of the Wyatt pressure tensor $\nabla \underline{\underline{p}}^{W\textnormal{A}}(\vec q,t)$:
\begin{eqnarray}
\nabla \underline{\underline{p}}^{W\textnormal{A}}(\vec q,t) &=&  \nabla \underline{\underline{p}}^{W\textnormal{A},1}(\vec q,t) + \nabla \underline{\underline{p}}^{W\textnormal{A},2}(\vec q,t) \nonumber \\
                                                &=& \left( \frac {\partial p_{\rho  \rho }^{W\textnormal{A},1}}{\partial \rho}  +  \frac{1}{\rho} p_{\rho  \rho }^{W\textnormal{A},1} + 
                                                    \frac{\partial p_{\rho z}^{W\textnormal{A},1}}{\partial z}  + \frac {\partial P_\textnormal{A} }{\partial \rho}  \right) \vec e_{\rho } \nonumber \\
                                                &&  \left (  \frac {\partial p_{\rho  z }^{W\textnormal{A},1}}{\partial \rho} + \frac{1}{\rho} p_{\rho  z }^{W\textnormal{A},1} + 
                                                    \frac{\partial p_{z z }^{W\textnormal{A},1}}{\partial z} + \frac {\partial P_\textnormal{A} }{\partial z}  \right) \vec e_{z}.  \label{Result Nabla p Wyatt}
\end{eqnarray}
As a result, both  $\nabla \underline{\underline{p}}^{K\textnormal{A}}(\vec q,t)$ and $\nabla \underline{\underline{p}}^{W\textnormal{A}}(\vec q,t)$ have no $\vec e_\varphi$-component due to the symmetry properties described by Eqns.\ (\ref{SymmetryS}) and (\ref{SymmetryD}). Apart from this identical property of $\nabla \underline{\underline{p}}^{W\textnormal{A}}(\vec q,t)$ and $\nabla \underline{\underline{p}}^{K\textnormal{A}}(\vec q,t)$, for numerical applications -- where $\nabla \underline{\underline{p}}^{W\textnormal{A}}(\vec q,t)$ or $\nabla \underline{\underline{p}}^{K\textnormal{A}}(\vec q,t)$, respectively, are input quantities in the MPQCE (\ref{MQNSE for particle sort k}) -- the use of the Wyatt pressure tensor $\underline{\underline{p}}^{W\textnormal{A}}(\vec q,t)$ is advantageous: \newline
The reason for this is that the evaluation equation (\ref{tensor_divergences_scalar_coordinates_P 2}) for the calculation of the second-order part $\underline{\underline{p}}^{W\textnormal{A},2}(\vec q,t)$ of the Wyatt pressure tensor is more compact and less complicated than the corresponding Eqn.\ (\ref{tensor_divergences_cylindrical_coordinates_W 2}) for the calculation of the second-order part $\underline{\underline{p}}^{K\textnormal{A},2}(\vec q,t)$ of the Kuzmenkov pressure tensor.
So, the Wyatt pressure tensor is not only easier to interpret physically than the Kuzmenkov pressure tensor, but it is easier to apply numerically, too. \newline \newline 
As the last point in this section, we mention that cylindrical coordinate matrix elements for the parts $\underline{\underline{\Pi}}^{X\textnormal{A},1}(\vec q,t)$ and $\underline{\underline{\Pi}}^{X\textnormal{A},2}(\vec q,t)$ of the momentum flow density tensor $\underline{\underline{\Pi}}(\vec q,t)$ can be derived in an analogous manner like for the parts $\underline{\underline{p}}^{X\textnormal{A},1}(\vec q,t)$ and $\underline{\underline{p}}^{X\textnormal{A},2}(\vec q,t)$  of the pressure tensor  $\underline{\underline{p}}(\vec q,t)$: \newline 
By comparing Eqns.\ (\ref{Kusmenkov_K2}) and (\ref{Kusmenkov_K3}) for the Cartesian matrix elements $\Pi^{X\textnormal{A},1}_{\alpha \beta}(\vec q,t)$ with Eqn.\ (\ref{pWk1}) for the Cartesian matrix elements  $p^{X\textnormal{A},1}_{\alpha \beta}(\vec q,t)$, it becomes evident that the formula for the cylindrical coordinate matrix elements $\Pi^{X\textnormal{A},1}_{\alpha' \beta'}(\vec q,t)$ is found by substituting the vector components of the relative velocity $\vec u_1^{\textnormal{A}}(\vec Q,t)$ by the corresponding vector components of the velocity $\vec w_1^{\textnormal{A}}(\vec Q,t)$ in Eqn.\ (\ref{W1_cylindrical}) for the cylindrical coordinate matrix elements $p^{X\textnormal{A},1}_{\alpha' \beta'}(\vec q,t)$: 
\begin{eqnarray}
\Pi_{\alpha' \beta'}^{K\textnormal{A},1}(\vec q,t) &=& \Pi_{\alpha' \beta'}^{W\textnormal{A},1}(\vec q,t) \nonumber \\
&=& N(\textnormal{A}) \int \textnormal{d}Q \; \delta (\vec q - \vec q_1^{\hspace{0.05 cm} \textnormal{A}}) \; D \; m_\textnormal{A} \left ( w_{1 \alpha'}^\textnormal{A} w_{1 \beta'}^\textnormal{A} + d_{1 \alpha'}^\textnormal{A} d_{1 \beta'}^\textnormal{A}  \right).     \label{Pi_W1_cylindrical}
\end{eqnarray}
In addition, because of Eqns.\ (\ref{Pi_part_2_Kusmenkov}) and (\ref{Pi_part_2_Wyatt}), it holds 
\begin{eqnarray}
\Pi_{\alpha \beta}^{X\textnormal{A},2}(\vec q,t) &=& p_{\alpha \beta}^{X\textnormal{A},2}(\vec q,t).
\end{eqnarray}
So, the cylindrical coordinate matrix elements $\Pi_{\alpha' \beta'}^{X\textnormal{A},2}(\vec q,t)$ are given by 
\begin{eqnarray}
\Pi_{\alpha' \beta'}^{X\textnormal{A},2}(\vec q,t) &=& p_{\alpha' \beta'}^{X\textnormal{A},2}(\vec q,t).
\end{eqnarray}
Thus, to calculate $\Pi_{\alpha' \beta'}^{X\textnormal{A},2}(\vec q,t)$, we can just use the results which we derived above for $p_{\alpha' \beta'}^{X\textnormal{A},2}(\vec q,t)$. Therefore, the use of the Wyatt momentum flow density tensor $\underline{\underline{\Pi}}^{W\textnormal{A}}(\vec q,t)$ is advantageous compared to the use of the Kuzmenkov momentum flow density tensor $\underline{\underline{\Pi}}^{K\textnormal{A}}(\vec q,t)$ because the second-order 
part $\underline{\underline{\Pi}}^{W\textnormal{A},2}(\vec q,t)$ of the Wyatt tensor can be calculated more easily than the corresponding second-order part 
$\underline{\underline{\Pi}}^{K\textnormal{A},2}(\vec q,t)$ of the Kuzmenkov tensor. 
\section{Summary}
In this paper, we derived MPQHD in detail for an exact wave function describing an ensemble of several particle sorts. For this task, we first derived the MPCE related to the conservation of mass for each of the particle sorts. One can also derive an MPCE for the total particle ensemble by summing up the MPCEs for all the different sorts of particles. Moreover, we derived for each sort of particles two different equations of motion. The first one of these equations of motion is the MPEEM; it describes the temporal change of the mass flux density of the particles of the analysed sort -- and one can derive the MPEEM by applying the Ehrenfest theorem for the calculation of this temporal change. The second one of these equations of motion is the MPQCE; it is closely related to Cauchy's equation of motion, which is well-known in classical hydrodynamics and is related to the momentum balance in fluids. The MPEEMs for the different sorts of particles are linear differential equations, so one can get an MPEEM for the total particle ensemble just by adding up the MPEEMs for all the different sorts of particles. The MPQCEs for the different sorts of particles are non-linear, so adding up these equations does not lead to an MPQCE for the total particle ensemble. However, a derivation of an MPQCE for the total particle ensemble is still possible. \newline
In all the MPQCEs, both for a certain sort of particles and for the total particle ensemble, a quantity appears which is called the divergence of the pressure tensor. Similar to a potential, this pressure tensor is not defined uniquely. For an MPQCE related to a certain sort of particles, the properties of two different versions of this tensor are discussed: The first one is named the ``Wyatt pressure tensor'' because of the form of the momentum flow density tensor, which is another tensor closely connected to the pressure tensor, in \cite{Wyatt_2005}, p.\ 31. The second one is named the ``Kuzmenkov pressure tensor'' because it appears in \cite{Kuzmenkov_1999}. The terms contributing to the Wyatt pressure tensor can be interpreted physically better than the Kusmenkov pressure tensor. Moreover, we made a coordinate transformation of both tensor versions from Cartesian coordinates to cyclindrical coordinates and calculated the tensor divergence for both versions in cyclindrical coordinates. This calculation can be performed more easily for the Wyatt pressure tensor than for the Kusmenkov pressure tensor because a certain summand contributing to the Wyatt pressure tensor is just a scalar multiplied by the diagonal unit tensor, while the according summand contributing to the Kusmenkov pressure tensor is a full tensor with non-diagonal elements. \newline
In addition, in all the MPEEMs, a quantity called the divergence of the momentum flow density tensor appears, and for an MPEEM related to a certain sort of particles, we introduce both a Kuzmenkov version and a Wyatt version of this tensor. We analyzed these two versions of the momentum flow density tensor  
in an analogous manner like the two versions of the pressure tensor mentioned above. The results of the analysis of the Kuzmenkov and the Wyatt momentum flow density tensors are just analogous to that of the two corresponding pressure tensors -- so, the Wyatt momentum flow density tensor is more easily to interpret and to apply than the Kuzmenkov momentum flow density tensor. \newline 
These results show that the right choice of the pressure tensor can simplify quantum hydrodynamic calculations, and researchers doing quantum hydrodynamics should regard this point.  
\section*{Acknowledgement}
We acknowledge helpful discussions with T. AlBaraghtheh and her hints to interesting references. 

\end{document}